\documentclass{article}

\usepackage{lineno}

\usepackage{ulem}       % for crossed-put text with \sout{}

\usepackage[verbose]{placeins} 		% für Befehl \floatbarrier zur Verfügung. Dieser erzwingt die  Positionierung aller Gleitobjekte, die noch nicht platziert wurden. Erst anschließent geht es mit dem Fließtext, der nach der floatbarrier steht, weiter. [verbose] Schreibt viele zusätzliche Informationen in die Log-Datei

\usepackage{hyperref} % for \pdfbookmark and hyperlinks
\hypersetup{
    colorlinks=false, % Do not color the text
    pdfborder={0 0 1}, % Border width for links
    pdfborderstyle={/S /U /W 1}, % Underline style
    linktocpage, % underline pages instead of headings in tab of concents
}

\usepackage[a4paper, left=2.5cm, right=2.5cm, top=3cm, bottom=4cm]{geometry}
\usepackage{needspace} % for extra space
\usepackage{fancyhdr} % for custom headers and footers
\usepackage[square,numbers,sort&compress]{natbib}  % numbers, sorted and compressed citations
\usepackage{amssymb} % provides various useful mathematical symbols
\usepackage{amsmath} % provides various useful equation environments
\usepackage{amssymb} % for the set of natural, rational,... numbers. command e.g. \mathbb{N}

\usepackage{lmodern} % Zur Verwendung von Sonderzeichen (z.B allgemeine Gaskonstante)%%%früher:dsfont

\allowdisplaybreaks	% ermöglicht Seitenumbrüche bei langen align-Umgebungen

\usepackage{bm}     % Fette Buchstaben im Mathe-Modus mit \bm
\usepackage{mathtools} 		% für := (Definition) mittels \coloneqq

\newcommand{\mr}{\mathrm}		% shortcut für normale Schrift im Mathe-Modus
\newcommand{\dd}{\mr{d}}		% gerade geschriebenes d im Integral
\newcommand{\pd}{\partial}	% kursiv geschriebenes d in partiellen Ableitungen
\usepackage{siunitx}
\DeclareSIUnit\bar{bar}     % muss neu definitert werden, weil bar keine SI-Einheit mehr ist

\usepackage{float}	% für Gleitumgebung und erzwingt Positionierung von Floatobjekten mit "`H"'
\usepackage{graphicx}   % für Einbindung von Bildern
\usepackage[figuresright]{rotating}	% um Bilder, Tabellen, usw. in landscape Format anzuzeigen
\usepackage{subcaption} % für subfigures

\usepackage[labelfont=bf]{caption} 	% für fette Labels in float-Objekten

\usepackage{pdfpages} % to import PDFs with \includepdf
\usepackage{svg} % to import SVGs with \includesvg{image}

\usepackage{pgfplots} % for tikZ graphics
\pgfplotsset{compat=newest}
\usepackage{tikz}
\pgfplotsset{plot coordinates/math parser=false}

\usepackage{changepage} % for adjustwidth

\usepackage{longtable}	
\usepackage{supertabular}  % ermöglicht das Erzeugen von Tabellen die länger als eine Seite sind
\usepackage{tabularx}  % ermöglicht Tabellen mit automatischen Zeilenumbrüchen, wenn ein Zeileneintrag zu lang wird. Dafür eine Spalte mit X (statt l, r oder c) erstellen
\usepackage[para]{threeparttable}

\usepackage{multirow}   % mehrere Zeilen in Tabellen zusammenfassen

\usepackage{enumitem}
\usepackage{makecell} 		% für Zeilenumbrüche in Tabellen-Zellen
\usepackage{tablefootnote} 	% für Fußnoten in Tabellen. 
\usepackage{pdflscape} 		% für Querformat

\usepackage{booktabs}		% für \toprule und \bottomrule in Tabellen
\usepackage{caption}        % for better caption handling
\usepackage{boldline}       % for thick table grid lines
\usepackage{arydshln} 		% für gepunktete Linien in Tabellen	
\newcolumntype{C}[1]{>{\centering\arraybackslash}p{#1}} % centered with given width
\usepackage{dcolumn}
\newcolumntype{d}[1]{D{.}{.}{#1}} % alignment for decimal separator with specified number of decimal places; column definition e.g. d{2} for 2 decimal places
\newcolumntype{Y}{>{\centering\arraybackslash}X} % for stretching columns

\usepackage{xspace} 	% Abstände nach selbst definierten Befehlen können richtig gestaltet werden
\usepackage{ragged2e} 	% for \justifying command

\newcommand{\cotwo}{CO\textsubscript{2}\xspace}		% CO2
\newcommand{\chfour}{CH\textsubscript{4}\xspace}	% CH4
\usepackage{xcolor} % for colored text using e.g. \color{gray}

\usepackage{url}
\usepackage{xurl} % for long URLs that should be line-broken

\usepackage{lipsum} % for dummy text using \lipsum

\usepackage{hyphenat} % für angepasste Silbentrennung
\begin{document}

% adjust footer and header
\pagestyle{fancy}
\fancyhf{} % Clear default header and footer
\renewcommand{\headrulewidth}{0pt} % Remove the header rule
\fancyhead[L]{Extended preprint of a tutorial article in Journal of Process Control} % Left side of the header

\begin{center}
\LARGE A Tutorial to Multirate Extended Kalman Filter Design for Monitoring of Agricultural Anaerobic Digestion Plants \\ [5mm]

\large
Simon Hellmann\textsuperscript{1,2}, Terrance Wilms\textsuperscript{3}, Stefan Streif\textsuperscript{2}, Sören Weinrich\textsuperscript{4,1} \\ [3mm]

\normalsize
\textsuperscript{1} DBFZ Deutsches Biomasseforschungszentrum gGmbH, Biochemical Conversion Department, Leipzig, Germany. simon.hellmann@dbfz.de\\ [3mm]

\textsuperscript{2} Chemnitz University of Technology, Professorship for Automatic Control and System Dynamics, Chemnitz, Germany. stefan.streif@etit.tu-chemnitz.de\\ [3mm]

\textsuperscript{3} Technische Universität Berlin, Chair of Control, Berlin, Germany. terrance.wilms@tu-berlin.de \\ [3mm]

\textsuperscript{4} FH Münster University of Applied Sciences, Faculty of Energy, Building Services, Environmental~Engineering, Steinfurt, Germany. weinrich@fh-muenster.de
\end{center}

\vspace{1cm}

\setlength{\parindent}{0pt}
\Large
\copyright 2026. This is an extended preprint of a tutorial article in Journal of Process Control. Submitted on December 15, 2025, accepted on March 17, 2026. This preprint version is made available under a CC BY-NC-ND 4.0 license.

\vskip 1cm

The final published version is available at \url{https://doi.org/10.1016/j.jprocont.2026.103703}.

\normalsize

\setlength{\parindent}{15pt} % restore default indent
\clearpage

\section*{Highlights}
\begin{itemize}
\item Comprehensive derivation of sample state augmentation for delay handling
\item Application to a nonlinear anaerobic digestion (AD) model
\item Demand-driven operating scenario for full-scale agricultural AD plant
\item Investigation of delay, plant-model mismatch, and initialization on error convergence 
\item Systematic filter tuning for reliable state estimation
\end{itemize}
\section*{Abstract}
In many applications of biotechnology, measurements are available at different sampling rates, e.g., due to online sensors and offline lab analysis. Offline measurements typically involve time delays that may be unknown \textit{a priori} due to the underlying laboratory procedures. This multirate (MR) setting poses a challenge to Kalman filtering, where conventionally measurement data is assumed to be available on an equidistant time grid and without delays. 
This tutorial paper derives the MR version of an extended Kalman filter (EKF) based on sample state augmentation, and applies it to the anaerobic digestion (AD) process in a simulative agricultural setting. The performance of the MR-EKF is investigated for various scenarios including varying delay lengths, measurement noise levels, plant-model mismatch (PMM), and initial state error. Provided with an adequate tuning, the MR-EKF can reliably estimate the process state and, thus, appropriately fuse the delayed offline measurements and smooth the noisy online measurements. 

Because of the sample state augmentation approach, the delay length of offline measurements does not critically effect the performance of the state estimation, provided that observability is not lost during the delays. Poor state initialization and PMM affect convergence more than measurement noise levels. Furthermore, selecting an appropriate tuning was found to be critically important for successful application of the MR-EKF for which a systematic approach is presented. 

This tutorial provides implementation guidance for practitioners seeking to successfully apply state estimation for multirate systems. Thus, it contributes to the development of demand-driven operation of biogas plants, which may aid in stabilizing a renewable electricity grid. 
\vspace{1cm}

\textbf{Keywords:} Soft Sensor, ADM1, State Observer, Tuning, Biogas Technology, Flexibilization

\clearpage

\setcounter{page}{1}
\fancyfoot[L]{\thepage} % Page number on the left side of the footer

%% main text
\section{Introduction}
Anaerobic digestion (AD) metabolizes biogenic feedstocks into biogas and organic digestate. Biogas can be converted to electricity and heat or upgraded as a substitute for natural gas \cite{Theuerl2019}, while the digestate can be used as nutrient-rich fertilizer \cite{Steindl.2025} or further upgraded for nutrient recovery \cite{Petrovic.2021}. Demand-driven operation of AD plants can compensate fluctuating renewable energy sources like wind and solar, and thereby stabilize the electricity grid \cite{Purkus.2018}. While flexible feeding poses a financially viable opportunity for demand-driven AD operation \cite{Mauky.2017}, it remains a challenge to ensure stable and efficient process conditions \cite{Gaida2017,Kazemi2020}. Thus, sudden changes in substrate composition and significant overload can inhibit the microbial community and cause process instabilities and failure \cite{Jimenez2015,Theuerl2019}. Therefore, reliable process monitoring is required for demand-oriented biogas production using dynamic feeding. 

There exists extensive literature on monitoring of biological systems \cite{Dochain.2003,Lyubenova.2021}, and especially AD systems \cite{Jimenez2015,Cruz2021}, with a spectrum of methods that range from simple to very complex. Examples of simple algorithms are simulation-based observers \cite{Raeyatdoost2019}, linear observers \cite{Schaum.2015}, and Luenberger observers \cite{AlcarazGonzalez2021}. Kalman filters (KF) strike a balance between model-based predictions and data-based corrections. KF have been shown to deliver satisfactory state estimation accuracy given good model calibration and an adequate tuning \cite{Tuveri.2021,Haugen2014,Raeyatdoost.2023}. More complex algorithms include diverse nonlinear approaches that are usually tied to a specific model or a certain class of models \cite{Li.2004,Tatiraju.1999}, such as high-gain observers \cite{MendezAcosta.2005}, adaptive observers \cite{Dochain.2003},  asymptotic observers \cite{Bastin.1990, Petre.2020}, and super-twisting observers \cite{Sbarciog.2014}. Furthermore, moving horizon estimation (MHE) defines an optimization problem based on a (non)linear process model, which is repeatedly solved over a fixed-length window into the past \cite{Kim.2023}. Depending on the window size and the consideration of measurement and process noise, MHE usually involves a high computational load \cite{Elsheikh.2021}. Conversely, soft sensors based on machine learning were shown to deliver promising results for monitoring of AD systems when provided with appropriate training datasets \cite{Gaida2012,Kazemi2020}. However, their validity is generally limited to the training data, which impairs generalization and extrapolation \cite{Ewering.62520246282024}. Finally, data-driven approaches based on whitening filtering and/or correlation analysis can be used to detect process anomalies, and thereby provide pragmatic monitoring schemes for multiple input multiple output (MIMO) processes in open and closed-loop \cite{Chen.2021,Huang.1997}.

Given the usually scarce measurement availability at full-scale AD plants, approaches based on KF were found to deliver an adequate compromise between estimation accuracy and implementation complexity \cite{RochaCozatl.2015}. Kalman filtering provides a pragmatic and universally applicable approach that is not limited to a specific plant model such as many of the tailor-made nonlinear approaches \cite{Lyubenova.2021}. Common extensions to the conventional linear KF include the extended Kalman filter (EKF), which linearizes a nonlinear process model around the currently estimated operating points \cite{RochaCozatl.2015,Tuveri.2021}, or the unscented Kalman filter (UKF), in which the nonlinearities are captured using sigma points \cite{Haugen2014,Raeyatdoost2019,Tuveri.2021,Hellmann.62520246282024}. 

The AD process, like many processes in biotechnology, involves measurements available at different time resolutions. These classes of systems are called multirate (MR) systems. While measurements like pH or gas production can be obtained online (at high time resolution with sample times ranging from seconds to minutes), others such as volatile fatty acids (VFA) require sampling and subsequent laboratory analysis (sample times ranging from days to weeks). They are, thus, available at irregular intervals and with \textit{a priori} unknown delays. 

Different approaches to adapt KF to a MR scenario have been proposed \cite{Gudi.1995,Durr.2021,Zhao.2015} and compared in simulative case studies \cite{Gopalakrishnan.2011}. The available approaches include filter recalculation upon return of the delayed offline measurements; the method of Alexander \cite{Alexander.1991}, which applies for linear KF; and state augmentation methods such as fixed-lag smoothing \cite{Simon.2006} and sample-state augmentation \cite{Gopalakrishnan.2011}. Among these, sample-state augmentation could be identified as the best compromise between estimation accuracy and computational effort \cite{Gopalakrishnan.2011}. However, the literature still lacks a comprehensive description on the implementation details and systematic investigation of the parameters that influence the sample-state augmentation for KF. 

The quality of model-based monitoring algorithms strongly depends on a suitable process model \cite{Lyubenova.2021}. This not only pertains to the model's accuracy and the appropriate calibration of model parameters involved \cite{Segura.2025} but especially relates to the model's observability, which is determined by available measurement signals \cite{Villaverde.2019,Hellmann.2023}.

AD is a highly nonlinear biological process, which has been modeled in various ways in the literature \cite{Kegl.2025,Ficara2012}. While there exist complex and accurate process models of the AD process such as the established Anaerobic Digestion Model No.~1 (ADM1) \cite{Batstone2002} and its extensions \cite{DonosoBravo2022,Carecci.2024}, many of them are unsuitable for monitoring and control purposes due to the multitude of tunable parameters \cite{Segura.2025}, which contrasts with the limited measurement availability in full scale \cite{Jimenez2015}, as well as the lack of observability \cite{Nimmegeers.2017,Hellmann.2023}. 

Simpler models have been developed specifically for monitoring and control applications \cite{Angelidaki.1993,Rodriguez.2008}. The most prominent AD model used in this context has been proposed by Bernard et al. \cite{Bernard2001}, which has been applied to state estimation \cite{Raeyatdoost.2023} and control of AD \cite{Ahmed.2020}. More recently, Weinrich and Nelles \cite{Weinrich2021b} systematically simplified the complex ADM1 and transformed it into five different mass-based variations. Notably, these simplifications maintain the fundamental modeling procedures and underlying stoichiometry of the original ADM1 \cite{Weinrich2021}. At the same time, they can be applied to model-based monitoring and control in an agricultural setting thanks to their lower system order and mass-based reference unit \cite{Mauky.2016,Tisocco.2024}. 

This tutorial derives the multirate extended Kalman filter (MR-EKF) based on sample-state augmentation for single and multiple augmentations, as well as out-of-sequence measurements, i.e., measurements returning in a different sequence than sampled. The MR-EKF is first systematically tuned and then applied to a mass-based ADM1 simplification. The complete workflow of MR-EKF application is covered in a case study relevant to practical operation of agricultural AD plants. State estimation performance of the MR-EKF is systematically discussed by investigating its sensitivity with respect to the MR-EKF parameters. This tutorial thus provides valuable guidance on critical implementation details of sample-state augmentation, and derives instructive advice for monitoring of biological systems, and especially AD.

\section{Materials and Methods}
\label{sec:methods}
The following section first defines the class of MR systems investigated, gives a brief summary of the conventional EKF, and describes the MR EKF based on sample-state augmentation. Afterwards, the AD model investigated and the case study considered are introduced. Finally, MR-EKF tuning and implementation details are provided.
\subsection{Class of systems investigated}
\label{sec:meth:class_of_systems}
The class of systems that will be investigated is described by a set of continuous-time ordinary differential equations (ODE) and discrete-time measurements in variable structure as defined in \cite{Kramer.2005b}
\begin{subequations}
\label{eq:system_model}
\begin{align}
    \dot x &= f(x(t),u(t),\theta) + w(t), \qquad x_0 =x(t_0), \label{eq:system_model_ode} \\
    y_k &= \begin{cases} \quad h^\mathrm{on}(x_k, \theta) + v^\mathrm{on}_k, & \,\, k \ne r,\\
    \left[\begin{array}{l} h^\mathrm{on}(x_k,\theta) + v^\mathrm{on}_k \\ h^\mathrm{off}(x_s,\theta) + v^\mathrm{off}_s
    \end{array}\right], & \,\, k = r, \\
    \end{cases}
    \label{eq:system_model_output}
\end{align}
\end{subequations}
with states\footnote{All variables used in this tutorial are summarized in the nomenclature section at the end of the manuscript.} $x \in \mathbb{R}^{n}$, control inputs $u \in \mathbb{R}^{n_\mathrm{u}}$ and time-varying\footnote{Time-invariant model parameters are described in Sec.~\ref{sec:meth:model_parameters}.} parameters $\theta \in \mathbb{R}^{n_\mathrm{\theta}}$. It further holds that $x_k = x(t_k)$ with $t_k = k \Delta t, k\ge 0$ and online sample time $\Delta t$. % the Since time and measurement updates occur only at discrete time instances, these instances are referred to as time}  
The nominal model is described by the dynamics $f$ and the output function $h$. Measurements are subdivided into online measurements $y^\mathrm{on}\in \mathbb{R}^{q_\mathrm{on}}$, and offline measurements $y^\mathrm{off}\in \mathbb{R}^{q_\mathrm{off}}$, denoted with superscript indices on and off, respectively. Online measurements are available instantaneously. Conversely, offline measurements are sampled at time $s$ and delayed by $N_\mathrm{d}$ time steps. They are thus only available at the return time $r = s + N_\mathrm{d}$, while the time delay is generally variable and unknown \textit{a priori}.\footnote{The delay length is implicitly limited by practical, numerical constraints, which are addressed in Sec.~\ref{sec:res:delay}.}

\begin{figure*}
    \centering
    \includegraphics[width=0.9\linewidth]{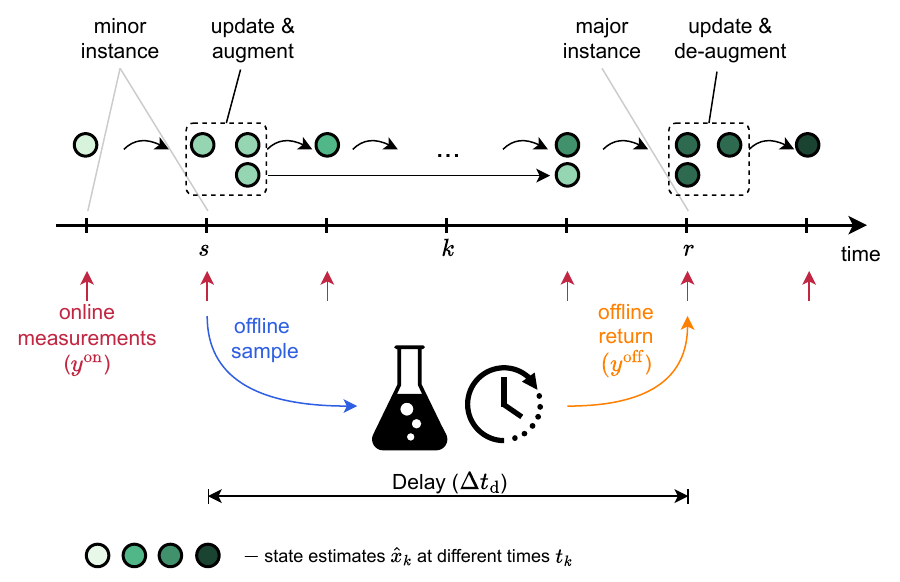}
    \caption{Time grid of multirate measurements (Sec.~\ref{sec:meth:class_of_systems}) and sample-state augmentation of the MR-EKF (Sec.~\ref{sec:meth:ss_augmentation}).}
\label{fig:meth:MR-EKF_timegrid}
\end{figure*}

Process and measurement noise $w$ and $v$ are assumed as additive, zero-mean, stationary Gaussian noise processes
\begin{subequations}
\begin{align}
    E\left\{w(t)\right\}&=0, \quad E\{w(t)\, w(\tau)^T\}=Q(t)\, \delta (t-\tau), \\
    E\left\{v_k\right\}&=0, \quad E\{v_k\, v_l^T\}=R_k \, \delta_{k,l}, 
% w \sim \mathcal{N}(0,Q) \\ v_k\sim \mathcal{N}(0,R_k)
\end{align}
\end{subequations}
with the Dirac and Kronecker operators $\delta(t)$ and $\delta_{k,l}$, measurement noise covariance matrix $R_k$ and spectral density matrix of the process noise $Q(t)$. 
% 
% \begin{subequations}
% \begin{align}
%     y_k^\mathrm{on} &= h^\mathrm{on}(x_k) + v_k^\mathrm{on} \\
%     y_r^\mathrm{off} &= h^\mathrm{off}(x_s) + v_s^\mathrm{off}, \quad \text{if} \quad r = s + N_\mathrm{d}.
% \end{align}
% \label{eq:meth:yOnOff}
% \end{subequations}
% 
Similar to Eq.~\eqref{eq:system_model_output}, both measurement noise vector and corresponding covariance matrix $R_k$ (usually diagonal) are subdivided, with dimensions of zero and identity matrices in subscript parentheses %into corresponding time-invariant block matrices: 
\begin{subequations}
\label{eq:meth:partitioning_noise_and_R}
\begin{align}
    v_k &= \begin{cases} \quad v^\mathrm{on}_k, & k \ne r, \\
    \left[\begin{array}{l} v^\mathrm{on}_k \\ v^\mathrm{off}_s
    \end{array}\right], & k = r, \\
    \end{cases} \\ % 
    R_k &= \begin{cases} \hspace{7.3mm} R^\mathrm{on}_k, & k \ne r, \\
    \left[\begin{array}{cc} R^\mathrm{on}_k & 0_{(q_\mathrm{on} \times q_\mathrm{off})} \\ 0_{(q_\mathrm{off} \times q_\mathrm{on})} & R^\mathrm{off}_s
    \end{array}\right], & k = r. \\
    \end{cases}
\end{align}
\end{subequations}
It is assumed that both sampling and return of offline measurements take place at discrete time steps of the online time grid, as illustrated in Fig.~\ref{fig:meth:MR-EKF_timegrid}. 

\subsection{Multirate extended Kalman filter}
The EKF can be adapted in multiple ways to deal with MR measurements, as described by Gopalakrishnan et al. \cite{Gopalakrishnan.2011}. These authors identified sample-state augmentation as the best compromise between estimation accuracy and computational effort. Sample-state augmentation was first proposed by van der Merwe \cite{vanderMerwe.2004b} and is described in more detail by Zhao et al. \cite{Zhao.2015} for a cubature Kalman filter (CKF). More recently, it as been applied to an UKF \cite{Kemmer.2023} and is also used in this tutorial. The necessary modifications specifically for the EKF are described in Sec.~\ref{sec:meth:ss_augmentation}, which is preceded by a brief recapitulation of the continuous-discrete EKF for the singlerate case. 
\subsubsection{Continuous-discrete extended Kalman filter}
\label{sec:EKF_singlerate}
The continuous-discrete EKF considers a continuous-time formulation of the dynamics with measurements available at discrete times. 
%linearizes the nonlinear model equations \eqref{eq:system_model} around the current state estimate. 
% $\hat x_{k|k}$ denotes the estimate of the true state $x_k$ given measurements up to time $k$. 
$\hat x_{k}$ denotes the estimate of the true state $x_k$ after incorporating measurements up to time $k$. The corresponding symmetric covariance matrix of the estimation error is denoted as $P_{k}$ (referred to as the $P$-matrix in the following). At the initialization of the EKF, the following holds:
\begin{equation}
    \hat x_{0} = E\left\{x_0\right\}, \quad P_{0} = E\left\{\left(x_0 - \hat x_{0}\right) \left(x_0 - \hat x_{0}\right)^T\right\} .
\end{equation}
% 
% Schneider and Georgakis \cite{Schneider.2013} proposed a pragmatic tuning of the initial conditions.
Both state estimate and $P$-matrix are updated recursively in two steps: time update and measurement update. 

During the time update, the estimate based on the last measurements $\hat x_{k-1}$ %$= \hat x(t_{k-1}^+)$ 
is propagated to the next time step~$k$ and delivers the \textit{a priori} estimate $\hat x_{k}^-$ %$= \hat x(t_k^-)$ 
before incorporation of the new measurements. For brevity, \textit{a priori} and \textit{a posteriori} estimates are referred to as the \textit{priors} and \textit{posteriors}. The propagation for $x$ is carried out according to the nominal model, while process noise is considered for the $P$-matrix
\begin{subequations}
\begin{align}
    \dot{\hat x} (t) &= f(\hat x(t),u(t),\theta), \label{eq:meth:time_update_x_single_rate} \\
    % 
    % \dot P (t) &= F\left(\hat x(t)\right) P(t) + P(t) F^\mathrm{T}\left(\hat x(t)\right) + Q(t) 
    \dot P (t) &= F(t) P(t) + P(t) F^{T}(t) + Q(t) \label{eq:meth:time_update_P_single_rate}.
\end{align}
\end{subequations}
$F$ is the linearization of the ODE \eqref{eq:system_model_ode} around the current state estimate\footnote{$F$ actually depends on $x, u, \theta, c, $ and $a$. For simpler notation, only time-dependence is stated and all other independent variables are suppressed.}
\begin{equation}
    % F\left(\hat x(t)\right) = \left. \frac{\pd f}{\pd x} \right|_{\hat x(t),u(t),t}.
    F(t) = \left. \frac{\pd f}{\pd x} \right|_{\hat x(t),u(t),\theta}.
    \label{eq:meth:F}
\end{equation}
This results in the following priors for the state and covariance matrix: 
\begin{subequations}
\begin{align}
    \hat x_{k}^- % = \hat x_{k|k-1}
    &= \hat x_{k-1} + \int_{t_{k-1}}^{t_{k}} \dot{\hat x} (\tau) \, \dd \tau, 
    \label{eq:meth:TU_x} \\ % \int_{t_{k-1}}^{t_k} f\left( \hat x(\tau), u(\tau), \tau \right) \dd \tau \\
    P_{k}^- %= P_{k|k-1} 
    &= P_{k-1}^- + \int_{t_{k-1}}^{t_{k}} \dot P(\tau) \, \dd \tau .
    \label{eq:meth:TU_P} % \int_{t_{k-1}}^{t_k} F\left(\hat x(\tau)\right) P(\tau) + P(\tau) F^\mathrm{T}\left(\hat x(\tau)\right) + Q(\tau) \, \dd \tau 
\end{align}
\end{subequations}

During the measurement update, the available measurements $y_{k} = y(t_{k})$ are compared with the model predictions (yielding innovations $\Delta y_k$) which are used to correct \mbox{priors} to posteriors of $x$ and $P$-matrix using the Kalman gain matrix $K_{k}$ %\footnote{A-priori and \textit{a posteriori} estimates are also referred to as \textit{prior} and \textit{posterior}.} 
% = \hat x(t_k^+)$. 
% 
\begin{subequations}
\label{eq:EKF_MU}
\begin{align}
    \hat x_{k} % = \hat x_{k|k} 
    &= \hat x_{k}^- + K_{k} [ \underbrace{y_{k} - h\left(\hat x_{k}^- \right)}_{\Delta y_k} ], \label{eq:meth:MU_x} \\
    P_{k} %= P_{k|k} 
    &= \left[ I_{(n \times n)} - K_{k} \, H_{k}\right] P_{k}^- \left[ I_{(n \times n)} - K_{k} \, H_{k}\right]^\mathrm{T} + K_k \, R_k \, K_k^\mathrm{T} \label{eq:meth:Joseph} \\ 
    &= \left[ I_{(n \times n)} - K_{k} \, H_{k}\right] P_k^- \notag,
\end{align}
\end{subequations}
where $I_{(n \times n)} \in \mathbb{R}^{n \times n}$ denotes an identity matrix of $n$ rows and columns. The Kalman gain is computed as 
\begin{subequations}
\begin{align}
    K_k &= P_{k}^- H_k^\mathrm{T} \big[ \underbrace{H_k P_{k}^- H_k^\mathrm{T} + R_k }_{S_k}\big]^{-1}, \label{eq:meth:KalmanGain} \\
    H_k &= \left. \frac{\pd h}{\pd x} \right|_{\hat x_{k}^-},
\end{align}
\end{subequations}
where $H$ is the linearization of the model output $h$ around the prior.\footnote{For the ordinary continuous-discrete EKF, only online measurements are assumed.} The auxiliary matrix $S_k$ is used in Kalman filter tuning described in Sec.~\ref{sec:meth:boulkroune}.  

Note that the first variant in Eq.~\eqref{eq:meth:Joseph} applies the Joseph formula for the update of covariance matrix $P$ to ensure symmetry and positive definiteness \cite{Simon.2006}. It reduces numerical round-off errors which the conventional formulation (second variant in Eq.~\eqref{eq:meth:Joseph}) is more prone to \cite{vanderMerwe.2004b}.  

In the case that only a subset of entries $q_\mathrm{av} < q$ of the full measurement vector $y_k \in \mathbb{R}^{q \times 1}$ is available (denoted with subscript av), the output-related variables $y_k$, $h_k$, $H_k$ and $R_k$ must be reduced %to the available channels 
by cancelling out the columns and rows corresponding to the measurement signals that are not available. This results in the reduced dimensions of the Kalman gain $K_k \in \mathbb{R}^{n \times q_\mathrm{av}}$ in Eq.~\eqref{eq:meth:KalmanGain}. 
\subsubsection{Sample-state augmentation}
\label{sec:meth:ss_augmentation}
Sample-state augmentation incorporates delayed and infrequent measurements into Kalman filters. The procedure is illustrated in Fig.~\ref{fig:meth:MR-EKF_timegrid}. 
%
% It is briefly described by [Quelle Gopalakrishnan (2011)] for the EKF, and illustrated in detail for the cubature Kalman filter (CKF) by [Quelle Zhao (2015)]. In the present study, the algorithm by Zhao (2015) [Quelle] has been adapted for use in the EKF and is layed out in the following. 
For this purpose, the outputs are partitioned into online and delayed offline measurements, Eq.~\eqref{eq:system_model_output}. 
% {\color{lightgray} Erstmal mündlich beschreiben, was passiert, dann zwischen delay period und major instance unterscheiden; delay period: Beginn der augmentation, dann propagation, dann major instance}
Upon sampling at time step $k=s$, the state vector $\hat x$ is augmented by a sample state. % $\hat x_{s|s}$, i.e. the current a-posteriori state estimate at the sample time. 
Likewise, the $P$-matrix is augmented. During the delay period of length $\Delta t_\mathrm{d}$ (i.e., before return of the offline sample), the entries in $\hat x$ and $P$ that correspond to the sample time remain unchanged. As soon as the offline measurement value is returned at $k=r$ with $r=s + N_\mathrm{d}$, all of the augmented entries of state estimate and $P$-matrix are updated by using the entire available array of online and delayed offline measurements. The augmentation can then be removed (de-augmentation) until the next sampling point. Fig.~\ref{fig:meth:MR-EKF_timegrid} shows the sample-state augmentation assuming that only one sample is taken at a time and the corresponding offline measurements are returned before the next sampling. However, it may happen in practice that multiple samples are drawn before the previous samples have returned. Therefore, the more general case of multiple samples and corresponding augmentations is explained in Sec.~\ref{sec:meth:multiple_augmentations}.

During the delay period $s \le k < r$, the system state is augmented with the state at the sample time \cite{Gopalakrishnan.2011} 
\begin{equation}
    x_k^\mathrm{a} =  \left[\begin{array}{c}  x_k\\ x_s
    \end{array}\right], \label{eq:x_augmentation}
    %
    % x_{k+1}^\mathrm{a} &= \left[\begin{array}{c} f_k(x_k,u_k) + w_k\\ x_s 
    % \end{array}\right] = f^\mathrm{a}_k\left(x_k^\mathrm{a},u_k,w_k\right),
    % 
\end{equation}
where the superscript index a denotes the augmentation. % \paragraph{Initialization at sampling}
At sample time $k=s$, state vector and $P$-matrix are initialized with the latest posteriors at the sample time, i.e., the estimates after the corresponding online measurements of time $k=s$ have been fused \cite{Gopalakrishnan.2011}
% Note that augmentation is done after the corresponding online measurements at time $k=s$ have been used for the update of the a-posteriori estimate $\hat x_{s|s}$.   
% 
\begin{equation}
    \hat x_{k}^\mathrm{a} % = \hat x_{s}^\mathrm{a} 
    = \left[\begin{array}{c} \hat x_{s}\\ \hat x_{s}
    \end{array}\right], 
    \quad P_{k}^\mathrm{a} %= P_{s}^\mathrm{a} 
    = \left[\begin{array}{cc} P_{s} & P_{s} \\ P_{s} & P_{s}
    \end{array}\right].
    \label{eq:augmentation_single}
\end{equation}

For the time update, the sample state must be held constant, so the corresponding state differential equations of the nominal model are augmented as \cite{Zhao.2015}
\begin{equation}
    \dot{\hat x}^\mathrm{a} (t) = \left[\begin{array}{c} f(\hat x(t),u(t),\theta)\\ 0_{(n \times 1)}
    \end{array}\right] = f^\mathrm{a}\left(\hat x^\mathrm{a}(t),u(t),\theta\right),
    \label{eq:x_ODE_augmented}
\end{equation}
where $\hat x$ are only the first $n$ entries of $\hat x^\mathrm{a}$.\footnote{In the following, these entries are referred to as the \textit{online} entries as they are updated only through the online measurements during the delay period.} Likewise, the augmented $P$-matrix is propagated according to %the following matrix Riccati differential equation: 
\begin{equation}
    % \dot P^{\mathrm{a}}(t) = F^\mathrm{a}\left(\hat x^\mathrm{a}(t)\right) P^\mathrm{a}(t) + P^\mathrm{a}(t) {F^\mathrm{a}}^{T}\left(\hat x^\mathrm{a}(t)\right) + Q^\mathrm{a}(t).
    \dot P^{\mathrm{a}}(t) = F^\mathrm{a}(t) P^\mathrm{a}(t) + P^\mathrm{a}(t) {F^\mathrm{a}}^{T}(t) + Q^\mathrm{a}(t).\label{eq:P_dot_aug}
\end{equation}
Herein, the augmented forms of the matrices $F^\mathrm{a}$ and $Q^\mathrm{a}$ must be used \cite{Gopalakrishnan.2011,Zhao.2015}
\begin{subequations}
\label{eq:meth:aug_of_F_and_Q}
\begin{align}
    F^\mathrm{a}(t) &= \frac{\pd f^\mathrm{a}}{\pd x^\mathrm{a}} = \left[\begin{array}{ll} \left.\frac{\pd f}{\pd x} \right|_{\hat x(t),u(t),\theta} & 0_{(n \times n)} \\ \quad 0_{(n \times n)} & 0_{(n \times n)}
    \end{array}\right] , \\
    Q^\mathrm{a}(t) &= \left[\begin{array}{cc} Q(t) & 0_{(n \times n)} \\ 0_{(n \times n)} & 0_{(n \times n)}
    \end{array}\right].
    \label{eq:F_matrix_single_aug}
\end{align}
\end{subequations}
For the computation of $F^\mathrm{a}$, only the online entries of $\hat x^\mathrm{a}$ are used, which directly follows from Eqs.~\eqref{eq:x_augmentation} and \eqref{eq:x_ODE_augmented}. 
%$\hat x^\mathrm{a}_{(1:n)}$ 
% where the subscript index $(1:n)$ %in $F^\mathrm{a}$ %in $\hat x^\mathrm{a}_{(1:n)}$ denotes the first $n$ entries of the augmented state vector. 

% \paragraph{Time Update}
The time update is again computed in continuous time but with the augmented forms of %both sets of differential equations, i.e. those for the state transition and the $P$-matrix: 
the state differential equations and matrix differential equations of the $P$-matrix, which gives \cite{Zhao.2015}
\begin{subequations}
\label{eq:time_update_augmented}
% \small
\begin{align}
    \hat x^{\mathrm{a} -}_{k+1} % = \hat x^\mathrm{a}_{k+1|k} 
    &= \hat x^\mathrm{a}_{k} + \int_{t_{k}}^{t_{k+1}} \dot{\hat x}^\mathrm{a}(\tau) \, \dd \tau%f^\mathrm{a}\left( \hat x^\mathrm{a}(\tau), u(\tau), \tau \right) \dd \tau  
    = \left[\begin{array}{c} \hat x_{k+1}^-\\ \hat x_{s}
    \end{array}\right],\\
    P^{\mathrm{a} -}_{k+1} % = P^{\mathrm{a} -}_{k+1} 
    &= P^\mathrm{a}_{k} + \int_{t_{k}}^{t_{k+1}} \dot P^{\mathrm{a}}(\tau) \, \dd \tau  = %\notag \\
    % 
    % &\quad 
    \left[\begin{array}{cc} P_{k+1}^- & P_{k+1,s}^- \\ P_{s,k+1}^- & P_{s} %F^\mathrm{a}\left(\hat x^\mathrm{a}(\tau)\right) P^\mathrm{a}(\tau) + P^\mathrm{a}(\tau) {F^\mathrm{a}}^{T}\left(\hat x^\mathrm{a}(\tau)\right) + Q^\mathrm{a}(\tau) \, \dd \tau
    \end{array}\right] \label{eq:TU_P_augmented}.
\end{align}
\end{subequations}
Symmetry of the matrix Riccati differential equation in \eqref{eq:P_dot_aug} ensures that the $P$-matrix remains symmetric and its bottom-right block matrix entry (associated only to the sample state) remains unchanged. %during time update in the delay period. 
At the same time, the off-diagonal block entries describe the cross-covariance between the online states and the sample state \cite{vanderMerwe.2004b}. 

% \paragraph{Measurement Update}
During the measurement update, the two cases \textit{minor instance} and \textit{major instance} must be distinguished, cf. Fig.~\ref{fig:meth:MR-EKF_timegrid}. % Unlike for the time update, 
Van der Merwe \cite{vanderMerwe.2004b} proposed to use an indicator matrix $M_k$ %inspired by the Schmidt-Kalman filter [XY: Quelle] 
to ensure that the correct entries of the augmented state vector and error covariance matrix are updated. This indicator matrix contains zero or identity block matrices to mark certain states as \textit{ancillary}: as long as the offline measurements are not returned (minor instances during delay period), only the %first $n$ 
online entries of the augmented state may be updated. At a major instance, however, the entire augmented state vector must be updated using the online and delayed offline measurements. 

The measurement update corrects the prior $\hat x^{\mathrm{a} -}_{k+1}$ through the Kalman gain $K^{\mathrm{a}}_{k+1}$ 
and measurements $y^\mathrm{a}_{k+1}$. For better readability of the following equations, please note the index shift from $k+1$, Eq.~\eqref{eq:time_update_augmented}, to an arbitrary time point $s<k\le r$. 
% 
% \begin{equation}
%     \hat x^{\mathrm{a} -}_{k+1} 
%     % \hat x^\mathrm{a}_{k|k-1} 
%     = \left[\begin{array}{c} \hat x_{k|k-1} \\ \hat x_{s|s}
%     \end{array}\right]
% \end{equation}
%
The prior $\hat x^{\mathrm{a} -}_{k}$ and its corresponding $P$-matrix %\footnote{denoted with index shift applied}
$P^{\mathrm{a} -}_{k}$ are updated as \cite{Zhao.2015}
\begin{subequations}
\begin{align}
    \hat x^\mathrm{a}_{k} &= \hat x^{\mathrm{a} -}_{k} + K^\mathrm{a}_{k} \left[ y^\mathrm{a}_{k} - h^\mathrm{a} \left( \hat x^{\mathrm{a} -}_{k} \right) \right], \\
    P^\mathrm{a}_{k} &= \left[ I_{(2n \times 2n)} - K^\mathrm{a}_{k} \, H^\mathrm{a}_{k}\right] P^{\mathrm{a} -}_{k} \left[ I_{(2n \times 2n)} - K^\mathrm{a}_{k} \, H^\mathrm{a}_{k}\right]^{T} + K^\mathrm{a}_{k} R^\mathrm{a}_{k} {K^\mathrm{a}_{k}}^{T}.
\end{align}
\end{subequations}
While for measurement update in the ordinary EKF using the Joseph form of the $P$-matrix update is beneficial only for increased numerical stability, cf. Eq.~\eqref{eq:meth:Joseph}, the Joseph form \textit{must} be used to ensure symmetry in the augmented case \cite{vanderMerwe.2004,Zhao.2015}. The augmented Kalman gain $K^\mathrm{a}_{k}$ is then computed based on the output linearization $H^\mathrm{a}_{k}$ 
\begin{equation}
    K^\mathrm{a}_{k} = M_{k} P^{\mathrm{a} -}_{k} {H^\mathrm{a}_{k}}^{T} \left[ H^\mathrm{a}_{k} P^{\mathrm{a} -}_{k} {H^\mathrm{a}_{k}}^{T} + R^\mathrm{a}_{k} \right]^{-1}.
\end{equation}
Computation of the indicator matrix $M_{k}$ changes for minor and major updates during the delay period, as described separately in the following. The same holds for the output linearization $H_{k}^\mathrm{a}$, which is generally computed as 
\begin{equation}
    H_{k}^\mathrm{a} = \left. \frac{\pd h^\mathrm{a}}{\pd x^\mathrm{a}} \right|_{\hat x_{k}^{\mathrm{a} -}}.
\end{equation}
% 
% \textbf{XY}: evtl. Formel (38) von Zhao bringen, mit der klar wird, dass der untere rechte Blockeintrag von P nicht geupdatet wird. Dann musst du aber auch in die Indizes von P die states reinnehmen und nicht nur die Zeitpunkte k und s!
%
\paragraph{Minor instance} 
During the delay period $k<r$ with $r=s+N_\mathrm{d}$, only online measurements are available, so only minor instances occur. % as the offline sample has not yet returned. 
Therefore, the model output as well as the measurement covariance matrix are reduced according to Eqs.~\eqref{eq:system_model_output} and \eqref{eq:meth:partitioning_noise_and_R}. The indicator matrix $M_k$ enforces that the sample-state entries remain unchanged despite the generally nonzero Kalman gains \cite{vanderMerwe.2004,Zhao.2015}
\begin{subequations}
\begin{align}
    y^\mathrm{a}_k &= y^\mathrm{on}_k = h^\mathrm{on} \left( x_k \right) + v^\mathrm{on}_k, \\
    % 
    %h^\mathrm{a} &= 
    h^\mathrm{a} &\left( \hat x^{\mathrm{a} -}_{k} \right) = h^\mathrm{on} \left( \hat x_{k}^- \right) \label{eq:model_output_aug_minor}, \\
    R^\mathrm{a}_k &= R^\mathrm{on}_k, \\
    H^\mathrm{a}_k &= \left[ \left. \frac{\pd h^\mathrm{on}}{\pd x} \right|_{ \hat x_k^- }
    % {\hat x^\mathrm{a}_{k|k-1 \, (1:n)}}
    \,
    0_{(q_\mathrm{on} \times n)} \right] \label{eq:H_mat_aug_minor}, \\
    M_k &= \left[\begin{array}{cc} I_{(n \times n)} & 0_{(n \times n)} \\ 0_{(n \times n)} & 0_{(n \times n)}
    \end{array}\right].\label{eq:meth:M_during_delay}
\end{align}
\end{subequations}
Only the online entries of the prior $\hat x_k^{\mathrm{a} -}$ are used in Eqs.~\eqref{eq:model_output_aug_minor} and \eqref{eq:H_mat_aug_minor}. After the measurement update at a minor instance, the augmented state and the $P$-matrix result as 
\begin{equation}
    \hat x_{k}^\mathrm{a} = \left[\begin{array}{c} \hat x_{k} \\ \hat x_{s}
    \end{array}\right], 
    \quad P^{\mathrm{a}}_{k} = \left[\begin{array}{cc} P_{k} & P_{k,s} \\ P_{s,k} & P_{s} \end{array}\right].
    \label{eq:MU_minor_single_aug}
\end{equation}
\paragraph{Major Instance} 
At the major instance $k=r=s+N_\mathrm{d}$, the offline sample has returned, so both online and \mbox{offline} measurements are available.\footnote{For didactic reasons, the equations are shown for the full major instance only, i.e., all online and offline measurements are available, $q_\mathrm{av} = q_\mathrm{on} + q_\mathrm{off}$. If this is violated, rows and columns corresponding to the missing measurements must be deleted. The procedure is directly in line with the singlerate EKF, cf. Sec.~\ref{sec:EKF_singlerate}.} The full measurement vector, model output, and measurement covariance matrix must be used for the measurement update \cite{Zhao.2015}, that is, 
\begin{subequations}
\begin{align}
    y^\mathrm{a}_k &= \left[\begin{array}{c} y^\mathrm{on}_k \\ y^\mathrm{off}_s \end{array}\right] = \left[\begin{array}{c} h^\mathrm{on} (x_k) + v^\mathrm{on}_k \\ h^\mathrm{off} (x_s) + v^\mathrm{off}_s \end{array}\right] \label{eq:meth:y_major_instance}, \\
    % 
    % h^\mathrm{a} &= 
    h^\mathrm{a} \left( \hat x^{\mathrm{a} -}_{k} \right) &= \left[\begin{array}{c} h^\mathrm{on} \left( \hat x_{k}^- \right) \\ h^\mathrm{off} \left( \hat x_{s} \right) \end{array}\right] \label{eq:model_output_aug_major}, \\
    R^\mathrm{a}_k &= \left[\begin{array}{cc} R^\mathrm{on}_k & 0_{(q_\mathrm{on} \times q_\mathrm{off})} \\ 0_{(q_\mathrm{off} \times q_\mathrm{on})} & R^\mathrm{off}_s
    \end{array}\right] \label{eq:meth:R_major_instance}, \\
    H_k^\mathrm{a} &= \left[ \begin{array}{cc} \left. \frac{\pd h^\mathrm{on}}{\pd x} \right|_{ \hat x_k^- } & 0_{(q_\mathrm{on} \times n)} \\ 0_{(q_\mathrm{off} \times n)} & \left. \frac{\pd h^\mathrm{off}}{\pd x} \right|_{ \hat x_s }
    \end{array} \right] \label{eq:H_mat_aug_major}, \\
    % {\hat x^\mathrm{a}_{k|k-1 \, (n+1:2n)}}
    M_k &= \left[\begin{array}{cc} I_{(n \times n)} & 0_{(n \times n)} \\ 0_{(n \times n)} & I_{(n \times n)} \label{eq:M_matrix_single_major}
    \end{array}\right].
\end{align}
\end{subequations}
The predicted model output \eqref{eq:model_output_aug_major} and the output linearization \eqref{eq:H_mat_aug_major} are now based both on the online state and the sample state. 
%As in the single rate case, if only a subset of the entries of the entire measurement vector is available, the output-dependent variables in \eqref{eq:meth:y_major_instance} - \eqref{eq:H_mat_aug_major} must be reduced according to the available measurements, cf. Sec.~\ref{sec:EKF_singlerate}.
Further, the indicator matrix $M_k$ in \eqref{eq:M_matrix_single_major} enforces the sample-state entries to be updated \cite{Zhao.2015}, which is denoted with the superscript index $+$
\begin{equation}
    \hat x_{k}^\mathrm{a} = \left[\begin{array}{c} \hat x_{k} \\ \hat x_{s}^+
    \end{array}\right], 
    \quad P^{\mathrm{a}}_{k} = \left[\begin{array}{cc} P_{k} & P_{k,s}^+ \\ P_{s,k}^+ & P_{s}^+ \end{array}\right].
    \label{eq:MU_major_single_aug}
\end{equation}
The updated augmented entries are no longer used after the major update, and are thus deleted (de-augmentation). 
\subsubsection{Multiple augmentations and out-of-sequence measurements}
\label{sec:meth:multiple_augmentations}
\begin{figure}
    \centering
    \includegraphics[width=0.7\linewidth]{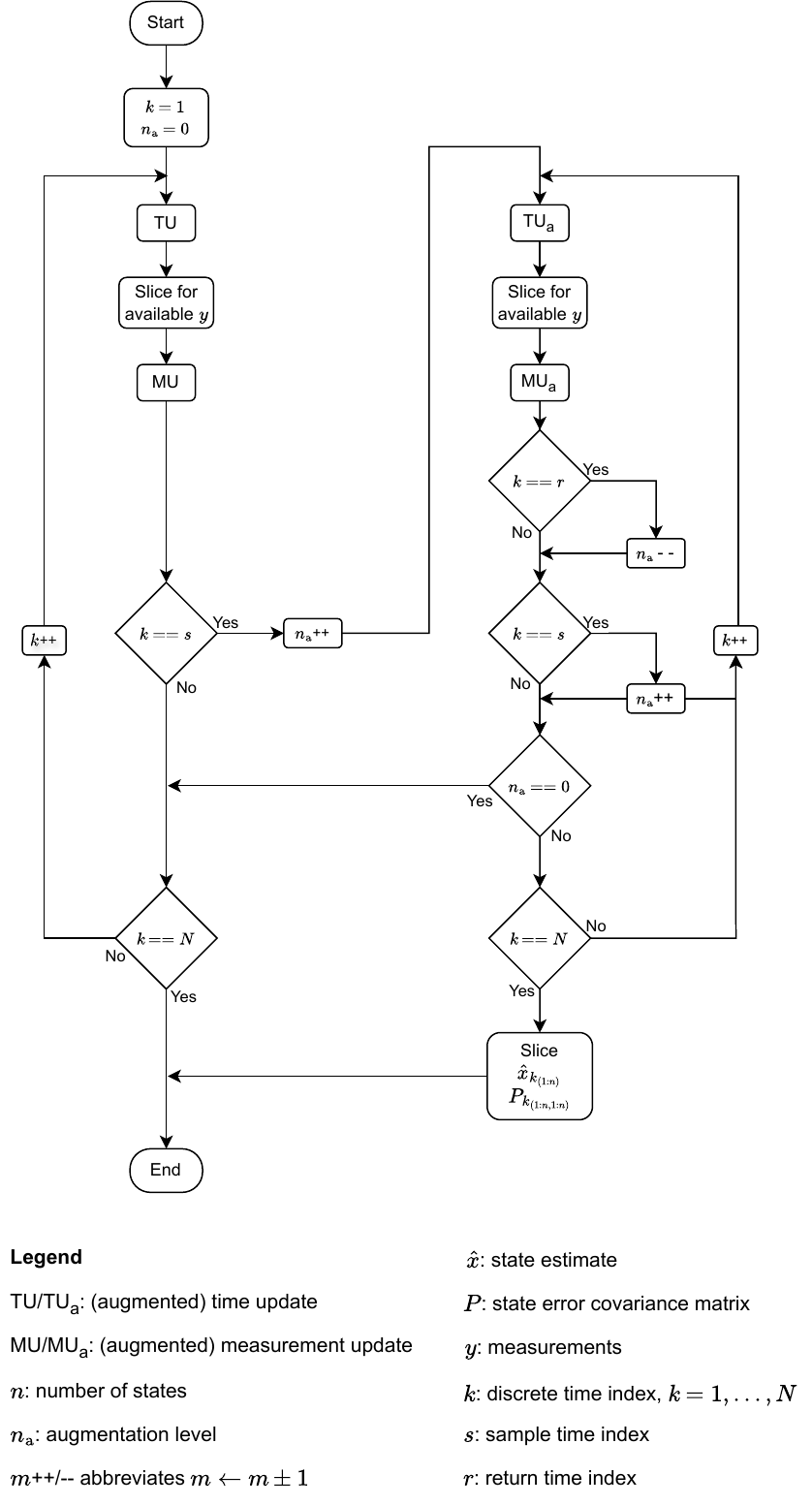}
    \caption{Block flow diagram for the MR-EKF separated into delay-free MR-EKF (left) and MR-EKF with sample-state augmentation for delay handling (right).}
\label{fig:blockflow_MR-EKF}
\end{figure}

The procedure described so far considered single samples and single returns, i.e., the delay period always ends before the next sample is taken \cite{vanderMerwe.2004b,Gopalakrishnan.2011,Zhao.2015}. In the case that multiple samples are drawn before the first return, %the procedure must be extended.
the information of the respective sample-state estimates %at each sample time 
must be maintained during the corresponding delay periods. Consequently, at each delayed sample, state vector and $P$-matrix must be further augmented, which increases the degree of augmentation $n_\mathrm{a}$ with each sample. For instance, two consecutive samples taken at $s_1$ and $k=s_2 > s_1$ require the following augmentation to ensure that the cross-covariance between online entries and individual sample states is considered like in the single augmentation case%with sample states of the respective sample times such that:
\begin{equation}
    % \small
    \hat x_{k}^\mathrm{a} 
    = \left[\begin{array}{c} 
    \hat x_{s_2} \\ \hat x_{s_1} \\ \hat x_{s_2} 
    \end{array}\right], 
    \quad
    P_{k}^\mathrm{a} = \left[\begin{array}{lll} 
    P_{s_2} & P_{s_2,s_1} & P_{s_2} \\ 
    P_{s_1,s_2} & P_{s_1} & 0_{(n \times n)} \\
    P_{s_2} & 0_{(n \times n)} & P_{s_2} \\
    \end{array}\right].
    \label{eq:augmentation_multiple}
\end{equation}
The lower right off-diagonal block entries of the $P$-matrix in Eq.~\eqref{eq:augmentation_multiple} are zero as the samples are considered to be independent of each other. %, hence there is no cross covariance between separate sample times. 
Time updates at time $k > s_2$ during the delay period with zero-augmentation as per Eqs.~\eqref{eq:x_ODE_augmented} and \eqref{eq:meth:aug_of_F_and_Q} deliver augmented priors of state and $P$-matrix
%
% \begin{subequations}
\begin{equation}
    % \small
    \hat x_{k}^{\mathrm{a} -}
    = \left[\begin{array}{c} 
    \hat x_{k}^- \\ \hat x_{s_1} \\ \hat x_{s_2} 
    \end{array}\right], 
    % \\
    \quad
    P_{k}^{\mathrm{a} -} = \left[\begin{array}{lll} 
    P_{k}^- & P_{k,s_1}^- & P_{k,s_2}^- \\ 
    P_{s_1,k}^- & P_{s_1} & 0_{(n \times n)} \\
    P_{s_2,k}^- & 0_{(n \times n)} & P_{s_2} \\
    \end{array}\right],
    \label{eq:TU_multiple_aug}
\end{equation}
% \end{subequations}
%
%with appropriate augmentation with zeros analogous to \eqref{eq:F_matrix_single_aug}. 
where the sample-state entries remain constant. The same holds for measurement updates  during minor instances. Appropriately augmenting the indicator matrix $M_k$ with zeros analogous to \eqref{eq:meth:M_during_delay} results in
\begin{equation}
    % \small
    \hat x_{k}^{\mathrm{a}}
    = \left[\begin{array}{c} 
    \hat x_{k} \\ \hat x_{s_1} \\ \hat x_{s_2} 
    \end{array}\right], 
    \quad
    P_{k}^{\mathrm{a}} = \left[\begin{array}{lll} 
    P_{k} & P_{k,s_1} & P_{k,s_2} \\ 
    P_{s_1,k} & P_{s_1} & 0_{(n \times n)} \\
    P_{s_2,k} & 0_{(n \times n)} & P_{s_2} \\
    \end{array}\right].
    \label{eq:MU_minor_multi_aug}
\end{equation}
For measurement updates at major instances, the indicator matrix $M_k$ must have identity block matrices in those blocks corresponding to sample states whose measurements return at the major instance. %For example, two consecutive samples taken at $k=s_1$ and $k=s_2 > s_1$ result in double augmentation with sample states of the respective sample times such that
%
% \begin{equation}
%     \hat x^\mathrm{a}_{k|k} = \left[\begin{array}{l} \hat x_{k|k} \\ \hat x_{s_1|s_1} \\ \hat x_{s_2|s_2} \end{array}\right] \in \mathbb{R}^{3n \times 1}.   
% \end{equation}
%
%During the delay period, the indicator matrix only contains an identity matrix in the top-left $n \times n$ block analogous to \eqref{eq:meth:M_during_delay}. 
For example, if the measurements of the second sample $s_2$ return \textit{before} those of the first sample $s_1$ (known as \textit{out-of-sequence} measurements \cite{Gopalakrishnan.2011}), the indicator matrix $M_k$ ensures an update of the online entries and sample-state entries of $s_2$, but avoids an update of sample-state entries of $s_1$ through
\begin{equation}
    M_k = \left[\begin{array}{ccc} I_{(n \times n)} & 0_{(n \times n)} & 0_{(n \times n)} \\ 0_{(n \times n)} & 0_{(n \times n)} & 0_{(n \times n)} \\ 0_{(n \times n)} & 0_{(n \times n)} & I_{(n \times n)} \end{array}\right].
\end{equation}
This results in partially updated posteriors
\begin{equation}
% \small
    \hat x_{k}^{\mathrm{a}}
    = \left[\begin{array}{c} 
    \hat x_{k} \\ \hat x_{s_1} \\ \hat x_{s_2}^+ 
    \end{array}\right], 
    \quad
    P_{k}^{\mathrm{a}} = \left[\begin{array}{lll} 
    P_{k} & P_{k,s_1} & P_{k,s_2}^+ \\ 
    P_{s_1,k} & P_{s_1} & 0_{(n \times n)} \\
    P_{s_2,k}^+ & 0_{(n \times n)} & P_{s_2}^+ \\
    \end{array}\right].
    \label{eq:MU_major_multi_aug}
\end{equation}
Updated sample-state entries of $s_2$ are removed from the state and the $P$-matrix after the major update. Time and measurement updates continue with one degree less in augmentation, i.e., like for single augmentation in the example above. Fig.~\ref{fig:blockflow_MR-EKF} summarizes this as a block flow diagram of the MR-EKF procedure for multiple augmentations. 
\subsubsection{Observabilty of multirate systems}
In MR systems, ideally structural observability is satisfied from online measurements alone \cite{Gudi.1995}. For nonlinear systems, this can be shown by means of an observability matrix \cite{Tatiraju.1999,Villaverde.2019,Hellmann.2023}. If structural observability cannot be shown based on online measurements alone, offline measurements can additionally be considered \cite{LopezNegrete.2012,Elsheikh.2021}. In the context of sample-state augmentation, Gopalakrishnan et al. noted that observability is retained for the augmented system, provided that the original system is observable \cite{Gopalakrishnan.2011}. Structural observability of the model used in this work is discussed in Sec.~\ref{sec:meth:observability_adm1_r3}.
\subsection{Modeling} 
\label{sec:meth:modeling}
For this tutorial, a modified version of the ADM1-R3 proposed by Weinrich and Nelles \cite{Weinrich2021b} was used. It describes AD of organic substrates by means of degradable macronutrients (carbohydrates, proteins, and lipids) into methane (\chfour) and carbon dioxide (\cotwo), via the intermediate formation of acetic acid (AC). The state vector is composed of mass concentrations of the following components
\begin{equation}
% \small
x = \big[S_\mathrm{ac}, S_\mathrm{ch4}, S_\mathrm{IC}, S_\mathrm{IN}, X_\mathrm{ch}, X_\mathrm{pr}, X_\mathrm{li}, X_\mathrm{bac}, X_\mathrm{ac},  S_\mathrm{ac^-}, S_\mathrm{hco3^-}, S_\mathrm{nh3}, S_\mathrm{ch4,gas}, S_\mathrm{co2,gas}\big]^T, \label{eq:R3-Core:stateVector}
\end{equation}
where $S_i$, $X_i$ and $S_{i,\mathrm{gas}}$ denote concentrations of soluble, particulate, and gaseous compounds, respectively. Details on model derivation and stoichiometric degradation pathways are included in \ref{sec:apx:model_equations} \cite{Weinrich2021b}.

\begin{figure}
    \centering
    \includegraphics[width=0.9\linewidth]{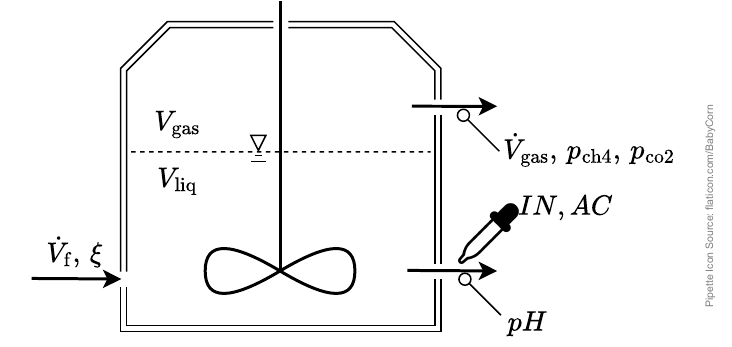}
    \caption{Setup of the AD plant with online and offline measurements, feed volume flow, and operating volumes.}
    \label{fig:measurement_setup}
\end{figure}

Two modifications compared to \cite{Weinrich2021b} were made in this tutorial. First, the state of water was removed as it represents a quasi-autonomous state which only becomes observable through direct measurement \cite{Hellmann.2023}. Second, the free residual ion concentration was considered a time-variant parameter $\theta_8$ instead of separate (nonmeasurable) states for cations and anions. % XY: macht für dieses Paper keinen Sinn, das hinzuzunehmen; einfach im Paper von Leanders MA als Modifikation erwähnen: Third, the inflowing concentration of inorganic nitrogen was amplified by a substrate-specific, time-varying parameter $\theta_9$ to adequately calibrate ammonium nitrogen measurements in the reactor.
The system dynamics are described by a set of ODEs of the form
\begin{equation}
    \label{eq:ADM1-R3-ode-class}
    %\dot x = f(x,u,\theta), \quad x_0 \text{ given}. 
    \dot x = A(u,\theta,a)\, x + \Tilde{f}(x,u,\theta,\xi,c,a),
\end{equation}
where $A \in \mathbb{R}^{n \times n}$ and $\Tilde f \in \mathbb{R}^{n}$ denote separable linear and nonlinear parts of the equations. The system input $u$ is the feed volume flow $\dot V_\mathrm{f}$ with influent concentrations $\xi$. Parameters $\theta$, 
%$\theta \in \mathbb{R}^{n_\mathrm
$c$,  
% $\xi \in \mathbb{R}^{n}$
and $a$ are described in Sec.~\ref{sec:meth:model_parameters}. The full set of model equations is provided in \ref{sec:apx:model_equations}, cf. Eq.~\eqref{eq:R3-Core:x} for the ODEs, and Eq.~\eqref{eq:R3-Core:y} for the output equations. Measurable output variables are biogas production $\dot V_\mathrm{gas}$, partial pressures of \chfour and \cotwo ($p_\mathrm{ch4}$ and $p_\mathrm{co2}$), the pH, total inorganic nitrogen (IN) and AC, as shown in Fig.~\ref{fig:measurement_setup}. The first three outputs describe the gas phase of the AD process, the latter three the liquid phase. $\dot V_\mathrm{gas}$, $p_\mathrm{ch4}, p_\mathrm{co2}$, and pH were considered online measurements, IN and AC as offline measurements, cf. Sec.~\ref{sec:meth:measurements}. 

\subsubsection{Observability of the ADM1-R3} 
\label{sec:meth:observability_adm1_r3}
Local observability and structural identifiability of the ADM1-R3 model has already been shown in a previous investigation \cite{Hellmann.2023}. The model used in this study involves a subset of states of the ADM1-R3 and two additional parameters $\theta_8$ and $\theta_9$. By applying the differential geometric approach as proposed by \cite{Villaverde.2016}, the model was shown to be locally observable and structurally identifiable given the measurements in Eq. \eqref{eq:R3-Core:y}. Local observability holds also in the case when only online measurements are available.
\subsubsection{Model parameters}
\label{sec:meth:model_parameters}
% 
% \paragraph{Time-invariant Parameters}
%  
Time-invariant parameters involve stoichiometric constants $a$, substrate-specific influent concentrations $\xi$, and miscellaneous time-invariant parameters $c$. Note that the default value of $k_\mathrm{p}$ (Table~\ref{tab:model_parameters_soeren}) had to be increased by one order of magnitude to obtain about atmospheric head space pressures \cite{Rosen2005}. 
% whose absolute value is different from 0 or 1. and \ref{tab:R3-param-nomenclature}
% Note that index $i$ denotes the component (column) and $j$ the reaction (row). 
% Table~\ref{tab:R3-param-nomenclature} shows the calculation of the time-invariant parameters $c$ from the typical ADM1 notation of \cite{Weinrich2021b}. 
Table~\ref{tab:R3-param-nomenclature} explains the aggregation of ADM1 model parameters of \cite{Weinrich2021b} to parameters $c$ used in the state-space representation in Eqs.~\eqref{eq:ADM1-R3-ode-class}, \eqref{eq:R3-Core:x}, and \eqref{eq:R3-Core:y}. Table~\ref{tab:petersen-ADM1-R3-Core} summarizes the numeric values of stoichiometric constants $a$. 
Numerical values of the influent concentrations $\xi$ are described in Sec.~\ref{sec:case_study}.  

% 
% \paragraph{Time-varying Parameters}
%
Time-varying parameters $\theta$ involve hydrolysis constants $k_\mathrm{ch}, k_\mathrm{pr}, k_\mathrm{li}$ and first-order degradation constant $k_\mathrm{dec}$, the Monod growth parameters of acetoclastic methanogens $\mu_{\mathrm{m,ac}}$ and $K_{\mathrm{S,ac}}$, the ammonia inhibition constant $K_{\mathrm{I,nh3}}$, the effective residual ion concentration $\Delta S_\mathrm{ion,eff}$, and a correction factor $\varphi_{\mathrm{IN,in}}$ for influent IN. Numerical values for individual time-varying parameters are summarized in Table~\ref{tab:meth:time_variant_params} and further detailed in Sec.~\ref{sec:meth:PMM}. 
%Synthetic data was generated with ground-truth values $\theta$.
%
\subsection{Case study: demand-oriented AD operation}
\label{sec:case_study}
The MR-EKF was applied to estimate the AD process state under demand-oriented biogas production. To this end, the substrate supply was varied dynamically to meet time-varying gas production requirements of a combined heat and power (CHP) unit operated according to a fluctuating electricity prize. 

\begin{figure}
    \centering
    \includegraphics[width=0.9\linewidth]{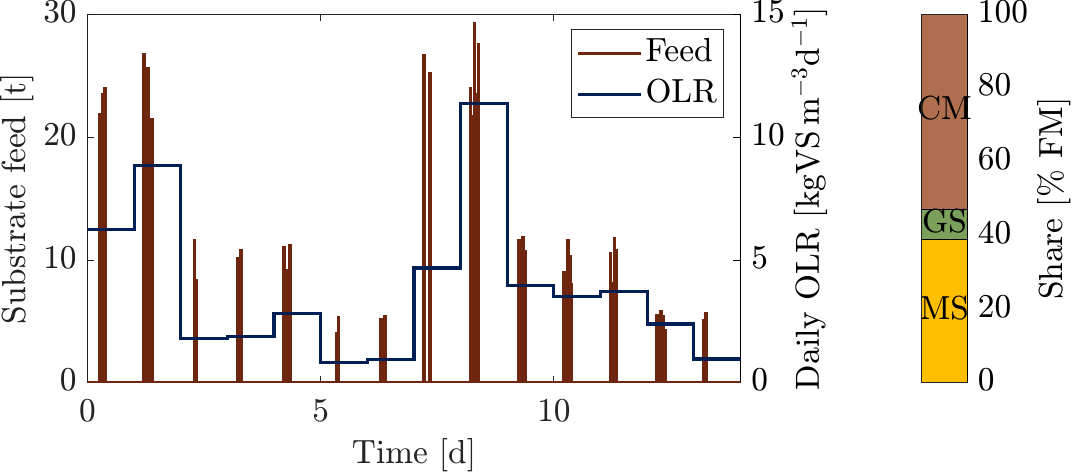}
    \caption{Dynamic feeding pattern for \SI{14}{\day} in metric tons and organic loading rate (OLR) and constant substrate composition (right, MS: maize silage, GS: grass silage, CM: cattle manure). Feeding durations were set to \SI{15}{\minute}. Day 1 and 8 are Mondays.}
\label{fig:feeding_pattern}
\end{figure}

Substrate feed, daily organic loading rate (OLR), and substrate composition are shown in Fig.~\ref{fig:feeding_pattern}. 

Feedings were set on an hourly time grid, which is commonly applied at a full-scale AD plant \cite{Dittmer.2022}. Next, the feeding events were concentrated in the morning hours from 5 am to 9 am to induce pronounced AC peaks. Conversely, feeding breaks between 9 am and 5 am of the following day were chosen to sustain reactor stability by allowing for the intermediates to settle \cite{Bonk.2018, Mauky.2017}. Moreover, the feeding was lowered before the weekends and increased before the start of the new week to account for lower electricity prices on the weekend \cite{Mauky.2017}. Furthermore, the hourly feed amount was randomized with \SI{20}{\percent} variation to mimic practical demand-oriented operation \cite{Dittmer.2022}. Feedings were assumed to last \SI{15}{\minute} each. Finally, a constant substrate composition of agricultural AD plants was assumed in line with the maximum of 40\,\% fresh matter (FM) of maize and grains according to the German Renewable Energy Sources Act (German: Erneuerbare Energien Gesetz, EEG 2023) \cite{GermanFederalMinistryofJustice.2023}. 

System dimensions and gross operating conditions are summarized in Table~\ref{tab:system_dimensioning}. They were inspired by average values of full-scale operational German AD plants \cite{DanielGromke.2018}. It was assumed that the AD reactor is constantly filled to around \SI{87}{\percent} (\SI{13}{\percent} headspace), which is in the range reported by \cite{Mauky.2016} and \cite{Naegele.2012}. 

\begin{table}%[hbt]
\centering
\begin{threeparttable}
\centering
    \renewcommand{\arraystretch}{1.3}
    \caption{System dimensions and operating conditions described by daily organic loading rate and daily hydraulic retention time.}
    \label{tab:system_dimensioning}
    \begin{tabularx}{0.55\linewidth}{lcY}
        \toprule
        \multicolumn{3}{l}{Fermenter volumes and process temperature} \\
        \cdashline{1-3}
        \multicolumn{3}{l}{$V_\mathrm{liq}=$ \SI{2000}{\cubic\meter} \qquad \quad $V_\mathrm{gas}=$ \SI{300}{\cubic\meter} \qquad \quad $T=$ \SI{38}{\celsius}} \\
        \midrule
         & average & min$-$max \\
        \cdashline{1-3}
        OLR$^{\mathrm{a,b}}$ [kg VS \si{\per\cubic\meter\day}] & 3.86 & $0.8-11.4$ \\ 
        \cdashline{1-3} 
        HRT$^{\mathrm{c}}$ [\si{\day}] & 46.8 & $15.9-212.8$ \\ 
        \bottomrule
    \end{tabularx} 
    \begin{tablenotes}
    \footnotesize
        \item[${\mathrm{a}}$] Organic loading rate
        \item[${\mathrm{b}}$] Volatile solids 
        \item[${\mathrm{c}}$] Hydraulic retention time
    \end{tablenotes}
\end{threeparttable}
\end{table}
% 

% These feedings comprise a constant substrate mix of maize silage, grass silage and cattle manure, with fresh matter based shares of 50.3/16.0/\SI{33.7}{\percent}, respectively. This composition resulted from average values of German agricultural AD plants \cite{FachagenturNachwachsendeRohstoffee.V..2021}. 
Substrate-specific influent concentrations were assumed as summarized in Table~\ref{tab:inlet_concentrations}. 
% For maize silage, values of acetic acid were taken from \cite{Weibach.2008b}. Influent inorganic nitrogen of maize silage was taken from from \cite{Schlattmann.2011}. The other 
Concentrations were derived from averaged in-house measurements at the German Biomass Research Center (German: \textit{Deutsches Biomasseforschungszentrum}, DBFZ). The measurement procedures involved in substrate characterization can be reviewed in \cite{Liebetrau.2020}. %and are qualitatively consistent with Hellmann et al. (2025). 
Substrates were assumed to contain no microbial biomass. Influent concentrations were otherwise computed as in \cite{Hellmann.2025}. 
\subsection{MR-EKF parameters}
\label{sec:ekf_hyperpamams}
Parameters considered for the different MR-EKF scenarios include synthetic MR  measurements, plant-model mismatch (PMM), and initial state estimates. 
\subsubsection{Online and offline measurements}
\label{sec:meth:measurements}
Synthetic measurement data was derived from ground-truth output trajectories and perturbed by means of additive noise, online and offline sample times, and delays. Outputs signals $y_1$ to $y_4$ ($\dot V_\mathrm{gas}, p_\mathrm{ch4}, p_\mathrm{co2}$, and pH) were considered as delay-free online measurements, output signals $y_5$ and $y_6$ (IN and AC) as delayed offline measurements according to Table~\ref{tab:meth:measurements}.

\begin{table*}
\centering
\begin{threeparttable}
    \renewcommand{\arraystretch}{1.3}
    \caption{Measurement signals and their assumed standard deviations $\sigma$, sample times and delays.}
    \label{tab:meth:measurements}
    \begin{tabular}{l*{5}{c}} % *{n}{column(s) pattern}
        \toprule
        Signal & Symbol & $\sigma$  & Unit & Sample time$^{\mathrm{a}}$ & Delay$^{\mathrm{a,b}}$\\
        \midrule
        Gas production & $\dot V_\mathrm{gas}$ & 25$^{\mathrm{c}}$ & \si{\cubic\meter\per\day} & \SI{1}{\hour} & $-$ \\
        Partial pressure of methane & $p_\mathrm{ch4}$ & 0.001$^{\mathrm{d}}$ & \si{\bar} & \SI{1}{\hour} & $-$ \\
        Partial pressure of carbon dioxide & $p_\mathrm{co2}$ & 0.001$^{\mathrm{d}}$ & \si{\bar} & \SI{1}{\hour} & $-$ \\
        pH value & $pH$ & 0.02$^{\mathrm{e}}$ & $-$ & \SI{1}{\hour} & $-$ \\
        \cdashline{1-6}
        Inorganic nitrogen & $S_\mathrm{IN}$ & 0.12$^{\mathrm{f}}$ & \si{\kilogram\per\cubic\meter} & \SI{0.87}{\day} $-$ \SI{1.13}{\day} & 0/6/12/\SI{24}{\hour} \\
        Acetic acid & $S_\mathrm{ac}$ & 0.05$^{\mathrm{f}}$ & \si{\kilogram\per\cubic\meter} & \SI{0.8}{\day} $-$ \SI{1.2}{\day} & 0/12/24/\SI{36}{\hour} \\
        \bottomrule
    \end{tabular} 
    \begin{tablenotes}
    \footnotesize
    %
    %\item[${\mathrm{a}}$] Offline sample times were assumed to be irregular to account for changing operational schedules of lab staff, and are thus given as ranges. 
    \item[${\mathrm{a}}$] Offline sample and return times were rounded up to the next online time grid entry, cf. Sec.~\ref{sec:meth:ss_augmentation}.
    \item[${\mathrm{b}}$] Values given for no, short, medium, and long delay.
    \item[${\mathrm{c}}$] Value derived from datasheet of Ritter drum-type gas counter TG50 assuming nominal gas production of \SI{5000}{\cubic\meter\per\day}.
    \item[${\mathrm{d}}$] Value derived from datasheet of Awite AwiFlex assuming standard conditions.
    \item[${\mathrm{e}}$] Value taken from datasheet of Hack Knick pH electrode Stratos Pro A2.
    \item[${\mathrm{f}}$] Values based on DBFZ measurements: multiple samples were drawn from lab-scale AD reactor operated at steady-state conditions and analyzed using in-house procedures \cite{Liebetrau.2020}. 
    \end{tablenotes}
\end{threeparttable}
\end{table*}
\paragraph{Measurement methods}
Biogas production was assumed to be measured with a drum-type gas meter, while similar measurement frequencies and accuracies can be obtained with alternative measurement techniques such as ultrasonic or vortex-based methods \cite{Liebetrau.2020}. Gas composition was assumed to be measured using infrared sensors \cite{Liebetrau.2020}, and pH with an in-line electrode \cite{Jimenez2015}.

IN was assumed to be determined as the total ammonium nitrogen using photometric measurements, and AC using gas chromatography (GC) \cite{Liebetrau.2020}. 
\paragraph{Additive noise}
Additive measurement noise was imposed on the simulated output trajectories, with standard deviations $\sigma$ given in Table~\ref{tab:meth:measurements}. To account for different measurement noise levels, standard deviations were amplified with a constant measurement noise level factor $k_\mathrm{\sigma}$. It was chosen as 0.5, 1, and 2, resulting in low, medium, and high measurement noise. Equal noise was considered for samples and returns.
\paragraph{Sample times}
Output signals $y_1$ to $y_4$ were assumed to be available online with a sample time of \SI{60}{\minute}, which holds for a well-equipped full-scale agricultural AD plant \cite{Madsen.2011,Nguyen2015,Spanjers.2006}. For offline signals $y_5$ and $y_6$, approximately daily sampling was assumed according to Table~\ref{tab:meth:measurements}. Sample times were randomly placed between 6 am to 9 am to account for fluctuating operational schedules of the lab and staff.

%In practical AD operation, offline samples at such high resolutions are rather uncommon. These resolutions, however, illustrate well the performance of the MR-EKF in the light of multiple augmentations and within relatively short simulated time frames.
% 
\paragraph{Delays}
In real-life applications, offline measurements require to draw digestate samples and to process them in a laboratory. The resulting delay depends on transport and laboratory handling, and can vary accordingly. Different delay lengths in the range of values typically observed at DBFZ were investigated and referred to as short, medium, and long, cf. Table~\ref{tab:meth:measurements}. Additionally, the case of no offline delay was considered to investigate its effect on state estimation performance. For measurements of AC, longer sample times were assumed than for IN to account for the higher effort associated with analytical procedures \cite{Liebetrau.2020}. 
%
% \begin{table}[htb]
% \centering
% \small
%     \renewcommand{\arraystretch}{1.3}
%     \caption{Diagonal entries of measurement noise covariance matrices for low, medium and high noise levels. Values are stated in the squared unit of the respective measurement signal, cf. Table~\ref{tab:meth:measurements} XY: Werte anpassen!}
%     \label{tab:measurement_noise}
%     \begin{tabular}{clll}
%         \toprule
%         Measurement & low & medium & high \\
%         %
%         \midrule
%         % 
%         $\dot V_\mathrm{gas}$ & 1 & 1.5 & 2 \\
%         $p_\mathrm{ch4}$ & 1 & 1.5 & 2 \\
%         $p_\mathrm{co2}$ & 1 & 1.5 & 2 \\
%         $pH$ & 1 & 1.5 & 2 \\
%         $S_\mathrm{IN}$ & 1 & 1.5 & 2 \\
%         $S_\mathrm{ac}$ & 1 & 1.5 & 2 \\
%         \bottomrule
%     \end{tabular} 
%     % 
% \end{table}
% % 
% 
\subsubsection{Plant-model mismatch}
\label{sec:meth:PMM}
The time-varying model parameters $\theta$ are subject to calibration as they describe variable biological process conditions, e.g., caused by changing substrate compositions. Different levels of PMM were implemented by adding a multiplicative error on ground-truth values $\theta$ \cite{Li.2004,Tatiraju.1999}
\begin{equation}
\label{eq:meth:PMM_choice}
    \hat \theta = \theta \, (1 + k_\theta),
\end{equation}
where $k_\theta$ denotes the PMM factor. Its values were chosen as 0, 10, 20, and \SI{30}{\percent}, corresponding to no, low, medium, and high PMM, as shown in Table~\ref{tab:meth:variants_overview}. The resulting values of $\hat \theta$ are summarized in Table~\ref{tab:meth:time_variant_params}. Synthetic data was generated with ground-truth values $\theta$.

\begin{table}[htb]
\centering
\begin{threeparttable}
    \renewcommand{\arraystretch}{1.3}
    \caption{Time-variant model parameters, their notation in the ADM1-R3 \cite{Weinrich2021b} as well as their true and perturbed values to account for parametric plant-model mismatch.}
    \label{tab:meth:time_variant_params}
    \begin{tabular}{C{1mm}C{10mm} cd{4}llc} 
        \toprule
        \multirow{2}{*}{$i$} & \multirow{2}{*}{ADM1$^{\mathrm{a}}$} & \multirow{2}{*}{true $\theta$} & \multicolumn{3}{c}{Estimates $\hat \theta$ for diff. PMM$^{\mathrm{b}}$} & \multirow{2}{*}{Unit} \\
        \cdashline{4-6}
        & & & \multicolumn{1}{l}{low} & medium & high & \\ 
        \midrule
        1 & $k_\mathrm{ch}$ & 1.25  & 1.375     & 1.5   & 1.625 & \si{\per\day}\\
        2 & $k_\mathrm{pr}$ & 0.20  & 0.22      & 0.24  & 0.26  & \si{\per\day} \\
        3 & $k_\mathrm{li}$ & 0.10  & 0.11      & 0.12  & 0.13  & \si{\per\day} \\
        4 & $k_\mathrm{dec}$ & 0.020 & 0.022     & 0.024 & 0.026 & \si{\per\day} \\
        5 & $\mu_{\mathrm{m,ac}}$ & 0.40  & 0.44      & 0.48  & 0.52  & \si{\per\day} \\
        6 & $K_{\mathrm{S,ac}}$ & 0.14  & 0.154     & 0.168 & 0.182 & \si{\kilogram\per\cubic\meter} \\
        7 & $K_{\mathrm{I,nh3}}$ & 0.0306 & 0.0337   & 0.0367& 0.0398& \si{\kilogram\per\cubic\meter} \\
        8 & $\Delta S_\mathrm{ion,eff}$ & 0.0528 & 0.0581   & 0.0634& 0.0686& \si{\kilo\mole\per\cubic\meter} \\
        9 & $\varphi_\mathrm{IN,in}$ & 1.00  & 1.1       & 1.2   & 1.3   & $-$ \\
        %
        % \midrule
        % % 
        % \multicolumn{2}{l}{L1-norm $\Delta \bar \theta$} & $-$ & & g & m & p & $-$ \\
        % % 
        \bottomrule
    \end{tabular} 
    \begin{tablenotes}
    % \tiny
        \item[${\mathrm{a}}$] Notation of parameters according to the ADM1 simplifications of Weinrich and Nelles \cite{Weinrich2021}. Parameters $\theta_8$ and $\theta_9$ were added to account for calibration of pH and ammonium nitrogen, respectively, cf. model equations in \ref{sec:apx:model_equations}.
        \item[${\mathrm{b}}$] Plant-model mismatch. For no PMM, it holds that $\hat \theta = \theta$.
    \end{tablenotes}
\end{threeparttable}
\end{table}

\begin{figure}
\centering
\begin{subfigure}[b]{0.7\linewidth}
    \centering    
    \includegraphics[width=\linewidth]{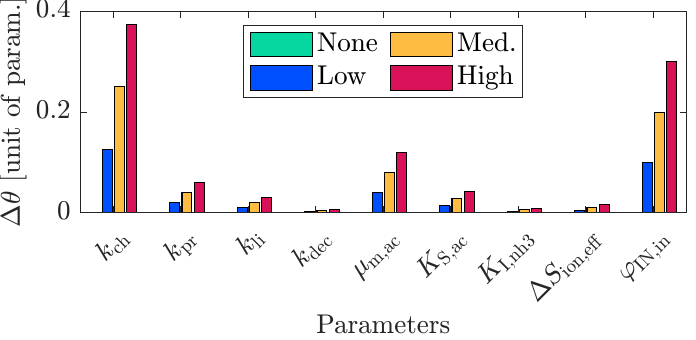}
    \caption{Absolute Parameter errors $\Delta \theta = \hat \theta - \theta$ 
    % for different levels of plant-model mismatch. Units of the parameters are given in Table~\ref{tab:meth:time_variant_params}. \textit{Med.} abbreviates \textit{Medium}.
    }
    \label{fig:meth:meaning_pmm}
% \end{figure}
\end{subfigure}
\vskip 0.2cm
\begin{subfigure}[b]{0.7\linewidth}
    \centering
% \begin{figure}
    \includegraphics[width=\linewidth]{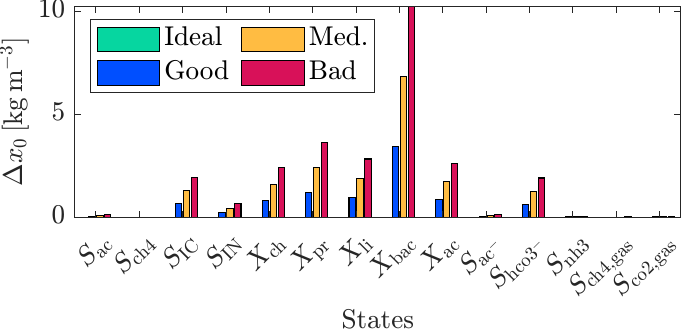}
    \caption{Absolute initial state errors $\Delta x_0 = \hat x_0 - x_0$. 
    % \textit{Med.} abbreviates \textit{Medium}.
    }
    \label{fig:meth:meaning_init_error}
\end{subfigure}
\caption{Absolute parameter errors $\Delta \theta = \hat \theta - \theta$ for different levels of plant-model mismatch, and absolute initial state errors $\Delta x_0 = {\hat x_0 - x_0}$. Units of the parameters are given in Table~\ref{tab:meth:time_variant_params}. \textit{Med.} abbreviates \textit{Medium}.}
\label{fig:meth:meaning_pmm_and_init}
\end{figure}

Relative parameter errors are shown in Fig.~\ref{fig:meth:meaning_pmm}.

\subsubsection{Initial state estimates}
To investigate the convergence behavior of the MR-EKF, the initial state estimate $\hat x_0$ was perturbed from the true initial value $x_0$ as
\begin{equation}
\label{eq:initial_state_estiate_perturbation}
    \hat x_0 = x_0 + k_\mathrm{x} \, \Delta x.
\end{equation}
Ground-truth values of $x_0$ and $\Delta x$ are summarized in Table~\ref{tab:initial_estimate}. $k_\mathrm{x}$ denotes the perturbation factor. It was chosen as 0, 0.5, 1 and 1.5, which is referred to as ideal, good, medium, and bad initial estimates. Values of perturbations $\Delta x$ were derived from Delory et al. \cite[Scenario G]{Delory.2025} by multiplying the coefficients of variation (CV) given therein with the respective steady state value $x_\mathrm{ss}$.

\begin{table}
\centering
\begin{threeparttable}
\centering
    \renewcommand{\arraystretch}{1.3}
    \caption{True initial states $x_0$ and perturbations$^\mathrm{a}$ $\Delta x_0$ of initial state estimates $\hat x_0$ in [\si{\kilogram\per\cubic\meter}].}
    \label{tab:initial_estimate}
    % *{n}{column(s) pattern}
    \begin{tabularx}{0.75\linewidth}{XX*{2}{d{4}}XXd{4}Y}
        \toprule
        $i^\mathrm{b}$ & St.$^\mathrm{c}$ & \multicolumn{1}{c}{$x_{0,i}$} & \multicolumn{1}{c}{$\Delta x_i$} & $i^\mathrm{b}$ & St.$^\mathrm{c}$ & \multicolumn{1}{c}{$x_{0,i}$} & \multicolumn{1}{c}{$\Delta x_i$} \\
        \midrule
        % mind: in Matlab, Delta x_0 is subtracted from x_0, I inverted the signs!
        1 & $S_\mathrm{ac}$ & 0.0935 & 0.0753  & 8 & $X_\mathrm{bac}$      & 10.8126& 6.8336 \\
        2 & $S_\mathrm{ch4}$& 0.0152 & 0.0007 & 9  & $X_\mathrm{ac}$       & 2.7521 & 1.7393 \\
        3 & $S_\mathrm{IC}$ & 8.5259 & 1.2959 & 10 & $S_\mathrm{ac^-}$     & 0.0933 & 0.0752 \\
        4 & $S_\mathrm{IN}$ & 2.3051 & 0.4334 & 11 & $S_\mathrm{hco3^-}$   & 7.9940 & 1.2710 \\
        5 & $X_\mathrm{ch}$ & 2.4604 & 1.6140 & 12  & $S_\mathrm{nh3}$     & 0.0877 & 0.0274 \\
        6 & $X_\mathrm{pr}$ & 2.7327 & 2.4212 & 13  & $S_\mathrm{ch4,gas}$ & 0.3891 & 0.0117 \\
        7 & $X_\mathrm{li}$ & 1.7016 & 1.8854 & 14  & $S_\mathrm{co2,gas}$ & 0.9143 & 0.0357 \\
        \bottomrule
    \end{tabularx} 
    \begin{tablenotes}
    \footnotesize
        \item[$\mathrm{a}$] Absolute perturbations $\Delta x_i$ of initial states were derived from Delory et al. \cite[Scenario G]{Delory.2025}.
        \item[$\mathrm{b}$] Index 
        \item[$\mathrm{c}$] State 
    \end{tablenotes}
\end{threeparttable}
\end{table}
Fig.~\ref{fig:meth:meaning_init_error} illustrates the resulting absolute initial estimation errors. All combinations of MR-EKF parameters are summarized in Table~\ref{tab:meth:variants_overview}. 

\begin{table}
\centering
\begin{threeparttable}
    \renewcommand{\arraystretch}{1.3}
    \caption{MR-EKF parameters for individual scenarios.}
    \label{tab:meth:variants_overview}
    \begin{tabularx}{0.6\linewidth}{*{7}{Y}}
        \toprule %
        \multirow{2}{*}{Scenario} & \multicolumn{2}{c}{Delay [h]} & Noise$^{\mathrm{a}}$ & PMM$^{\mathrm{b}}$ & Init.$^{\mathrm{c}}$ \\
        \cdashline{2-7}
        & AC & IN & $k_\sigma$ & $k_\theta$ & $k_\mathrm{x}$ \\
        \midrule
        1 & 0$^*$ & 0$^*$ & 0.5 & 0         & 0 \\
        2 & 12  & 6     & 1$^{*,\dagger}$ & 0.1$^\dagger$       & 0.5$^\dagger$ \\
        3 & 24  & 12    & 1.5     & 0.2$^*$   & 1$^*$ \\
        4 & 36  & 24    & $-$   & 0.3       & 2 \\
        \bottomrule 
    \end{tabularx} 
    \begin{tablenotes}
    \footnotesize
        \item[$*$] Configuration used for grid search tuning and medium MHE run. 
        \item[$\dagger$] Configuration used for good MHE run.
        \item[${\mathrm{a}}$] Measurement noise level 
        \item[${\mathrm{b}}$] Plant-model mismatch
        \item[${\mathrm{c}}$] Initial state error
    \end{tablenotes}
\end{threeparttable}
\end{table}
\subsection{Kalman filter tuning}
\label{sec:Kalman_filter_tuning}
Finding an ideal tuning for a Kalman filter is not trivial. While there exists a pragmatic and prominent tuning approach proposed by Schneider and Georgaskis \cite{Schneider.2013}, this approach did not result in satisfactory results during initial tests, both for the tuning of the $R$ and $Q$ matrix. Instead, a multitude of tunings were assessed in a grid search and evaluated \textit{ex post} by means of different criteria. To this end, tuning variables were limited to the diagonal entries of both $Q$ and $R$ \cite{Chen.2019, Boulkroune.2023}. The initial estimate of the error covariance matrix $P_0$ (in normalized coordinates, cf. Sec.~\ref{sec:meth:numerical_robustness}) was chosen as an identity matrix.  

As suggested by \cite{Schneider.2013}, tuning of the measurement uncertainty was modified by an amplification factor $k_\mathrm{R}$ of a diagonal matrix of variance values of individual measurement signals, cf. Table~\ref{tab:meth:measurements}. For process uncertainty, individual values of a diagonal matrix $Q$ were modified (in normalized coordinates, cf. Sec.~\ref{sec:meth:numerical_robustness}). All values were varied by means of Latin hypercube sampling \cite{McKay.1979} with 10,000 samples in logarithmic (log) scale between $10^{-2}$ and $10^{2}$. 

For all grid search runs, medium MR-EKF parameters were selected, cf. Table~\ref{tab:meth:variants_overview}, except for delays, which were set to zero. This choice allowed to accelerate computations and ensured a fair consideration of Kalman filter innovations from online and offline outputs, cf. Eq.~\eqref{eq:meth:nis}. For each tuning, the MR-EKF was applied to the same synthetic dataset described in Sec.~\ref{sec:case_study}. The grid search was run in a parallelized for-loop with a cutoff time of \SI{30}{\second} per for-loop iteration which was determined as sufficiently long after initial trials. 

Individual tunings were ranked according to three criteria: L1-norm of the normalized root mean squared error (NRMSE) between all true and estimated state trajectories ($\mathrm{NRMSE_x}$, criterion 1), L1-norm of NRMSE between all true and estimated outputs ($\mathrm{NRMSE_y}$, criterion 2), and the error function proposed by Boulkroune et al. \cite{Boulkroune.2023} (criterion 3), which is described in Sec.~\ref{sec:meth:boulkroune}. 

% Only the simulation results of the second week were considered for error computation to separate the long-term EKF performance from the initial transient convergence (day 7 $-$ 14 of the total data of \SI{14}{\day}). 

The NRMSE between ground-truth values $z$ and its estimates $\hat z$ was computed as\footnote{Normalizing only with the mean or median value of $z$ did not deliver a fair weighting of individual states and outputs as it disregards their dynamics.} 
\begin{equation}
\label{eq:meth:nrmse}
    \mathrm{NRMSE_z} = \frac{\sqrt{\frac{1}{N}\sum_{k=1}^{N}(\hat z_k - z_k)^2}}{\max(z) - \min(z)}.
\end{equation}
To avoid overfitting of an initial transient state error by individual tunings, only the data of the second week was considered for error computations. 
\subsubsection{Error function of Boulkroune}
\label{sec:meth:boulkroune}
The error function proposed by Boulkroune et al. \cite{Boulkroune.2023} adds multiple error metrics based on measurable properties to a single scalar value, which combines filter performance and consistency. It involves the normalized innovation squared (NIS) $\epsilon_{\mathrm{y},k}$, which is computed from innovations $\Delta y_k$ for each Kalman filter time step as a dimensionless scalar quantity
\begin{equation}
\label{eq:meth:nis}
    \epsilon_{\mathrm{y},k} = \Delta y_k^T \, S_k^{-1} \, \Delta y_k, \quad k=1, \ldots, N,
\end{equation}
with an auxiliary matrix\footnote{As for singlerate EKF described in Sec.~\ref{sec:EKF_singlerate}, $\Delta y_k$ and $S_k$ need to be filtered for columns and rows corresponding to available measurement signals in the case that offline measurements are missing at a particular time step $k$.} $S_k$ from Eq.~\eqref{eq:meth:KalmanGain}. For consistent filtering, the NIS is expected to follow a $\chi^2$ distribution with $q$ degrees of freedom \cite{Boulkroune.2023}, where $q$ is the number of measurement signals, cf. Sec.~\ref{sec:meth:class_of_systems}. 

In this work, the error function of Boulkroune was slightly modified to be independent of the number of time steps considered and to account for output signals of different orders of magnitude. The modified error function combines the output errors as a performance measure, and as consistency measures the trace of normalized $P$-matrices as well as the mean $\mu$, variance $\sigma^2$, and a two-sided confidence interval of the NIS. It gives a dimensionless number which reads 
\begin{equation} 
    \small
    J = \omega_1 \, || \mathrm{NRMSE_{\hat y}} ||_2 + \omega_2 \, \mathrm{RMS} \left( \mathrm{tr}\left(P_k\right) \right) + \omega_3 \, \left| \frac{\mu(\epsilon_{\mathrm y,k})}{q} - 1 \right| \omega_4 \, \left| \frac{\sigma^2(\epsilon_{\mathrm y,k})}{2q} - 1 \right| + \omega_5 \, \left| \frac{N_{\chi^2 \ge \gamma}(\epsilon_{\mathrm y,k})}{\alpha \, N} - 1 \right|,
\label{eq:boulkroune_error_fun}
\end{equation}
where $\omega$ denote weighting factors satisfying $0 < \omega_i < 1$ and $\sum_i \omega_i = 1$. Note that $\mathrm{NRMSE_{\hat y}}$ denotes the normalized root mean squared error between estimated and \textit{measured} outputs (normalized with the range of the measured signals). 
%\footnote{$NRMSE_\mathrm{\hat y} = \frac{\sqrt{\frac{1}{N}\sum_{k=1}^{N}(y_{\mathrm{m},k} - \hat y_k)^2}}{\max(y_\mathrm{m}) - \min(y_\mathrm{m})}$} 
It can thus be computed in real-life applications as opposed to the metric in Eq.~\eqref{eq:meth:nrmse}, which requires ground-truth values only known in simulative studies. $\mathrm{RMS}$ denotes the root mean square\footnote{$\mathrm{RMS}(z) = \sqrt{\frac{1}{N} \sum_{k=1}^{N} z_i^2}$ for a generic scalar signal $z$} computed based on normalized $P$-matrices, cf. Sec.~\ref{sec:meth:implementation}. $N_{\chi^2 \ge \gamma}$ denotes the number of NIS samples falling outside of the two-sided confidence interval\footnote{Since a two-sided confidence interval is considered, the precise notation of the number of NIS samples falling outside the confidence interval would be $N_{\chi^2 \notin [\gamma^- ,\gamma^+]}$ with lower and upper critical values $\gamma^-$ and $\gamma^+$. For easier readability and to be consistent with the original formulation of Boulkroune, this detail was omitted here.} of the $\chi^2$ distribution with significance level $\alpha$ and corresponding critical values $\gamma$. $\alpha$ was chosen as \SI{5}{\percent}. Individual weights $\omega_i$ were tuned such that the medians of all five error function summands resulting from the grid search lie in the same range of magnitude, which gives $\omega = [0.328, \, 3\cdot 10^{-4}, \, 0.328, \, 0.328, \, 0.164]$. A boxplot for all 10,000 grid search runs is shown in \ref{sec:apx}, Fig.~\ref{fig:boxplot_boulkroune_error_fun_parts}. Details on the theoretical motivation of considering NIS and a $\chi^2$ distribution can be reviewed in \cite{Boulkroune.2023, BarShalom.2002, Chen.2019}.  
\subsection{Moving horizon estimation}
\label{sec:meth:mhe}
A complementary approach to Kalman filtering for state estimation is moving horizon estimation (MHE) \cite{Kim.2023}. As a comparison to the MR-EKF approach, an MHE was implemented in casadi \cite{Andersson.2019} as inspired by Elsheikh et al. \cite{Elsheikh.2021}. The ODEs were discretized using orthogonal collocation with quadratic polynomials and two collocation points per step size of \SI{1}{\hour}. Due to the discrete time grid required by MHE, feeding events were distributed equally throughout the time intervals of \SI{1}{\hour}. This conserved the total feeding amount and ensured consistency with the MR-EKF implementation for which feeding durations were assumed to last \SI{15}{\minute}, cf. Sec.~\ref{sec:meth:implementation}. A hindsight horizon of \SI{24}{\hour} was considered sufficient after initial tests. The resulting nonlinear program was solved using IPOPT \cite{Wachter.2006}, which was accelerated by the linear solver ma27 \cite{CoinOr.2025}. Numerical stability could be drastically increased by assuming non-negative states.

Three differences to Elsheikh et al. \cite{Elsheikh.2021} should be noted: First, no process noise was considered, i.e., only states were subject to optimization. This increased numerical stability during the integration. Second, the arrival cost was considered in the objective function using the same initial $P$-matrix as for the MR-EKF\footnote{A Hessian-based arrival cost update as proposed by Fiedler et al. \cite{Fiedler.2020} was tested but delivered practically identical estimates.}, cf. Sec.~\ref{sec:Kalman_filter_tuning}. Third, the MHE was provided with online measurements only to obtain a conventional MHE implementation as described in \cite{Kim.2023,Tuveri.2023,Fiedler.2020}, and to avoid losing offline samples with delays longer than the horizon \cite{Elsheikh.2021}.  

The underlying optimization problem solved at each time step $k$ of the horizon of length $N$ reads
\begin{subequations}
    % \small
    \begin{align}
    \min_{\hat x_{k-N}, \ldots, \hat x_k} &\sum_{i=k-N+1}^{k} ||y_i^\mathrm{on} - h^\mathrm{on}(\hat x_i)||^2_{(R^\mathrm{on})^{-1}} + \Vert \hat x_{k-N} - \bar x_{k-N} \Vert ^2_{P_0^{-1}} \\
    \mathrm{s.t.} \quad \hat x_{i+1} &= \bar f(\hat x_i, u_i, \theta), \quad i=k-N, \ldots, k-1 % \\
    % \hat x_{k-N} &= \bar x_{k-N}, % SH: dont set when you ignore process noise, otherwise you fix the arrival state to the prior and thus ignore the arrival cost!
\end{align}
\end{subequations}
where $||x||_P$ denotes the weighted 2-norm $x^T P x$. Furthermore, $\bar x_{k-N}$ is the arrival state prior, i.e., the prior state estimate at the beginning of the horizon, and $\bar f$ is the discretized state transition function of the continuous-time ODEs $f$, cf. Eq.~\ref{eq:system_model_ode}.
\subsection{Implementation aspects}
\label{sec:meth:implementation}
The code was implemented in Matlab R2024a using the symbolic math and statistics and machine learning toolboxes \cite{TheMathWorksInc..2024}. All simulations were run on a standard laptop with an intel Core-i5-1335U processor, 10 cores, \SI{32}{GB} of RAM under Windows 10.
% macbook pro with macOS Sequoia~15, an M1 chip and \SI{8}{GB} of RAM.
%standard laptop was used with Windows 10, a 13th generation Intel Core i5-1335U processor (10 cores, \SI{1300}{\mega\hertz}) and \SI{16}{GB} of RAM. 
% 
\paragraph{Initialization}
Dynamic simulations were initialized in steady state (ss) using ground-truth model parameters $\theta$
\begin{equation}
    x_0 = x_\mathrm{ss}(x_{0,\mathrm{ss}}, u_\mathrm{ss}, \theta),
\end{equation}
cf. Table~\ref{tab:initial_estimate}. For this purpose, the average OLR of the dynamic feeding pattern in Fig.~\ref{fig:feeding_pattern} was applied as a constant feed volume flow $u_\mathrm{ss}$ for $\Delta t_\mathrm{ss}=\SI{500}{\day}$. This transition into steady state was in turn initialized with values $x_{0,\mathrm{ss}}$ as summarized in Table~\ref{tab:init_condition_ss}. 
\paragraph{Augmentation}
In the MR-EKF implementation, it was assumed that only a single sample is drawn and/or processed at every online time step. If multiple samples were drawn at the exact same time with separate returns, samples were distributed across consecutive online time steps. Likewise, if multiple returns of the same signal were to be processed simultaneously, they were distributed across consecutive online time steps. 

An alternative for handling multiple simultaneous samples lies in using the same sample state for multiple augmentations, which are then processed in separate major instances.
\subsubsection{Improvements to numerical robustness}
\label{sec:meth:numerical_robustness}
The following section details some aspects on successful numerical implementation of the MR-EKF. 
\paragraph{Normalization}
Numerical efficiency and robustness could be significantly increased by normalizing states $x$, outputs~$y$ and the input $u$ with constant diagonal matrices~$T$
\begin{subequations}
\label{eq:meth:normalization}
\begin{align}
    \bar x &= T_\mathrm{x}^{-1} \, x, \label{eq:meth:normalizationX} \\ %\quad T_x = \mathrm{diag}\{x_\mathrm{ss}\}, \\
    \bar y &= T_\mathrm{y}^{-1} \, y, \\ %\quad T_y = \mathrm{diag}\{y_\mathrm{ss}\}, \\
    \bar u &= T_\mathrm{u}^{-1} \, u, % \quad T_u = \mathrm{diag}\{u_\mathrm{ss}\},
\end{align}
\end{subequations}
where normalized coordinates are denoted with a superscript bar ( $\bar{}$ ). Influent concentrations $\xi$ were normalized like states in Eq.~\eqref{eq:meth:normalizationX}. Normalization factors, i.e., the diagonal entries of normalization matrices $T$, were chosen as steady-state values, cf. \ref{sec:apx:model_equations}, Table~\ref{tab:normalization}. This also means that tuning matrices $Q, R$ and $P_0$ were processed in normalized coordinates, cf. Sec.~\ref{sec:Kalman_filter_tuning}. Likewise, state estimates and $P$-matrix in absolute coordinates $\hat x$ and $P$ are obtained by de-normalization 
\begin{subequations}
\begin{align}
    \hat x &= T_\mathrm{x} \, \bar{\hat x}, \\
    P &= T_\mathrm{x} \, \bar P \, T_\mathrm{x}^T.
\end{align}
\end{subequations}
\paragraph{Clipping}
Numerical round-off errors could be significantly reduced by projecting the state prior onto its admissible values before each time update (clipping).\footnote{The implemented filter is, thus, a constrained extended Kalman filter.} The states of the given AD model describe nonnegative mass concentrations. Thus, negative state priors were clipped to a small positive value ($10^{-3}$, in normalized coordinates).
\paragraph{Numerical integration}
Numerical integration was implemented by means of Matlab's \texttt{ode15s} \cite{TheMathWorksInc..2024} with relative and absolute tolerances of $10^{-4}$ and $10^{-7}$, respectively, as well as imposing nonnegativity for all states. To prevent the integrator from skipping over relatively short feeding intervals, the integration was implemented sequentially across intervals of constant feeding regime, with nonzero feed volume flow only during feeding impulses. 
\paragraph{Jacobians for time update}
One way of implementing the integration of state and $P$-matrix during the time update is to integrate their ODEs simultaneously for optimal time step control through the ODE solver. For this purpose, the dynamics of the $P$-matrix must be vectorized and appended to the dynamics of the states. 

Numerical robustness of the time update can be increased by providing Jacobians of the ODEs \cite{TheMathWorksInc..2025}, i.e., both for the state ($\nabla f$) and the $P$-matrix ($\nabla \dot P$), cf. Eqs.~\eqref{eq:meth:time_update_x_single_rate} and \eqref{eq:meth:time_update_P_single_rate}. If Jacobians for the $P$-matrix are provided, the time update needs to be computed sequentially for $\hat x$ and $P$ since the dynamics of the $P$-matrix depend on the solution trajectory of $\hat x$, Eq.~\eqref{eq:meth:F}. This can be addressed by linearly interpolating the trajectory of $\hat x$ to the current time used by the ODE solver of the dynamics of $P$. 

Jacobians $\nabla f$ of the state ODE are already known from integration of the $P$-matrix and given by $F$ in Eq.~\eqref{eq:meth:F}. Jacobians $\nabla \dot P$ of the $P$-matrix can be derived symbolically and must be vectorized for usage in \texttt{ode15s}. Sample-state augmentation is considered by augmenting the state ODE $f$ and its Jacobian $F$ with zeros, Eqs.~\eqref{eq:x_ODE_augmented} and \eqref{eq:F_matrix_single_aug}. 

In the case that analytic Jacobians cannot be provided, it can be exploited that the Jacobians are often sparse \cite{Curtis.1974}. Therefore, Jacobian patterns for $\nabla f$ and $\nabla \dot P$ can be provided to inform the ODE solver on the locations of nonzero entries in $\nabla f$ and $\nabla P$ \cite{TheMathWorksInc..2025}. 
% 
% \section{Results and Discussion}
% \label{sec:results}
% %
% The following section discusses the results of the grid search tuning and underlines the importance of an adequate tuning. 
%
\section{Results of the grid search tuning}
\label{sec:grid_search}
\begin{figure}
    \centering
    \includegraphics[width=0.75\linewidth]{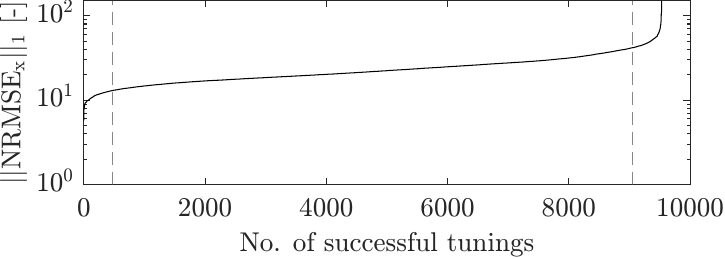}
    \caption{Ranking of tunings from grid search by means of the L1-norm of the NRMSE of all states, as well as limits of top and bottom \SI{5}{\percent} of successful tunings.}
    \label{fig:res:ranking_tuning}
\end{figure}

%
% No satisfactory tunings were obtained by choosing $R$ simply on the basis of variances according to sensor data sheets (as suggested by \cite{Schneider.2013}) and tuning only an amplification factor $k_\mathrm{R}$. Despite the wide range of the LHS search space, this resulted in differently strong smoothing of AC, and slightly different smoothing of IN. All online output variables were directly tracked and not smoothed (plots not shown). The reason is the large difference in resulting normalized matrices $R$, in which AC is more uncertain than all other output variables by several orders of magnitude, resulting in an overly reliance on the other output variables during the EKF corrections. 
The 10,000 different tunings for diagonal entries of $Q$ and $R$ resulted in vastly different state estimations. Only such tunings were considered that did not exceed the time threshold of \SI{30}{\second}. All other tunings were omitted because these usually led to filter divergence, and thus very long time updates. This applied for 468 out of 10,000 tunings (around \SI{4.7}{\percent}). The other \SI{95.3}{\percent} satisfied the time threshold, and are referred to as the successful tunings. To illustrate the effect of widely different tunings, Fig.~\ref{fig:res:ranking_tuning} shows the ranking of the successful tunings according to the L1-norm of the NRMSE of all states ($\mathrm{NRMSE_x}$) in log scale.

% 

%\footnote{NRMSEs were computed based on the entire trajectories \textit{including} the initial condition at $t_0$; the initial error had no such dominant effect on the entire NRMSE, so all is good!}
The ranking in Fig.~\ref{fig:res:ranking_tuning} can be roughly subdivided into three parts, which is indicated by vertical dashed lines. For the worst \SI{5}{\percent} of successful tunings (right), the error function decreases quickly. The central, almost horizontal part of Fig.~\ref{fig:res:ranking_tuning} encompasses around \SI{90}{\percent} of successful tunings, which deliver about the same quality of state estimation, i.e., the error nearly plateaus. Only for about the best \SI{5}{\percent} of successful tunings (left), the error decreased rapidly again with increasing slope towards the best tunings. A similar tilted S-shape can also be observed for the L1-norm of NRMSE of all outputs ($\mathrm{NRMSE_y}$, plots not shown). By contrast, the ranking plot of Boulkroune's error function does not show three characteristic parts, cf. Fig.~\ref{fig:ranking_Boulkroune}. Instead, many tunings deliver similar performance and only very few with really poor performance. 

\begin{figure*}
    \centering
    \includegraphics[width=0.9\linewidth]{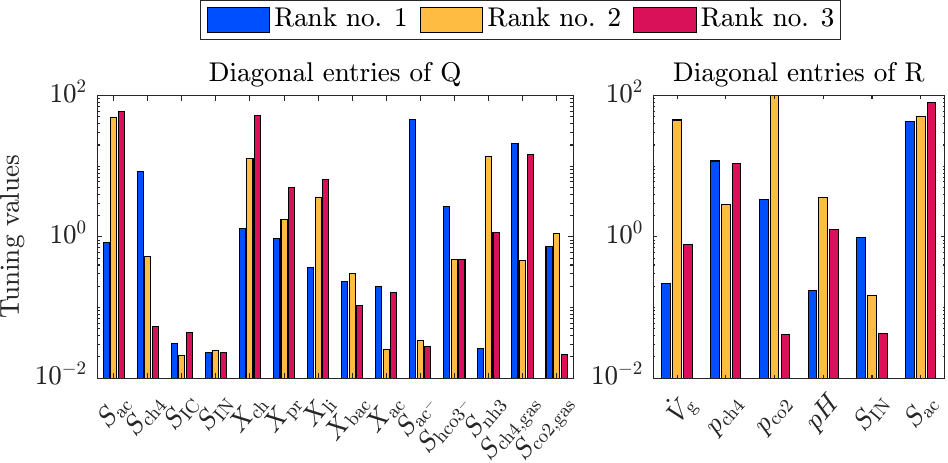}
    \caption{Top 3 tuning factors for $Q$- and $R$-matrix for criterion L1 norm of NRMSE of all states ($\mathrm{NRMSE_x}$) with resulting values of 5.79, 7.57, and 7.86, respectively.}
    \label{fig:res:top_tunings}
\end{figure*}

Fig.~\ref{fig:res:top_tunings} shows the diagonal entries of $Q$ and $R$ of the three best-ranked tunings of Fig.~\ref{fig:res:ranking_tuning} in absolute coordinates.
Their numeric values differ substantially (note the log scale), while their corresponding error function gives very similar values ($\mathrm{NRMSE_x}=$ 5.79, 7.57, and 7.86 for best ranked tunings no. 1, 2, and 3). This suggests that there exist many locally optimal tunings \cite{Boulkroune.2023,Chen.2024} instead of a unique global optimum. Comparing the best-ranked tuning values according to different criteria, e.g., $\mathrm{NRMSE_x}$, $\mathrm{NRMSE_y}$, and Boulkroune's error function shows that widely different tunings persist, cf. Fig.~\ref{fig:res:topX_tuning_factors}.

\begin{figure}
\centering
\includegraphics[width=0.7\linewidth]{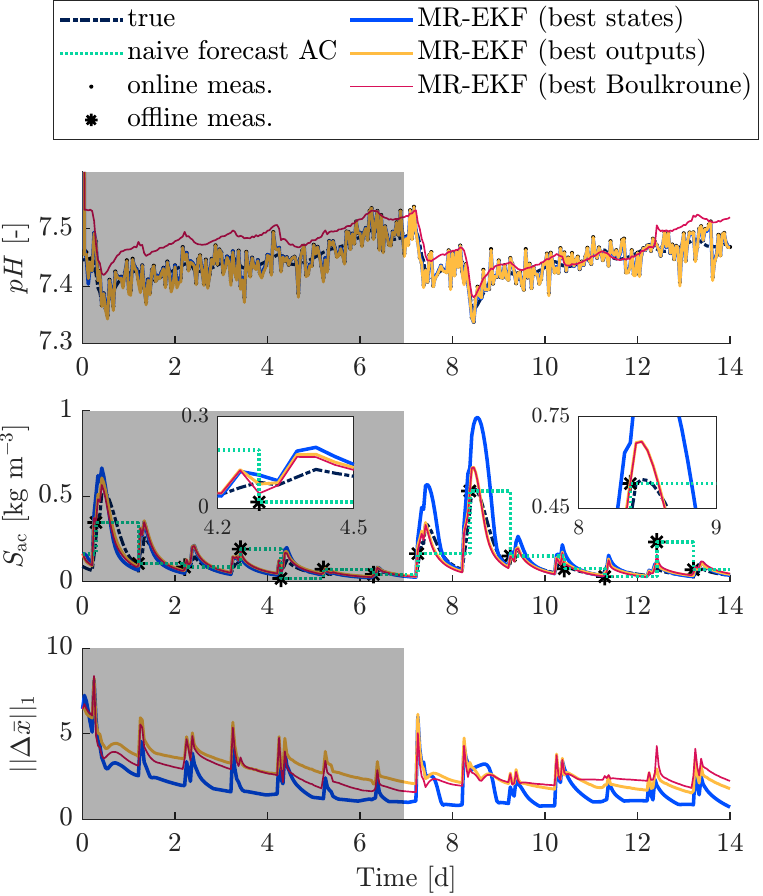}
\caption{MR-EKF estimates of pH (top), acetic acid (center) and L1-norm of normalized state estimation error (bottom) for three best-ranked tunings according to $\mathrm{NRMSE_x}$ (blue), $\mathrm{NRMSE_y}$ (yellow), and Boulkroune's error function (red). For the offline output acetic acid, the naïve forecasts (zero-order hold, ZOH) are shown. The first \SI{7}{\day} were ignored for error function calculations and are shaded out in gray. Zero delay was assumed, otherwise medium MR-EKF parameters as per Table~\ref{tab:meth:variants_overview}.}
\label{fig:res:gs_results_EKF_perf}
\end{figure}

Fig.~\ref{fig:res:gs_results_EKF_perf} shows the MR-EKF performance for the best-ranked tunings according to the three criteria $\mathrm{NRMSE_x}$ (blue), $\mathrm{NRMSE_y}$ (yellow), and Boulkroune's error function (red) by means of pH (top), acetic acid (AC, center), and the L1-norm of normalized state errors.\footnote{Computed as $\frac{|\hat x_k - x_k|}{\frac{1}{N} \sum_{k=1}^{N} |x_k|}$, with the number $N$ of online time steps.}

%

% %% BULLSHIT: 
% \begin{figure}
% \centering
% \begin{subfigure}[b]{\linewidth}
%     \centering    
%     \includegraphics[width=0.9\linewidth]{images/results/gridsearch_EKF_perf_pH.pdf}
%     \caption{pH}
%     \label{fig:res:gs_results_EKF_perf_pH}
% \end{subfigure}
% %
% \begin{subfigure}[b]{\linewidth}
%     \centering    
%     \includegraphics[width=0.9\linewidth]{images/results/gridsearch_EKF_perf_ac.pdf}
%     \caption{AC}
%     \label{fig:res:gs_results_EKF_perf_ac}
% \end{subfigure}
% % 
% \begin{subfigure}[b]{\linewidth}
%     \centering
%     \includegraphics[width=0.9\linewidth]{images/results/gridsearch_EKF_perf_L1.pdf}
%     \caption{L1 norm}
%     \label{fig:res:gs_results_EKF_perf_L1}
% \end{subfigure}
% % 
% \caption{MR-EKF estimates of pH (top), acetic acid (center) and L1-norm of normalized state estimation error (bottom) for three best-ranked tunings according to $\mathrm{NRMSE_x}$ (blue), $\mathrm{NRMSE_y}$ (yellow), and Boulkroune's error function (red). For the offline output acetic acid, the naïve forecasts (zero-order hold, ZOH) are shown. The first \SI{7}{\day} were ignored for error function calculations and are shaded out in gray.}
% \label{fig:res:gs_results_EKF_perf}
% \end{figure}
%
The pH filtering differs strongly depending on the tuning. The tunings of the best $\mathrm{NRMSE_x}$ and $\mathrm{NRMSE_y}$ show little smoothing and follow the noisy measurements almost directly, which implies overfitting. On the contrary, Boulkroune's tuning shows strong smoothing despite an initially high estimation error and good convergence to the true pH (unknown to the filter). The positive bias in pH estimation during the last 2 days is due to a strong positive correction in inorganic nitrogen (IN, plot not shown), which is plausible in light of the strong coupling of pH and IN in the ADM1-R3 model.

In the center plot of Fig.~\ref{fig:res:gs_results_EKF_perf} showing AC estimates, the green dashed line additionally shows naïve forecasts of the delayed offline output AC by means of a zero-order hold (ZOH). Clearly, the model-based estimation of the MR-EKF is superior to the naïve forecasts, and anticipates much better the dynamic course of acid concentrations resulting from substrate feeding events, cf. Fig.~\ref{fig:feeding_pattern}. Furthermore, the AC estimates clearly show filter corrections at measurement times (stars), especially at day 4 (for the best Boulkroune tuning) and day 8 (for best $\mathrm{NRMSE_x}$ tuning), cf. the insets in Fig.~\ref{fig:res:gs_results_EKF_perf}. The tunings of the best $\mathrm{NRMSE_y}$ and Boulkroune shows almost identical AC estimates close to ground-truth values, while the tuning of the best $\mathrm{NRMSE_x}$ shows higher deviations especially after high substrate feedings (cf. day 1 and 8 in Fig.~\ref{fig:feeding_pattern}).

Lastly, the bottom graph of Fig.~\ref{fig:res:gs_results_EKF_perf} shows the L1-norm of normalized state estimation errors, which indicates the overall state estimation performance. All tunings show a decreasing estimation error over time, but retain small and fluctuating permament state errors. The tuning according to $\mathrm{NRMSE_x}$ (blue graph) of course shows the lowest values in this metric. As for AC, the tunings of best $\mathrm{NRMSE_y}$ (yellow) and best Boulkroune (red) show similar performance with slightly lower permanent estimation errors for $\mathrm{NRMSE_y}$ as of day 10. 

Overall, the best-ranked tuning of Boulkroune's error function (red graphs in Figs.~\ref{fig:res:gs_results_EKF_perf} and \ref{fig:res:topX_tuning_factors}) was selected for the systematic investigation of MR-EKF parameters in the following section, because it delivers a good compromise between appropriate smoothing and correction of online and offline measurements. Further, it exhibits convergence for all state errors. Finally, this metric can be applied in real applications where ground truth is unknown.

While many applications of state estimation in the literature simply use a tuning based on trial and error \cite{Gudi.1995,Zhao.2015,Kemmer.2023}, the tuning presented in this tutorial was chosen in a systematic and pragmatic fashion. Considering only the NIS for Kalman filter tuning has also been reported by others \cite{Chen.2024,Chen.2019}. While Boulkroune et al. (2023) used a gradient-based optimizer for improving their tuning \cite{Boulkroune.2023}, they considered a vastly smaller search space. Additionally they acknowledge that their objective function does not show a unique global optimum \cite{Boulkroune.2023}. 

On the other hand, gradient-free optimizers such as genetic algorithm could help improving the tuning \cite{Zhang.2020}. Moreover, Bayesian optimization could be a suitable tool for improving the grid search since the Kalman updates are highly stochastic \cite{Chen.2019}, and because the recursive nature of the Kalman filter updates prohibits an analytical closed form solution of the error function, which renders gradient-based optimizers rather unsuitable. In any case, an appropriate choice of the dimensionality of the search space \cite{Chen.2024} as well as the error function are critical aspects for optimal tuning of the MR-EKF. 
\section{Sensitivity study of the MR-EKF \mbox{parameters}}
\label{sec:res:systematic_influence_parameters}
The following section systematically discusses the influence of the four MR-EKF parameters (delay, measurement noise, PMM, and initial state error) on state estimation performance. Furthermore, their effects on state estimation performance are summarized and compared. The best tuning according to criterion 3 was assumed, and MR-EKF parameters as per Table~\ref{tab:meth:variants_overview}.  
\subsection{Delay}
\label{sec:res:delay}
\begin{figure}
\centering
\includegraphics[width=0.7\linewidth]{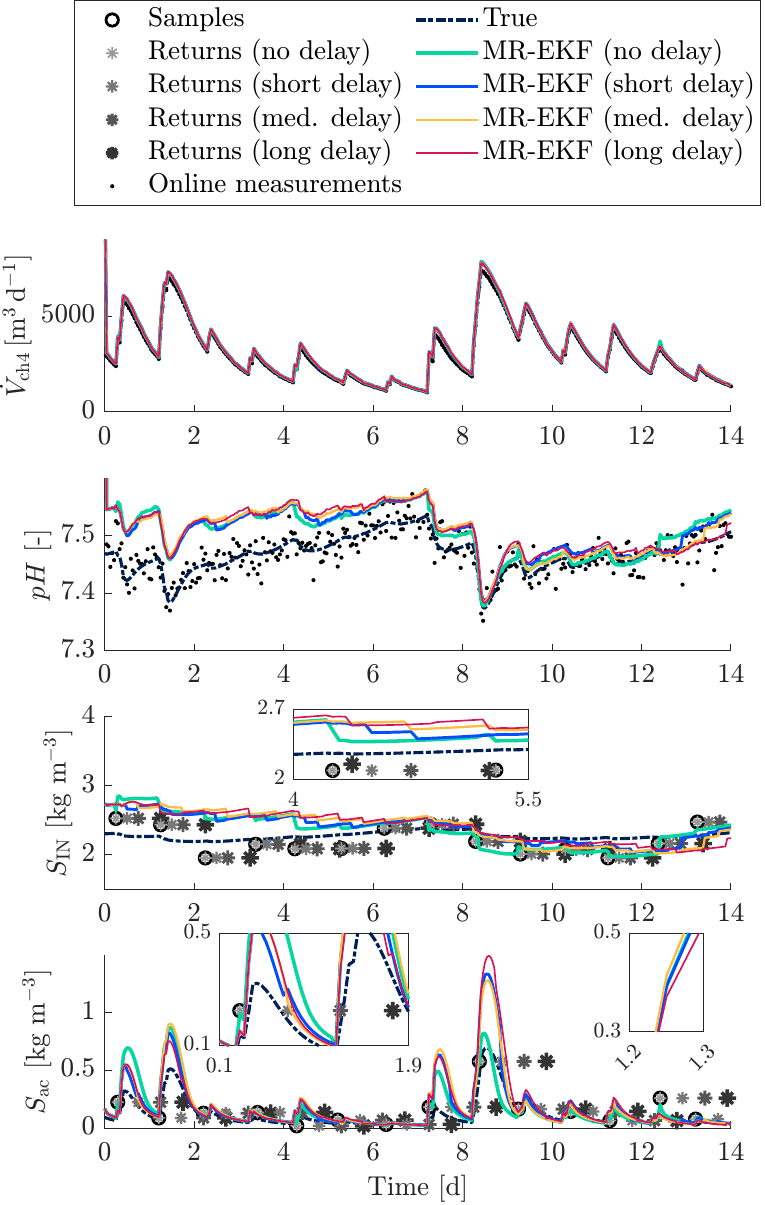}
\caption{Ground truth and MR-EKF estimates for different delay lengths with tuning according to criterion 3: online outputs methane production (first) and pH (second), offline outputs IN (third) and AC (fourth). Offline samples (circles) are fixed, while offline returns (stars) are subject to different delays. \textit{med.} abbreviates \textit{medium}.}
\label{fig:res:diff_delays_y}
\end{figure}

Fig.~\ref{fig:res:diff_delays_y} shows the state estimation performance of the MR-EKF for different delay lengths (cf. Table~\ref{tab:meth:variants_overview}) by means of subplots for the two online outputs \chfour production (first) and pH (second), and the offline outputs IN (third) and AC (fourth). %Legend entries apply for all graphs in the subplots. 

%

% %% BULLSHIT:
% \begin{figure}
% \centering
% \begin{subfigure}[b]{\linewidth}
%     % \centering 
%     \raggedleft   
%     \includegraphics[width=0.9\linewidth]{images/results/rel_outputs_for_diff_delays_criterion_3_ch4.pdf}
%     \caption{Methane production}
%     \label{fig:res:diff_delays_y_ch4}
% \end{subfigure}
% %
% \begin{subfigure}[b]{\linewidth}
%     % \centering    
%     \raggedleft
%     \includegraphics[width=0.9\linewidth]{images/results/rel_outputs_for_diff_delays_criterion_3_pH.pdf}
%     \caption{pH}
%     \label{fig:res:diff_delays_y_pH}
% \end{subfigure}
% %
% \begin{subfigure}[b]{\linewidth}
%     % \centering 
%     \raggedleft   
%     \includegraphics[width=0.9\linewidth]{images/results/rel_outputs_for_diff_delays_criterion_3_IN.pdf}
%     \caption{IN}
%     \label{fig:res:diff_delays_y_IN}
% \end{subfigure}
% % 
% \begin{subfigure}[b]{\linewidth}
%     % \centering
%     \raggedleft
%     \includegraphics[width=0.9\linewidth]{images/results/rel_outputs_for_diff_delays_criterion_3_ac.pdf}
%     \caption{AC}
%     \label{fig:res:diff_delays_y_ac}
% \end{subfigure}
% % 
% \caption{Ground truth and MR-EKF estimates for different delay lengths: online outputs methane production (top) and pH (second); offline outputs NH4-N (third) and AC (bottom). Offline samples (circles) are fix, while offline returns (stars) are subject to different delays. \textit{med.} abbreviates \textit{medium}.}
% \label{fig:res:diff_delays_y}
% \end{figure}
%

While the delay does not show a significant influence on the online outputs, different estimates become particularly evident for IN. Naturally, with increasing delay, offline measurements are returned later, which results in later corrections. Appropriate tuning according to Boulkroune's error function delivers strong corrections which made in the direction of the innovations. This is frequently reported for multirate systems, both for Kalman filtering \cite{Gudi.1995,Zhao.2015} and MHE \cite{Elsheikh.2021}, specifically for variable-structure outputs \cite{Kramer.2005b} which apply for this tutorial, cf. Eq.~\eqref{eq:system_model_output}. 

Since all estimated trajectories of IN start from the same initial condition with values higher than ground truth, and synthetic measurements are scattered around ground truth, all filters systematically correct towards lower values. After about 7$-$\SI{8}{\day}, the estimates of all four delay scenarios have closely approached ground truth, where the green graph (no delay) tends to converge faster than the others. However, the influence of the delay length is less clear than expected intuitively, since for a MIMO system Kalman gains are also affected by innovations in the other outputs, the given tuning, and the PMM. For some samples (e.g., at around day 4, cf. the inset in Fig.~\ref{fig:res:diff_delays_y}), longer delays result in postponed corrections in the same, true direction. Conversely, this does not hold for the sample drawn at day~8, where the no delay graph (green) shows the largest estimation error as opposed to the long delay graph (red), which shows almost no estimation error. 

After the initial convergence at around day 7$-$8, all estimates keep oscillating around ground truth depending on the sign of innovation as well as the corrections in the other outputs.

For AC, the influence of different delays is less pronounced. However, corrections upon offline returns are well visible, e.g., for returns of the first drawn AC sample, returning at days 0.29, 0.75, 1.25 and 1.75, cf. the insets in Fig.~\ref{fig:res:diff_delays_y}.\footnote{The return of AC at day 1.25 for medium delay (yellow plot) does not show a strong correction as it coincides with a return of IN (negative innovation) and high substrate feed, cf. Fig.~\ref{fig:feeding_pattern}.} As the innovation is positive, i.e., the measurement is larger than the estimate, upward corrections are made. With increasing delay, the corrections are executed later. However, this does not mean automatically that earlier corrections result in better estimates, because the initial error was already positive, and is further amplified by an upward correction, cf. the insets in Fig.~\ref{fig:res:diff_delays_y}. Yet, once the initial estimation error has been corrected (at around day 7), indeed earlier corrections tend to result in better estimates, especially for high ground-truth AC peaks, well visible, e.g., for the sample drawn at day 8. Here, clearly the red graph (long delay) shows the largest estimation error and the no delay graph (green) the smallest.

The corrections in offline outputs also affect the pH estimates (second subplot in Fig.~\ref{fig:res:diff_delays_y}). The higher the AC estimates and the lower the ones of IN, the lower the pH. Note, however, that IN contributes to the total alkalinity, which compensates the effect of high acids on the pH \cite{Weinrich2021}. Therefore, only for the last 2 days, clearly higher pH estimates become visible as the IN estimate sequentially increase for different delays, while the AC estimates do not differ substantially. Otherwise, the noisy pH measurements are smoothed well thanks to the tuning of Boulkroune, which puts a penalty on high innovation variances, cf. Eq.~\eqref{eq:boulkroune_error_fun}. 

Fig.~\ref{fig:res:L1_shares_delays_bestBoulkroune} summarizes the influence of different delays on overall state estimation performance. On the left, the L1-norm of normalized state estimation errors are shown over time, and the relative contribution of the three most influential states on the right. While for all delay lengths a clear convergence can be observed, shorter delays do not always reflect in lower estimation errors. This is plausible in light of the intertwined connection between the output innovations and the 14 model states, which makes it more challenging to directly link the delay of a specific offline measurement to its estimation, as was shown in Fig.~\ref{fig:res:diff_delays_y}. Yet, in Fig.~\ref{fig:res:L1_shares_delays_bestBoulkroune} the tendency of higher errors for longer delays can be observed after initial convergence, particularly at high feeding events (around day 7 and 8.5).  

At least two out of three macronutrients ($X_\mathrm{ch}$, $X_\mathrm{pr}$, or $X_\mathrm{li}$) are always among the top three contributors to the total estimation error, cf.~Fig.~\ref{fig:res:L1_shares_delays_bestBoulkroune}. Furthermore, the dissociated part of AC ($S_\mathrm{ac^-}$) appears in three out of four scenarios, cf. Table~\ref{tab:meth:variants_overview}. The remaining states are summarized as ``rest''. 
\subsubsection{Separate offline delays}
\begin{figure}[h]
    \centering
    \includegraphics[width=0.7\linewidth]{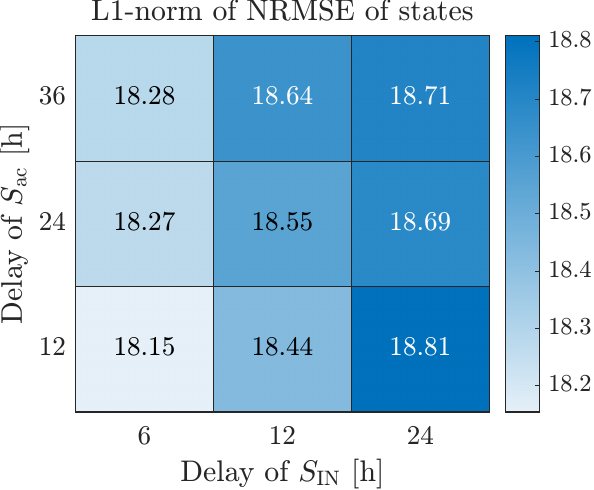}
    \caption{Heatmap of the L1-norm of NRMSE of states ($\mathrm{NRMSE_x}$) for individually varied offline delays for $S_\mathrm{ac}$ and $S_\mathrm{IN}$ with tuning according to criterion 3.}
    \label{fig:res:heatmap}
\end{figure}

The specific advantage of the MR-EKF over a conventional EKF is to handle delayed offline measurements. Therefore, Fig.~\ref{fig:res:heatmap} shows the influence of separate delays on offline measurements IN and AC.

Again, the $\mathrm{NRMSE_x}$ was chosen as an error metric. As opposed to Fig.~\ref{fig:res:diff_delays_y}, where both offline delays on IN and AC were varied in tandem, the delays of IN and AC were varied individually, with all other MR-EKF parameters ($k_\sigma, k_\theta, k_\mathrm{x}$) kept at default values, cf. Table~\ref{tab:meth:variants_overview}.

The longer the mean delay (diagonal of the heatmap in Fig.~\ref{fig:res:heatmap} from bottom left to top right), the higher the estimation error. However, longer IN delays increase the total estimation error more clearly than longer AC delays. This occurs because of the stronger corrections on innovations of IN compared with AC, which of course strongly depends on the chosen tuning. For the tuning according to criterion 1 ($\mathrm{NRMSE_x}$), the relative importance of AC delays is higher than for IN delays (plot not shown), which underlines the critical importance of finding an adequate Kalman filter tuning.

While longer delays clearly diminish the state estimation performance, the MR-EKF does not diverge even for long delays. The main reason is that the ADM1-R3 model is still observable even if only the online measurements are available, cf. Sec.~\ref{sec:meth:observability_adm1_r3}. Therefore, even delays longer than the ones considered do not cause the MR-EKF to diverge (plots not shown). 
However, the interplay between offline delays and sampling frequency is indeed limited numerically by the total number of pending offline samples due to the increasing computational load. With the given ADM1-R3 model and its 14 states, an augmentation of about 5 pending samples was identified as the maximum possible augmentation. This strongly depends on the model complexity and the available computational power. Moreover, for models which lose observability without offline measurements, extended delays could indeed cause filter divergence, if unobservable states are not detectable \cite{Tatiraju.1999}.   
\subsection{Measurement noise}
\begin{figure}
    \centering
    \includegraphics[width=0.7\linewidth]{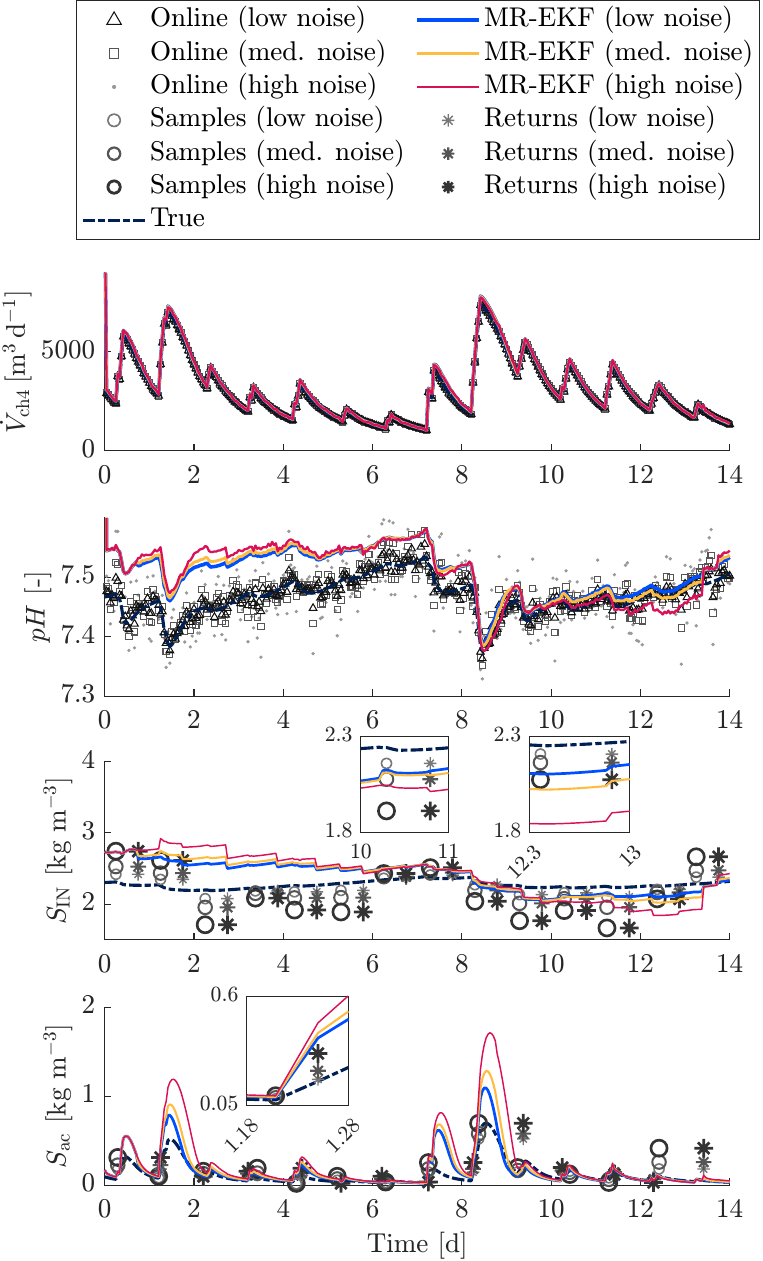}
    \caption{Ground truth and MR-EKF estimates for different measurement noise levels (all other MR-EKF parameters at default values) shown by means of output signals methane production (first), pH (second), and offline outputs ammonium nitrogen (third) and acetic acid (fourth). Noise levels of online measurements are shown by individual shapes. For offline measurements, samples (circles) and returns (stars) are indicated by individual marker sizes and shades (the bigger and darker, the higher the noise level). \textit{med.} abbreviates \textit{medium}. Tuning according to criterion 3.}
    \label{fig:res:diff_meas_noise_y}
\end{figure}

Levels of synthetic measurement noise were varied as low, medium, and high (cf. Table~\ref{tab:meth:variants_overview}). Corresponding online and offline measurements are indicated by different grey symbols in Fig.~\ref{fig:res:diff_meas_noise_y} (the darker, the higher the noise level). The corresponding estimates in Fig.~\ref{fig:res:diff_meas_noise_y} differ much less than for the different delays in Fig.~\ref{fig:res:diff_delays_y}. 

For \chfour production (first subplot), the signal-to-noise ratio is hardly affected, which explains that the three estimated trajectories almost overlap. For pH (second subplot), the estimates differ more, with less smoothing for higher measurement noise, and hence larger innovations. 

The innovations of the offline measurements naturally increase with the noise level. So do the amplitudes of the filter corrections, cf. Eq.~\eqref{eq:meth:MU_x}, which is shown in the right inset of the IN estimates in Fig.~\ref{fig:res:diff_meas_noise_y} (third subplot). Directions of the filter corrections then depend on the sign of the innovations, which are determined by the relative location of samples and sample time estimates: In the left inset of the IN subplots in Fig.~\ref{fig:res:diff_meas_noise_y}, the low noise graph shows a positive correction, whereas both the medium and high noise graphs show negative ones. This indicates that increasing measurement noise does not necessarily deliver worse estimates. 

A similar tendency can be observed for AC (fourth subplot in Fig.~\ref{fig:res:diff_meas_noise_y}). Stronger corrections but also higher estimation errors can be observed with increasing noise level, cf. the inset for AC in Fig.~\ref{fig:res:diff_meas_noise_y}. %\footnote{Mind that samples in the inset of AC do not belong to the returns, but already pertain to the next returns.} 
Even so, most of the time, ground-truth values are relatively low, and individual estimates hardly differ. This changes for periods of elevated feed mass (day 1 and 8, cf. Fig.~\ref{fig:feeding_pattern}) and high AC peaks. 

Overall, the state estimation error increases with increasing measurement noise \cite{Tatiraju.1999,Gopalakrishnan.2011}, especially during high feedings \cite{Tuveri.2021}, which is summarized in Fig.~\ref{fig:res:L1_shares_meas_noise_bestBoulkroune}. However, estimation performance was not severly diminished even for high noise levels \cite{Perea.08202007}. Notably, medium noise was assumed for the tuning grid search (cf. Table~\ref{tab:meth:variants_overview}). Yet, the tuning optimization did not overfit the synthetic measurement data, as medium noise estimates show no better performance than low noise estimates.

As for different delays, the macronutrient concentrations range among the most influential states with respect to total estimation error, especially toward more exact measurements, i.e., lower measurement noise, cf. right side of Fig.~\ref{fig:res:L1_shares_meas_noise_bestBoulkroune}. With worse measurements, i.e., towards higher noise, total and dissociated AC ($S_\mathrm{ac}$ and $S_\mathrm{ac^-}$, respectively) increasingly dominate the total estimation error. This highlights the importance of timely feedback especially for unreliable VFA measurements, as well as suitable tuning for reliable state estimation of a dynamically operated AD system. For online measurements in experimental and industrial applications, a re-tuning could be triggered if the sensor noise changes permanently \cite{Chen.2019}.  
\subsection{Plant-model mismatch}
\label{sec:res:pmm}
\begin{figure}
    \centering
    \includegraphics[width=0.7\linewidth]{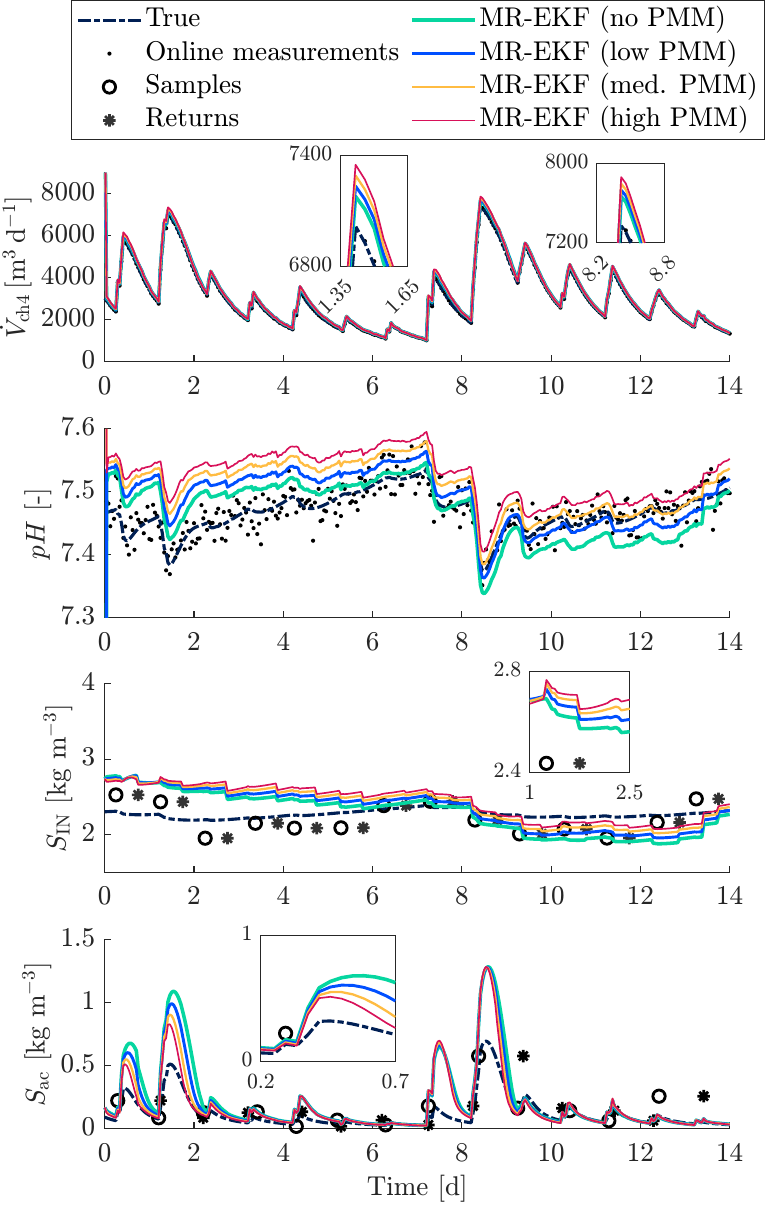}
    \caption{Estimation performance of the MR-EKF for different levels of plant-model mismatch by means of online outputs methane production (first) and pH (second), and the offline outputs IN (third) and AC (fourth). \textit{med.} abbreviates \textit{medium}. Tuning according to criterion 3.}
    \label{fig:res:diff_pmm_y}
\end{figure}

To show the effect of a PMM on estimation performance, all time-variant model parameters $\theta$ were incrementally increased relative to ground truth, cf. Table~\ref{tab:meth:time_variant_params}. The tuning was kept constant, which was optimized for medium PMM, cf. Table~\ref{tab:meth:variants_overview}. The positive PMM manifests itself in multiple aspects, which will be explained by means of Fig.~\ref{fig:res:diff_pmm_y}. 

The PMM delivers substantially different smoothings of \chfour production. The higher the PMM, the higher the trajectories obtained, cf. the insets in the top subplot of Fig.~\ref{fig:res:diff_pmm_y}. This is plausible since increasing the hydrolysis constants and the maximum growth rate of AC (cf. Table~\ref{tab:meth:time_variant_params}) leads to higher \chfour production rate \cite{Segura.2025,Weinrich2021}. However, this effect decreases as the AC estimates gradually overlap, cf. bottom subplot of Fig.~\ref{fig:res:diff_pmm_y}. To this end, higher AC concentrations result in lower pH, and thus stronger pH inhibition of acetoclastic methanogens \cite{Chen2008}. Yet, it is well visible that \chfour is formed almost instantaneously upon AC buildup, which is realistic for AD of agricultural substrates, where hydrolysis is known to be the rate-limiting step in \cite{Noike.1985,Weinrich2021}. 

For the pH, a strong bias effect of PMM can be observed in Fig.~\ref{fig:res:diff_pmm_y}, with high PMM resulting in clearly higher pH estimates. In the pH modeling of ADM1-R3, $\theta_8=S_\mathrm{ion,eff}$ acts as addtional cations, cf. Eqs.~\eqref{eq:R3-Core:y_ph} and \eqref{eq:R3-Core:SHPlus}, which lift the pH \cite{Bernard2001,PrasannaKumar.2024}. Within the first week, the differences between individual graphs grow slightly, but remain about constant thereafter.

This matches with the convergence behaviour of IN, where individual graphs in the third subplot of Fig.~\ref{fig:res:diff_pmm_y} diverge from another within the first week, and maintain their relative bias thereafter. All four IN estimates deliver decreasing trajectories, but steeper slopes and hence lower estimates are obtained with increasing PMM. This is plausible since 
$\theta_8=\varphi_\mathrm{IN}$ amplifies the influent concentration of IN, which reflects in higher IN estimates. However, measurement updates are hardly affected by the PMM, as similar innovations result in similar corrections, cf. the inset in the third subplot of Fig.~\ref{fig:res:diff_pmm_y}. 

Lastly, lower estimates of AC concentrations in the bottom subplot of Fig.~\ref{fig:res:diff_pmm_y} are obtained with higher PMM. Again, this can be explained with the underlying model dependencies, and underlines that the PMM mostly affects the time update. $\theta_5=\mu_\mathrm{m,ac}$ accelerates the degradation of AC and the formation of \chfour \cite{Segura.2025,Weinrich2021}, which consequently results in lower AC estimates, cf. the inset for AC in Fig.~\ref{fig:res:diff_pmm_y}. Differences between the graphs almost vanish after about 1 week, when permanent estimation errors remain in a similar range for individual PMM levels. 

Overall, the effect of the parametric PMM can be attributed to the model predictions during the time update, as differences between the trajectories accumulate gradually over time and measurement corrections do not differ noticeably for different PMM. 

Unlike for delay and noise, the total estimation error in Fig.~\ref{fig:res:L1_shares_pmm_bestBoulkroune} does not solely show the largest discrepancies at high feeding events. While differing trajectories during the AC peaks are still noticeable in the total estimation error over time and its top three contributors, the different permanent estimation errors are predominantly grounded in different estimates of the nonmeasurable macronutrients, cf. the right pane of Fig.~\ref{fig:res:L1_shares_pmm_bestBoulkroune}. As for AC and IN, the initial state error of macronutrients was chosen positive, cf. Table~\ref{tab:initial_estimate}. This positive estimation error is diminished faster for higher values of tunable model parameters such as hydrolysis constants, i.e., towards higher PMM. It is therefore reasonable that higher PMM deliver faster convergence, although this holds true only for the chosen setup. 

A PMM was reported to affect the bias in state estimation many times in the literature. This holds for Kalman filtering \cite{Gopalakrishnan.2011,Li.2004}, MHE \cite{Tuveri.2023,Bae.2021} and even observers with analytically proven global convergence \cite{Tatiraju.1999}. Depending on the parameter sensitivity of model predictions, highly influencial parameters could be jointly estimated along with the states \cite{Li.2004}. However, even for numerically advanced approaches, and particularly for structural PMM, persistent biases may remain \cite{Bae.2021,Li.2004,Tuveri.2023}. 
\subsection{Initial state estimation error}
\begin{figure}
    \centering
    \includegraphics[width=0.7\linewidth]{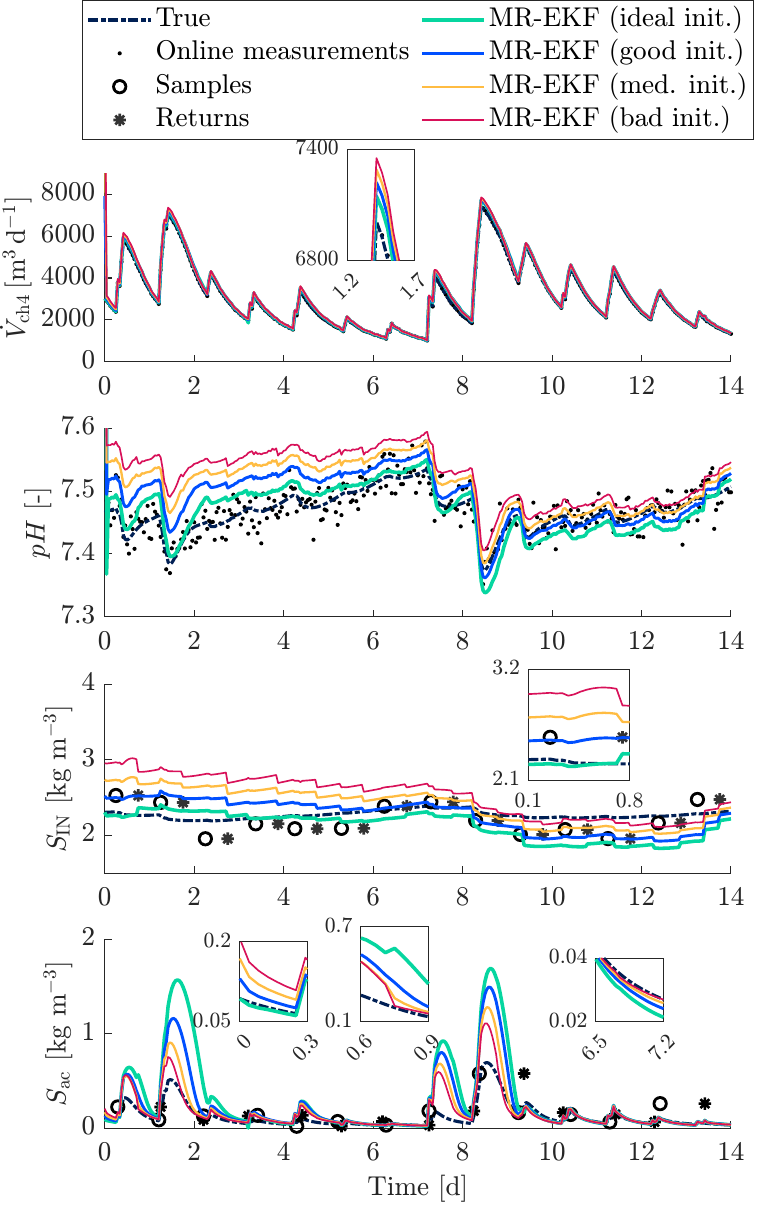}
    \caption{Ground truth and MR-EKF estimates for different initial state estimation errors (all other MR-EKF parameters at default values) by means of online outputs methane production (first) and pH (second), and the offline outputs IN (third) and AC (fourth). \textit{med.} abbreviates \textit{medium}. Tuning according to criterion 3.}
    \label{fig:res:diff_init_error_y}
\end{figure}

Fig.~\ref{fig:res:diff_init_error_y} shows the MR-EKF estimates for varying initial state error $\hat x_0$, where perturbations $\Delta x_0$ from ground truth were chosen according to Table~\ref{tab:initial_estimate}. 

The different initial values manifest in slightly different trajectories of \chfour production, which is most evident upon high feeding events (e.g., during day 2, cf. the inset of \chfour production in Fig.~\ref{fig:res:diff_init_error_y}). The tuning otherwise inclines towards strong smoothing, such that the model predictions during time updates clearly dominate the data-based corrections during measurement updates. 

The clearly different pH trajectories in Fig.~\ref{fig:res:diff_init_error_y} can be attributed to different initial states\footnote{The relative initial errors in ionic mass concentrations translate into clearly higher molar concentrations of total cations, Eq.~\eqref{eq:R3-Core:SHPlus}, which lifts the pH towards higher initial values.} (cf. Table~\ref{tab:initial_estimate}) mostly of IN, and the slow convergence of IN relative to AC. 

The slow convergence of IN has already been observed for the three previously altered MR-EKF parameters, and is rooted in the tuning and the variable structure measurements considered \cite{Kramer.2005}. The higher the initial state error, the larger the innovations, and the stronger subsequent corrections at returns, cf. the inset in the subplot for IN concentration. Again, the sign of the innovation at the sample time is critical for the direction of the correction.

AC shows a counterintuitive but plausible behavior. On the one hand, the initial errors are all nonnegative, cf. the left inset for AC in Fig.~\ref{fig:res:diff_init_error_y}, and larger initial state estimates are obtained for higher values of $k_\mathrm{x}$, cf. Eq.~\eqref{eq:initial_state_estiate_perturbation}. On the other hand, the feeding events change the order of the graphs, cf. the center inset and peak at around day~2. This can be explained with the relative dominance of convection over kinetic creation terms in the mass balance of AC, cf. Eq.~\eqref{eq:R3-Core:x_ac}. At fixed influent concentrations $\xi_1$ of AC, the effective influent mass flow $c_1 (\xi_1 - x_1)\, u$ of AC increases with higher state values $x_1$. This phenomenon gradually vanishes after the feedings, as can be seen in the right inset, where plots of increasing initial errors again result in higher estimates, although relative differences decrease over time, which indicates convergence. 

All four initial estimates eventually result in about equal trajectories after an initial convergence of around 10 days, as shown in Fig.~\ref{fig:res:L1_shares_init_errors_bestBoulkroune}. The decreasing dominance of feedings and associated convection terms of AC for higher initial errors is apparent at feeding events (in particular day 1 and 8), as well as the decreasing share of AC in total estimation errors. 

For Kalman filters, global convergence cannot be guaranteed in general \cite{Daid.2021}. This not only holds true for the EKF, which linearizes the nonlinear system locally, but also for higher-order approximations like the UKF \cite{Tuveri.2021}. While there exist strategies to increase the convergence robustness \cite{Kolas.2009,Perea.08202007}, most simulative studies consider only small initial perturbations \cite{Zhao.2015,Gopalakrishnan.2011}. However, even for nonlinear approaches such as MHE, cf. Sec.~\ref{sec:res:mhe}, local initial perturbations are frequently chosen \cite{Elsheikh.2021,Kramer.2005b} since the underlying arrival cost estimations are often based on linearizations \cite{LopezNegrete.2012}. In this light, the MR-EKF shows a reasonable robustness against increasing initial errors, which can be attributed to the appropriate tuning as well as the use of both online and offline measurements.
\subsection{Summary of MR-EKF parameter influence}
\begin{figure}[H]
\centering
\begin{subfigure}[b]{0.7\linewidth}
    \centering
    \includegraphics[width=\linewidth]{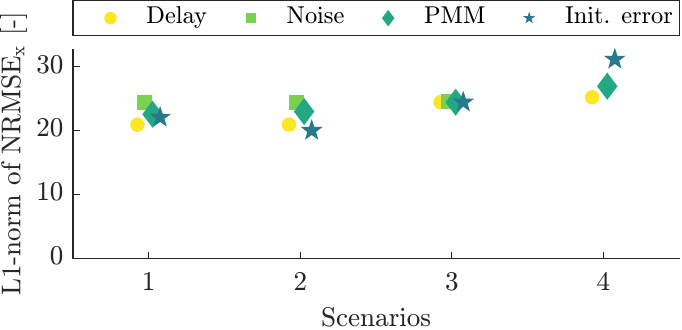}
    \caption{L1-norm of NRMSE of states}
    \label{fig:res:comparison_nrmse}
\end{subfigure}
\vskip 0.5cm
\begin{subfigure}[b]{0.7\linewidth}
    \centering
    \includegraphics[width=\linewidth]{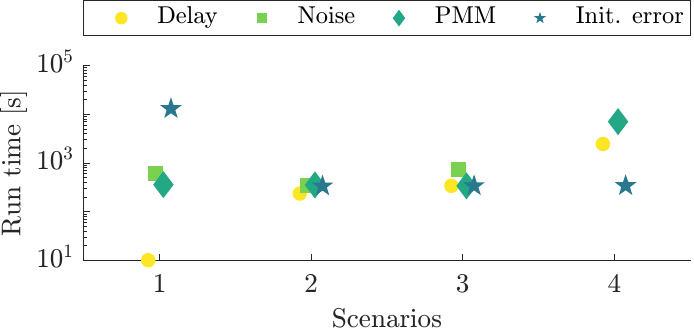}
    \caption{Run times}
    \label{fig:res:comparison_runtime}
\end{subfigure}
\caption{L1-norm of NRMSE of states and total run times for individual MR-EKF parameters. Markers were horizontally shifted from another to improve their visibility.}
\label{fig:res:comparison}
\end{figure}

Scenarios of the four MR-EKF parameters discussed (delay, measurement noise, PMM, and initial error) are summarized by means of total estimation errors (L1-norm of $\mathrm{NRMSE_x}$) in Fig.~\ref{fig:res:comparison_nrmse}, and total run times in Fig.~\ref{fig:res:comparison_runtime}. 

Clearly, the estimation performance under the given scenarios and tuning are predominantly influenced by initial state error (blue stars in Fig.~\ref{fig:res:comparison_nrmse}). By comparison, the other MR-EKF parameters show little effect \cite{Perea.08202007}. Note that despite the high peaks in AC concentrations during feeding events shown in Fig.~\ref{fig:res:L1_shares_init_errors_bestBoulkroune}, the slowly decreasing total initial error dominates the total value of $||\mathrm{NRMSE_x}||$.%\footnote{Note the different scales of the y-axes of the left subplots of Fig.~\ref{fig:res:L1_norm_and_shares}}. 

The run time is predominantly affected by the offline delays (yellow circles in Fig.~\ref{fig:res:comparison_runtime}), most notably by the level of augmentation during EKF runs. This is determined by the frequency of new samples before the return of previous samples. However, with a maximum runtime of about \SI{47}{\minute} for the longest delays, real-time capability is still ensured for the given ADM1-R3 model considering a total simulated time of \SI{14}{\day}.
\subsection{Comparison with alternative tuning of best $\mathrm{NRMSE_x}$}
\label{sec:res:alternative_tuning}
\begin{figure*}
\begin{adjustwidth}{-5mm}{-5mm} % widen both left and right by Xcm
    % row 1:
    \centering
    \begin{subfigure}[b]{0.51\textwidth}
        \centering
        \includegraphics[width=1\linewidth]{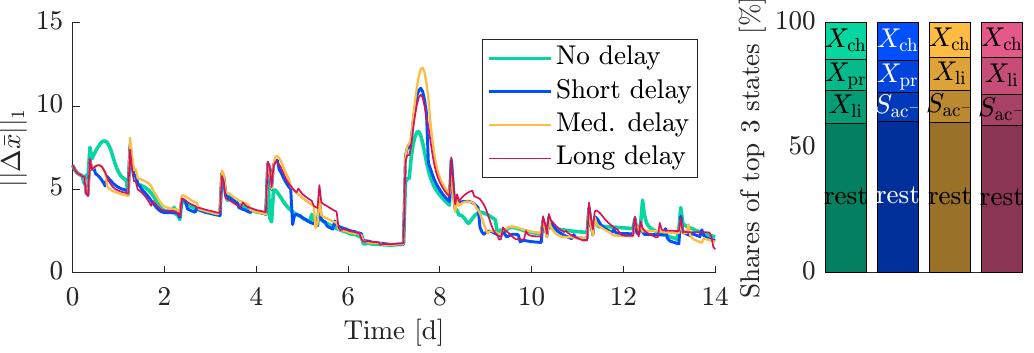}    
        \caption{offline delays (best Boulkroune)}
        \label{fig:res:L1_shares_delays_bestBoulkroune}
    \end{subfigure}
    \hfill
    \begin{subfigure}[b]{0.51\textwidth}  
        \centering 
        \includegraphics[width=1\linewidth]{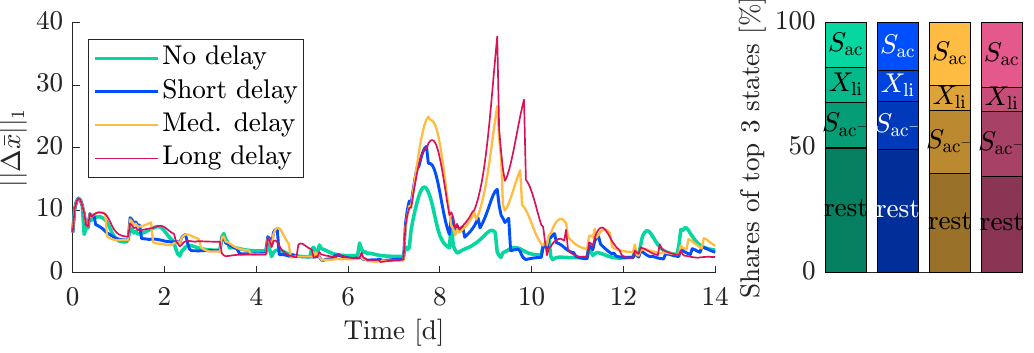}    
        \caption{offline delays (best states)}
        \label{fig:res:L1_shares_delays_bestStates}
    \end{subfigure}
    \vskip 1cm 
    % row 2:
    \begin{subfigure}[b]{0.51\textwidth}   
        \centering 
        \includegraphics[width=1\linewidth]{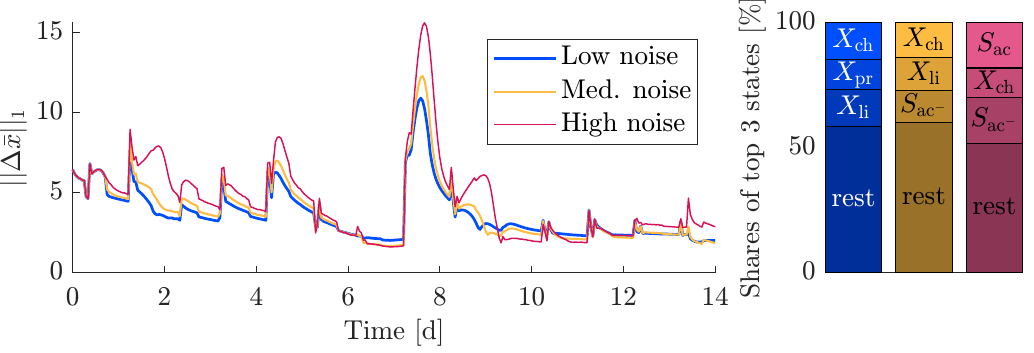}    
        \caption{measurement noise level (best Boulkroune)}
        \label{fig:res:L1_shares_meas_noise_bestBoulkroune} 
    \end{subfigure}
    \hfill
    \begin{subfigure}[b]{0.51\textwidth}   
        \centering 
        \includegraphics[width=1\linewidth]{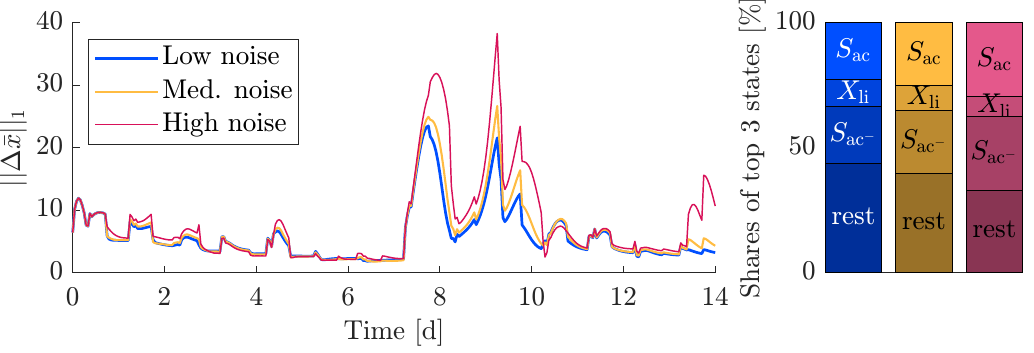}    
        \caption{measurement noise level (best states)}
        \label{fig:res:L1_shares_meas_noise_bestStates} 
    \end{subfigure}
    \vskip 1cm
    % row 3:
    \begin{subfigure}[b]{0.51\textwidth}   
        \centering 
        \includegraphics[width=1\linewidth]{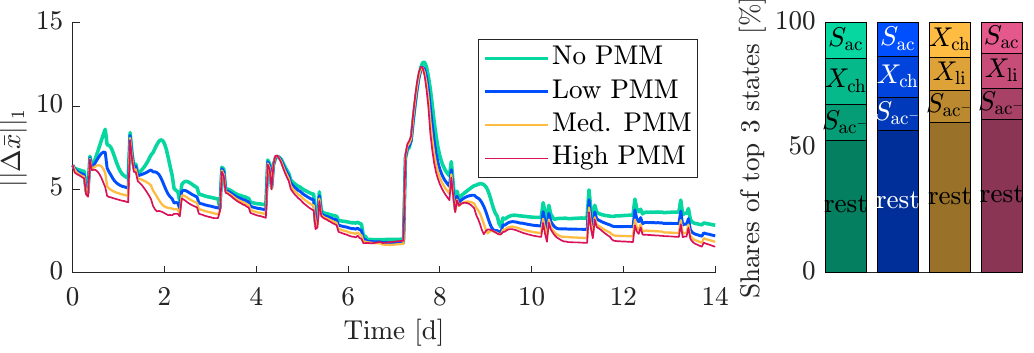}    
        \caption{plant-model mismatch (best Boulkroune)}
        \label{fig:res:L1_shares_pmm_bestBoulkroune} 
    \end{subfigure}
    \hfill
    \begin{subfigure}[b]{0.51\textwidth}   
        \centering 
        \includegraphics[width=1\linewidth]{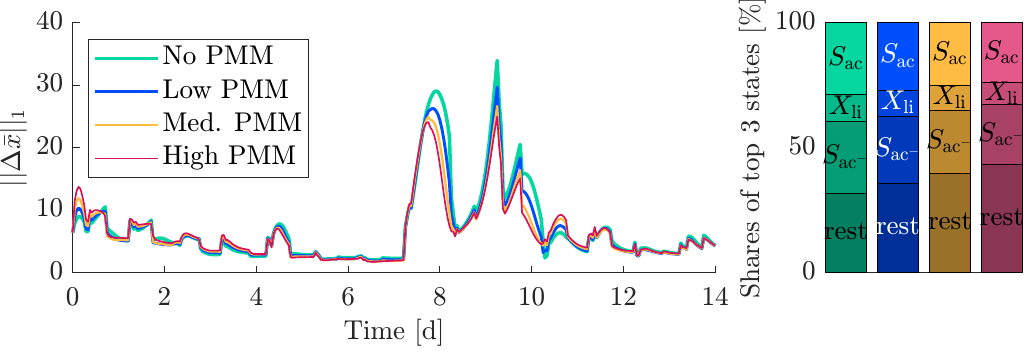}    
        \caption{plant-model mismatch (best states)}
        \label{fig:res:L1_shares_pmm_bestStates}
    \end{subfigure}
    \vskip 1cm
    % row 4:
    \begin{subfigure}[b]{0.51\textwidth}   
        \centering 
        \includegraphics[width=1\linewidth]{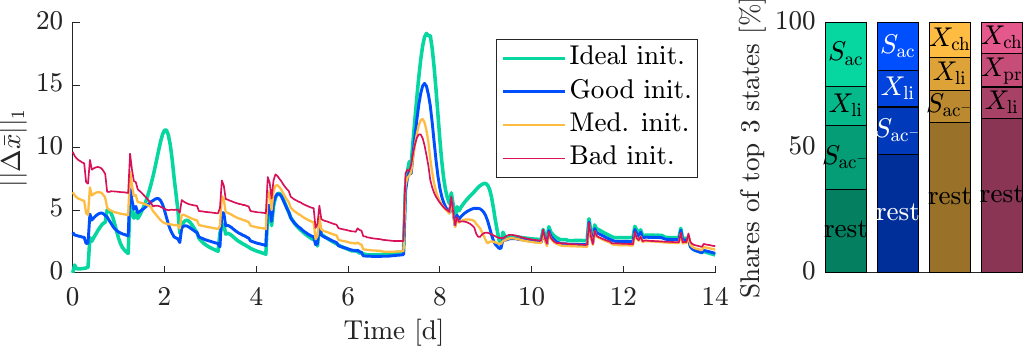}    
        \caption{initial estimation error (best Boulkroune)}
        \label{fig:res:L1_shares_init_errors_bestBoulkroune} 
    \end{subfigure}
    \hfill
    \begin{subfigure}[b]{0.51\textwidth}   
        \centering 
        \includegraphics[width=1\linewidth]{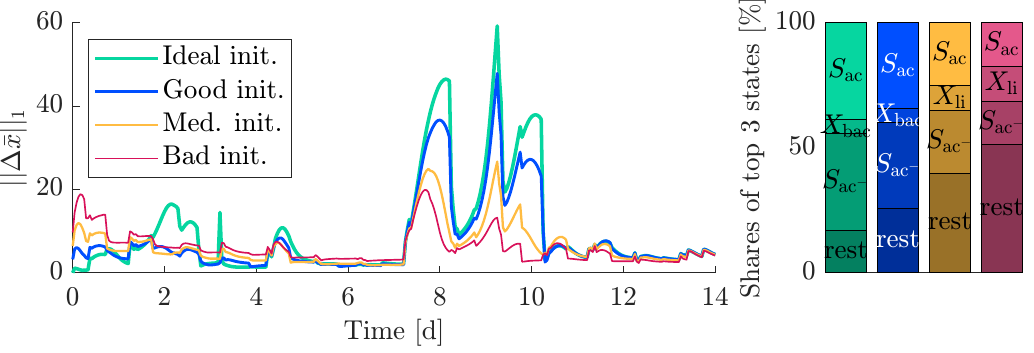}    
        \caption{initial estimation error (best states)}
        \label{fig:res:L1_shares_init_errors_bestStates} 
    \end{subfigure}
    \caption{L1-norm (in \si{\kilogram\per\cubic\meter}) of normalized estimation errors over time (left in each subplot pane) and top 3 shares of most contributing states in order of the state vector (right in each subplot pane). Rows pertain to all four investigated MR-EKF parameters (from top to bottom: delay, measurement noise, PMM, and initial error) and columns two different tuning criteria (left: criterion 3, i.e., best Boulkroune error, right: criterion 1, i.e., best $\mathrm{NRMSE_x}$). \textit{Med.} abbreviates \textit{Medium}. Note the different scales of the $y$-axes of the left and right subplots.}
    \label{fig:res:L1_norm_and_shares}
\end{adjustwidth}
\end{figure*}

In Section~\ref{sec:grid_search} the tuning of Boulkroune was chosen as the baseline tuning for the MR-EKF parameter investigations of Section~\ref{sec:res:systematic_influence_parameters}. It was pointed out many times that specific observations are strongly linked to the chosen tuning. To further illustrate this claim, Fig.~\ref{fig:res:L1_norm_and_shares} compares the total estimation errors as well as the relative contribution of the three most influential states of the best tuning according to Boulkroune (criterion 3, left) with those according to the best $\mathrm{NRMSE_x}$ (criterion 1, right). The color coding of the plots matches with those in Section~\ref{sec:res:systematic_influence_parameters}. 

Generally, much higher peak values of total estimation errors can be observed for the tuning according to criterion~1 (up to around \SI{60}{\kilogram\per\cubic\meter} compared to around \SI{20}{\kilogram\per\cubic\meter}). Moreover, the top three contributing states are mostly dominated by AC for criterion 1 instead of macronutrients and a more diverse group of other, lower ranked states (``rest'') for criterion 3.  

While the influence of the offline delay was not especially pronounced for Boulkroune's tuning (criterion 3), Fig.~\ref{fig:res:L1_shares_delays_bestBoulkroune}, it certainly matters for tuning criterion 1, Fig.~\ref{fig:res:L1_shares_delays_bestStates}. This applies both to the estimation error over time, with a tendency for larger errors for longer delays, especially upon feeding events, and for the most relevant states, which are increasingly dominated by AC for longer delays. This highlights that appropriate tuning is paramount especially for delayed states and outputs relevant to system stability.   

By comparison, different levels of measurement noise show a similar influence on total estimation error for both tuning criteria, Figs.~\ref{fig:res:L1_shares_meas_noise_bestBoulkroune} and \ref{fig:res:L1_shares_meas_noise_bestStates}. However, for criterion~1, estimation errors are more dominated by AC, which particularly occurs upon large feeding events. 

Clearly, the tuning according to criterion 1 leads to less bias among the different estimates, Fig.~\ref{fig:res:L1_shares_pmm_bestStates}, and again shows a stronger dominance of AC among the most influential states. This is due to a much stronger emphasis on innovations during measurement updates, which holds both for online and offline outputs (plots not shown). Similar to criterion 3, higher PMM tend to result in lower estimation errors, cf. the discussion in Sec.~\ref{sec:res:pmm}. 

% The relative contribution of states changes for different PMM: While $X_\mathrm{bac}$ contributes less for increasing PMM, the share of $S_\mathrm{ac}$ and $S_\mathrm{ac-}$ grows, which is in line with the strong associated measurement corrections at corresponding returns. Likewise, the cumulative share of all remaining states (\textit{rest}) decreases for growing PMM. Furthermore, for medium and high PMM, $X_\mathrm{ch}$ displaces $X_\mathrm{li}$ as the fifth most influential state.

Furthermore, Fig.~\ref{fig:res:L1_shares_init_errors_bestStates} shows that total estimation errors for different initialization are qualitatively similar for the tuning according to criteria 1 and 3, but that AC more clearly dominates the total estimation error for criterion~1.   

Finally, Fig.~\ref{fig:res:L1_norm_and_shares} illustrates that it is not instructive to state a maximum convergence time for the MR-EKF, as the estimation error is predominantly influenced by the feeding events, cf. Fig.~\ref{fig:feeding_pattern}, which cause the filters to exhibit large innovations and hence high estimation errors after an initial convergence during the first \SI{7}{\day}. The feeding-induced perturbation is least pronounced for the cases of no delay and bad initialization using the tuning according to criterion 1, cf. Fig.~\ref{fig:res:L1_shares_delays_bestStates} and \ref{fig:res:L1_shares_init_errors_bestStates}, but even then the initial estimation errors are exceeded during days 7-8. 
\section{Comparison with conventional MHE}
\label{sec:res:mhe}
The MR-EKF performance was put into perspective by comparing it to a conventional MHE which assumes singlerate online measurements. To ensure a fair comparison between both state estimation algorithms, the same $R$-matrix as for the MR-EKF was used to tune the MHE (based on criterion 3). Likewise, the MHE's arrival cost term was weighted with the initial $P$-matrix from the MR-EKF. Since process noise was not considered in the MHE, the $Q$-matrix of the MR-EKF remained unused in the MHE. 

\begin{figure}
    \centering
    \includegraphics[width=0.7\linewidth]{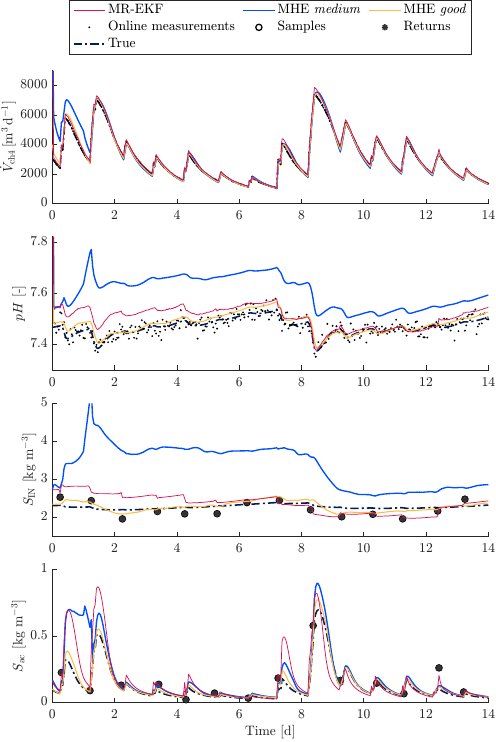}
    \caption{Comparison of multirate extended Kalman filter (MR-EKF) and moving horizon estimation (MHE) for no delay case and otherwise medium parameters (MHE \textit{medium}) or medium noise, low PMM and good initialization (MHE \textit{good}).}
    \label{fig:res:mhe_vs_ekf}
\end{figure}

Fig.~\ref{fig:res:mhe_vs_ekf} shows the estimates of both algorithms for zero delay and medium parameters (MHE \textit{medium}), as well as medium measurement noise, low PMM, and good initialization (MHE \textit{good}), cf. Table~\ref{tab:meth:variants_overview}. 

Both algorithms MR-EKF (red lines) and MHE \textit{medium} (blue lines) show similar performance for \chfour production (top subplot in Fig.~\ref{fig:res:mhe_vs_ekf}) despite a longer initial convergence phase of MHE \textit{medium}. The pH estimates (second subplot in Fig.~\ref{fig:res:mhe_vs_ekf}), however, differ significantly. The medium MHE clearly overestimates the pH, which likely indicates the strong influence of the initial state error \cite{LopezNegrete.2012}, especially given the positive PMM, cf. Fig.~\ref{fig:res:diff_init_error_y} on different initializations of the MR-EKF. Likewise, the IN estimates (third subplot in Fig.~\ref{fig:res:mhe_vs_ekf}) of the medium MHE clearly show a positive bias compared to ground truth with an initial divergence. On the contrary, the MR-EKF converges reasonably well due to offline corrections. Interestingly, both algorithms MR-EKF and MHE \textit{medium} show similar performance for AC (bottom subplot of Fig.~\ref{fig:res:mhe_vs_ekf}), after an initial convergence phase of the MHE before day~2. 

The medium MHE simulation took \SI{110}{\minute}, which is about three times as long as for the MR-EKF even in the worst case of long delays (cf. Fig.~\ref{fig:res:comparison_runtime}). This can be attributed to the repeated nonlinear optimization problems solved online in the MHE, as opposed to the sequential prediction-correction scheme of the MR-EKF. The computational load of the MHE would even increase if process noise were to be considered. Nevertheless, real-time capability was still ensured given a simulated time of \SI{14}{\day}.

%The tuning of both the MR-EKF and the medium MHE was based on a grid search evaluating the performance of the MR-EKF, which relied on nominal parameters, cf. Table~\ref{tab:meth:variants_overview}. Thus, directly applying this same tuning to the MHE may be unfair. Therefore, 
Fig.~\ref{fig:res:mhe_vs_ekf} additionally shows the MHE performance for good parameters (MHE \textit{good}, yellow lines), cf. Table~\ref{tab:meth:variants_overview}. In this setting, the same tuning delivers much better estimations, Fig.~\ref{fig:res:mhe_vs_ekf}. This holds especially for pH and IN estimates, cf. second and third subplot of Fig.~\ref{fig:res:mhe_vs_ekf}, which are much closer to ground truth than for the MHE with medium parameters, even if the offline measurements IN and AC were not used in the MHE formulation. This suggests that with good initialization and low PMM, \mbox{offline} corrections may not be strictly mandatory for good estimation performance. However, the biased IN estimates of MHE \textit{medium} indicate that offline corrections are indeed critical to avoid permanent estimations errors or divergence \cite{LopezNegrete.2012}. 

\begin{figure}
    \centering
    \includegraphics[width=0.7\linewidth]{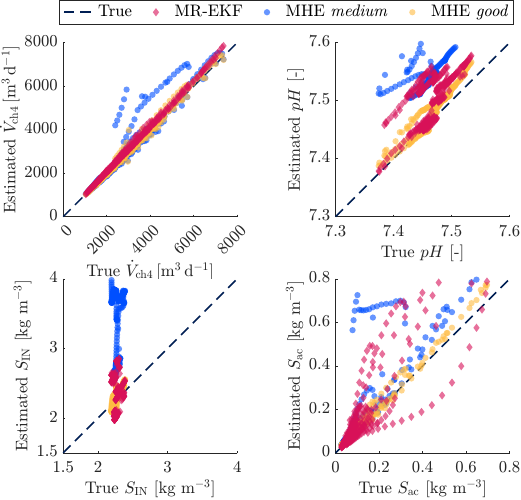}
    \caption{Parity plots of MR-EKF and MHE implementations \textit{medium} and \textit{good} for the delay-free case and the tuning according to criterion 3.}
    \label{fig:res:parity_plot}
\end{figure}

The comparison of MR-EKF and both MHE implementations is further illustrated by means of parity plots of four estimated outputs in Fig.~\ref{fig:res:parity_plot}. They show the individual estimates against their ground-truth counterparts, and thus illustrate bias and consistency of the estimates. Notably, the medium MHE shows a positive bias for all outputs, in particular for pH and IN, cf. top-right and bottom-left subplot of Fig.~\ref{fig:res:parity_plot}, respectively. By contrast, the MR-EKF shows a much lower bias for all outputs, although only the \chfour estimates (top-left subplot of Fig.~\ref{fig:res:parity_plot}), are close to the \SI{45}{\degree} line for the entire range of values, indicating consistency of the estimates. MR-EKF and medium MHE perform comparibly well for AC, with a slightly positive bias for the MHE at low ground-truth values, cf. bottom-right subplot of Fig.~\ref{fig:res:parity_plot}. In line with Fig.~\ref{fig:res:mhe_vs_ekf}, the MHE with good parameters shows a much better performance for all outputs, with estimates close to the \SI{45}{\degree} line and thus low bias and good consistency.

Finally, a reduced version of an MHE parameter study was carried out for the parameters measurement noise, PMM, and initial state error. It is shown in Fig.~\ref{fig:res:mhe_good} for a simulated time of the first \SI{7}{\day} of the feeding pattern of Fig.~\ref{fig:feeding_pattern}. For this purpose, the MHE was imposed with online measurement data and parameters summarized in Table~\ref{tab:meth:variants_good_mhe}. Within the mild parameter perturbations, the MHE showed good robustness and satisfactory state estimation performance. 
\section{Limitations}
The results presented in Sec.~\ref{sec:res:systematic_influence_parameters} and \ref{sec:res:mhe} were based on the tunings derived in Sec.~\ref{sec:grid_search}. Different tunings likely change the quantitative results, as discussed in Sec.~\ref{sec:res:alternative_tuning} and \ref{sec:res:mhe}. 

In practical AD operation, offline delays may be longer than the values considered in Table~\ref{tab:meth:measurements}, in particularly for AC. This is mostly due to sample transport to a laboratory that is not on-site. By contrast, sample preparation and sample analysis themselves require time ranging from minutes up to a few hours depending on the method of analyis \cite{Liebetrau.2020,Wagner.2017,Boe.2007}. Furthermore, immediate processing of incoming samples in commercial laboratories can be hindered by practical constraints, such as a minimum number of samples to analyse in a GC.  

These shorter delays are, however, compensated by sample times which are also shorter than in real life. In light of the MR-EKF framework, this effectively yields the same degree of sample-state augmentations. An interesting future research aspect could be to investigate different combinations of sampling frequency and delays in order to determine an optimal sampling strategy. 

Moreover, the results in Sec.~\ref{sec:res:systematic_influence_parameters} and \ref{sec:res:mhe} are strongly influenced by the default values of PMM and initial state errors. While this tutorial primarily addressed the MR-EKF and its specific parameters, future investigations should focus on more applied scenarios depending on the process and specific use case, such as AD process inhibition, sensor failure or uncertain substrate characterization to allow an exhaustive analysis and comparison of different scenarios. 

Furthermore, the results discussed largely rely on the NRMSE as an error metric. While this is a common choice in literature \cite{Horvath.2017,Ghani.2025,Condori.2019}, other error metrics should be considered to address different aspects of estimation performance, such as filter robustness, convergence, and consistency \cite{Boulkroune.2023}. 

Finally, this tutorial discussed \textit{in silico} results based on a simplified AD model where noise was assumed to be additive, bias-free, and Gaussian, as frequently assumed in literature \cite{Zhao2019,Elsheikh.2021,Tatiraju.1999,Li.2004}. Experimental validation and online application in different scales should be addressed in the future, with an emphasis on industrial constraints such as hardware limitations and failure, robust data handling, and limited computational resources. 

\section{Conclusion}
\label{sec:conclusion}
This tutorial derived the sample-state augmentation approach for a multirate extended Kalman filter (MR-EKF) for single and multiple augmentations. The MR-EKF was applied to a nonlinear model of the anaerobic digestion (AD) process, and an operating scenario relevant for practical, full-scale AD plants was assumed. By describing the complete workflow of its implementation, tuning, and application, it provided crucial implementation details that were missing in prior formulations.
 
An appropriate tuning was chosen systematically based on the tuning method proposed by Boulkroune \cite{Boulkroune.2023}, which was slightly refined to obtain a fair comparison of influencing error function components. The performance of the MR-EKF was investigated for different parameters relevant for state estimation of biological processes, i.e., \mbox{offline} delay lengths, measurement noise levels, plant-model mismatch, and initial state error.

The effect of the individual MR-EKF parameters were compared with respect to their estimation accuracy and run time. The most influential parameters were the initial state error and the delays of the offline measurements. Moreover, the MR-EKF showed superior performance compared with a conventional MHE implementation and at a lower computational cost. However, the tuning of the MR-EKF was shown to drastically impact the observations obtained. Finding an appropriate tuning requires to properly select the EKF tuning matrix entries as well an error function by which to compare tunings. 

This work systematically shows that a MR-EKF can deliver reliable state estimation of a nonlinear AD process with multirate measurement signals. For this purpose, it is paramount to have a well calibrated model and an appropriate tuning in place. This tutorial provides critical implementation guidance for practitioners working with multirate biological processes. Particularly, it derives insights for successfully applying the MR-EKF in monitoring and control schemes for demand-driven operation of full-scale AD plants.

%% epilogue
\section*{Acknowledgements}
S.H. acknowledges Michael Goldstein for his patient advice on practical questions regarding lab measurement procedures, Leander Lerch for the fruitful discussions on multiple state augmentations, Jean-Philippe Steyer for his constructive scrutiny of the methods, Daniel Stors for providing code for Boulkroune's error function, and Boulaid Boulkroune for patiently explaining his tuning approach. 
\section*{CRediT authorship contribution statement}
\textbf{Simon Hellmann:} Conceptualization, Data Curation, Investigation, Methodology, Software, Validation, Visualization, Writing - Original Draft. \textbf{Terrance Wilms:} Conceptualization, Methodology, Validation, Writing - Original Draft. \textbf{Stefan Streif:} Conceptualization, Methodology, Resources, Supervision, Writing - Review \& Editing. \textbf{Sören Weinrich:} Conceptualization, Funding acquisition, Methodology, Project administration, Supervision, Visualization, Writing - Review \& Editing. 

\section*{Declaration of competing interest}
The authors declare that they have no known competing financial interests or personal relationships that could have appeared to influence the work reported in this paper.

\section*{Funding sources}
The authors gratefully thank the German Federal Ministry of Agriculture, Food and Regional Identity for funding of the junior research group on simulation, monitoring and control of anaerobic processes (grant no 2219NR333). 
\section*{Declaration of generative AI and AI-assisted technologies in the manuscript preparation process}
During the preparation of this work the authors used github copilot for latex in vscode as well as ecosia chat in order to improve the English language usage and clarity of the writing style. Furthermore, ecosia chat, claude.ai, and chatGPT were used to draft individual code snippets. After using these tools/service, the authors reviewed and edited the content as needed and take full responsibility for the content of the published article.
\section*{Notation}
%
% \flushleft Table footnote text is given on the next page. 
% \raggedright
% The following enumeration refers to the comments in the notation table: 
%
% \flushleft
\begin{tabular}{p{0.2\linewidth}p{0.75\linewidth}}
\toprule
\textbf{Symbol} & \textbf{Description} \\
\midrule
$x, y, u$ & states, outputs, inputs \\
$\xi$ & influent concentrations\\
$c, a, \theta$ & model parameters \\
$f,\bar f, h$ & nonlinear dynamics and output equation \\
$t, \Delta t$ & time, time difference \\
$k, l, s, r$ & time instances$^\mathrm{a,b}$ \\ 
$n,q$ & number of states and outputs \\
$N$ & (total) number \\
$w, v$ & process and measurement noise \\ 
$Q,R$ & process and measurement noise matrices \\
$\delta$ & Kronecker delta \\ 
$\hat x, P$ & state estimate, error covariance matrix$^\mathrm{b,c}$ \\
$F,H$ & linearization matrices \\ 
$K,S$ & Kalman filter matrices \\ 
$I,0$ & identity and zero matrix \\
$M$ & indicator matrix \\
$T$ & normalization matrix \\ 
$E$ & expected value operator \\
$\epsilon_\mathrm{y}$ & normalized innovation squared (NIS) \\
$k, \mu, K_S, K_I$ & ADM1 model parameters \cite{Batstone2002,Weinrich.2021} \\
$V, T$ & volume, temperature \\ 
HRT, OLR & hydraulic retention time, organic loading rate \\
VS & volatile solids \\
RMS & root mean square \\
NRMSE & normalized root mean square error \\
$\omega, \gamma, \alpha, \chi$ & Boulkroune error function variables \\
\midrule
\textbf{Subscripts} & \\
\midrule
0 & initial \\ 
a & augmentation \\ 
d & delay \\
ss & steady state \\
x, y, u & state, output, input \\
\midrule
\textbf{Superscripts} & \\
\midrule
a & augmentation \\ 
av & available \\ 
off & offline \\
on & online \\ 
$\, \, \hat{}$ & estimate \\
$-$ & a priori estimate (prior)$^\mathrm{d}$ \\
\bottomrule
\end{tabular}
\justify
%
% Comments (continued on next page): 
\begin{enumerate}[label=\alph*]
    \footnotesize
    \item Time instances $t_k$ are referred to as time $k$ for easier readability.
    \item Subscript indices of P-matrices denote of P-matrices denote the corresponding times between which the covariance is considered, e.g., $P_{k1,k2} = E\left\{ (x_{k1} - \hat x_{k1}) (x_{k2} - \hat x_{k2})^T \right\}$. For identical times $k_1=k_2=k$, they are only stated once, i.e., $P_k = P_{k,k}$.
    \item Augmented entries are denoted by their sample time, e.g., $\hat x_{s1}$ for the augmented state entries belonging to the sample time $s1$.
    \item Posteriors are not specifically denoted.
\end{enumerate}

\appendix
% start counting tables, figures, equations from 1 again:
\setcounter{table}{0}   
\setcounter{figure}{0} 
\setcounter{equation}{0} 

% automatically add "Appendix" to section titles in the appendix:
\renewcommand{\thesection}{Appendix \Alph{section}}

\section{Boulkroune's error function}
\label{sec:apx}
\begin{figure}[hb]
    \centering
    \includegraphics[width=0.7\linewidth]{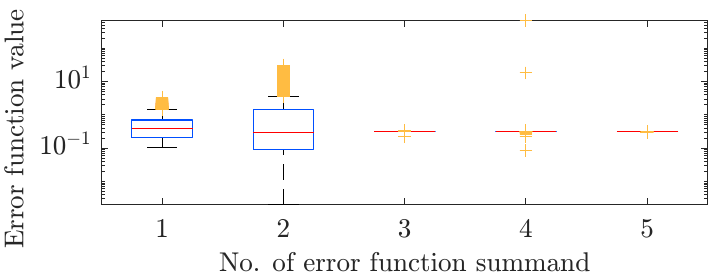}
    \caption{Boulkroune's error function components during grid search runs.}
    \label{fig:boxplot_boulkroune_error_fun_parts}
\end{figure}

Fig.~\ref{fig:boxplot_boulkroune_error_fun_parts} shows the distribution of Boulkroune's error function summands for individual tunings during grid search in the order of Eq.~\eqref{eq:boulkroune_error_fun}. Note that the medians lie in the same order of magnitude, which was achieved by tuning the weighting factors $\omega_i$ appropriately, cf. Eq.~\eqref{eq:boulkroune_error_fun}. 

\begin{figure}[hb]
    \centering
    \includegraphics[width=0.7\linewidth]{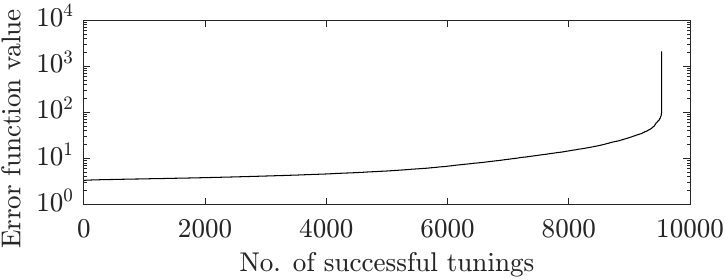}
    \caption{Ranking of tunings according to Boulkroune's error function.}
    \label{fig:ranking_Boulkroune}
\end{figure}

Fig.~\ref{fig:ranking_Boulkroune} shows the ranking of all successful tunings according to Boulkroune's error function.

\begin{figure*}
    \centering
    \includegraphics[width=0.9\linewidth]{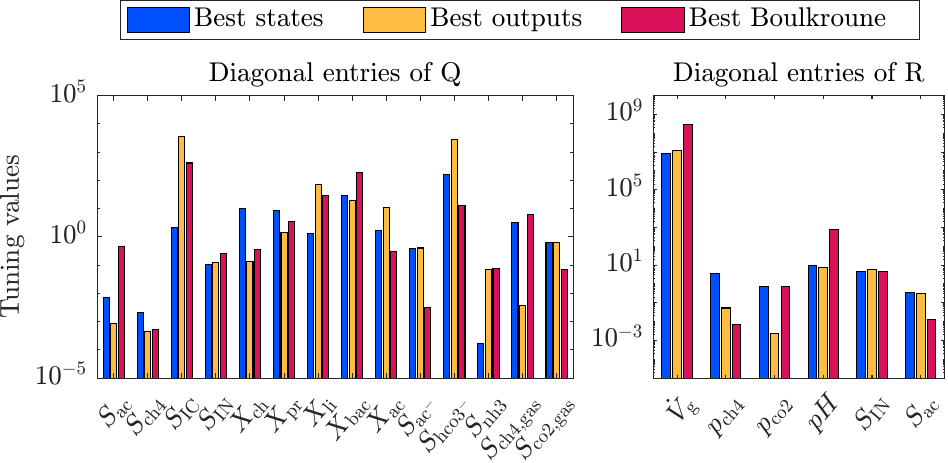}
    \caption{Tuning factors of best tunings according to 3 different criteria: best NRMSE of all states, best NRMSE of all outputs, best error function of Boulkroune. Tuning factors relate to diagonal entries of Q and R matrix. Note the wider range of the log scale compared to Fig.~\ref{fig:res:top_tunings}.}
    \label{fig:res:topX_tuning_factors}
\end{figure*}

Fig.~\ref{fig:res:topX_tuning_factors} shows the tuning factors of the best-ranked tunings according to the three criteria $\mathrm{NRMSE_x}$, $\mathrm{NRMSE_y}$ and Boulkroune's error function. 

\section{Moving Horizon Estimation}
% 
% start counting tables, figures, equations from 1 again:
\setcounter{table}{0}   
\setcounter{figure}{0} 
\setcounter{equation}{0} 
% 

% Fig.~\ref{fig:res:parity_plot} shows parity plots of the four output variables \chfour production rate, pH, IN and AC for the MR-EKF and the MHE implementations \textit{medium} and \textit{good}, cf. Table~\ref{tab:meth:variants_overview}, using the tuning according to criterion 3. The \SI{45}{\degree} line indicates perfect estimation. 

% % 
% \begin{figure}
%     \centering
%     \includegraphics[width=0.9\linewidth]{images/results/parity_rel_outputs_mhe_ekf_no_delay_criterion_3.pdf}
%     \caption{Parity plots of MR-EKF and MHE implementations \textit{medium} and \textit{good} for the delay-free case and the tuning according to criterion 3.}
%     \label{fig:res:parity_plot}
% \end{figure}
% % 

\begin{figure}
    \centering
    \includegraphics[width=0.7\linewidth]{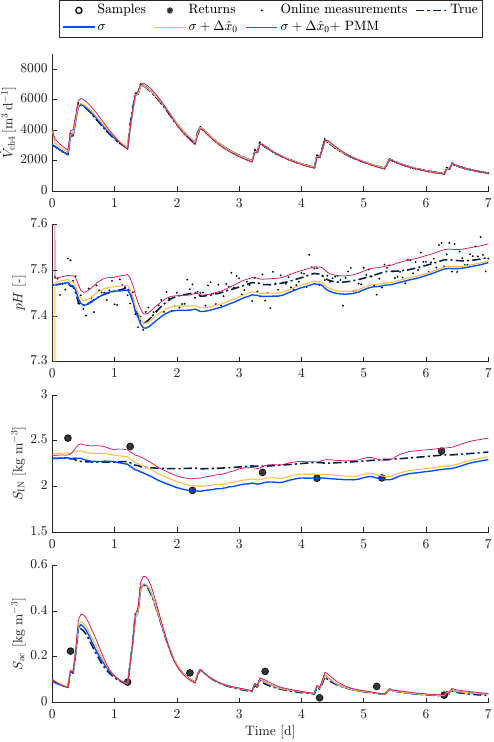}
    \caption{Comparison of moving horizon estimation (MHE) for three different combinations of scenarios: measurement noise $\sigma$, small initial state error $\Delta \hat x_0$, and plant-model mismatch (PMM).}
    \label{fig:res:mhe_good}
\end{figure}

Fig.~\ref{fig:res:mhe_good} shows three MHE runs for different mildly perturbed parameter settings summarized in Table~\ref{tab:meth:variants_good_mhe}. Note that the initial state error $\Delta \hat x_0$ was chosen small in all cases ($k_x \le 0.1$) due to its strong effect on the state error convergence of MHE.

\begin{table}
\centering
\begin{threeparttable}
    \renewcommand{\arraystretch}{1.3}
    \caption{Parameters used for reduced MHE parameter study.}
    \label{tab:meth:variants_good_mhe}
    \begin{tabularx}{0.6\linewidth}{c*{4}{Y}}
        \toprule %
        \multirow{2}{*}{Scenario$^{\mathrm{a,b,c}}$} & \multirow{2}{*}{Delay} & Noise$^{\mathrm{a}}$ & PMM$^{\mathrm{b}}$ & Init.$^{\mathrm{c}}$ \\
        \cdashline{3-5}
        & & $k_\sigma$ & $k_\theta$ & $k_\mathrm{x}$ \\
        \midrule
        $\sigma$                            & 0     & 0.5   & 0     & 0 \\
        $\sigma+\Delta \hat x_0$            & 0     & 0.5   & 0     & 0.1 \\
        $\sigma + \Delta \hat x_0 + $ PMM   & 0     & 0.5   & 0.1   & 0.1 \\
        \bottomrule 
    \end{tabularx} 
    \begin{tablenotes}
    \footnotesize
        \item[${\mathrm{a}}$] Measurement noise level ($\sigma$) 
        \item[${\mathrm{b}}$] Plant-model mismatch
        \item[${\mathrm{c}}$] Initial state error ($\Delta \hat x_0$)
    \end{tablenotes}
\end{threeparttable}
\end{table}
\section{Model equations of ADM1-R3} \label{sec:apx:model_equations}
% 
% start counting tables, figures, equations from 1 again:
\setcounter{table}{0}   
\setcounter{figure}{0} 
\setcounter{equation}{0} 
The system of ordinary differential equations (ODEs) of the modified ADM1-R3 reads
\begin{subequations} \label{eq:R3-Core:x}
\small
    \begin{align}
    \dot x_1 &= c_1 \left(\xi_1 - x_1\right) u + a_{11} \theta_1 x_5 + a_{12} \theta_2 x_6 + a_{13} \theta_3 x_7 + a_{14} \theta_5 \frac{x_1 \, x_9}{\theta_6 + x_1} I_\mathrm{ac}, \label{eq:R3-Core:x_ac}  \\
    \dot x_2 &= c_1 \left(\xi_2 - x_2\right) u + a_{21} \theta_1 x_5 + a_{22} \theta_2 x_6 + a_{23} \theta_3 x_7 - c_5 x_2 + c_6 x_{13} + a_{24} \theta_5  \frac{x_1 \, x_9}{\theta_6 + x_1} I_\mathrm{ac}, \\
    \dot x_3 &= c_1 \left(\xi_3 - x_3\right) u + a_{31} \theta_1 x_5 + a_{32} \theta_2 x_6 + a_{33} \theta_3 x_7 - c_5 x_3 + c_5 x_{11} + c_7 x_{14} + a_{34} \theta_5 \frac{x_1 \, x_9}{\theta_6 + x_1} I_\mathrm{ac}, \\
    \dot x_4 &= c_1 \left(\xi_4 \, \theta_9 - x_4\right) u + a_{41} \theta_1 x_5 + a_{42} \theta_2 x_6 + a_{43} \theta_3 x_7 + a_{44} \theta_5 \frac{x_1 \, x_9}{\theta_6 + x_1} I_\mathrm{ac}, \\
    \dot x_5 &= c_1 \left(\xi_5 - x_5\right) u - \theta_1 x_5 + a_{55} \theta_4 x_8 + a_{56} \theta_4 x_9, \\
    \dot x_6 &= c_1 \left(\xi_6 - x_6\right) u - \theta_2 x_6 + a_{65} \theta_4 x_8 + a_{66} \theta_4 x_9, \\ 
    \dot x_7 &= c_1 \left(\xi_7 - x_7\right) u - \theta_3 x_7 + a_{75} \theta_4 x_8 + a_{76} \theta_4 x_9, \\ 
    \dot x_8 &= c_1 \left(\xi_8 - x_8\right) u + a_{81} \theta_1 x_5 + a_{82} \theta_2 x_6 + a_{83} \theta_3 x_7 - \theta_4 x_8, \\
    \dot x_9 &= c_1 \left(\xi_9 - x_9\right) u + \theta_5 \frac{x_1 \, x_9}{\theta_6 + x_1} I_\mathrm{ac} - \theta_4 x_9, \\
    \dot x_{10} &= c_{29} \left(x_1 - x_{10}\right) - c_9 x_{10} S_\mathrm{H^+}(x,\theta), \\
    \dot x_{11} &= c_{30} \left(x_3 - x_{11}\right) - c_{10} x_{11} S_\mathrm{H^+}(x,\theta), \\
    \dot x_{12} &= c_{31} \left(x_4 - x_{12}\right) - c_{11} x_{12} S_\mathrm{H^+}(x,\theta), \\
    \dot x_{13} &= c_{22} x_{13}^3 + c_{23} x_{13}^2 x_{14} + c_{24} x_{13} x_{14}^2 + c_{25} x_{13}^2 + c_{26} x_{13} x_{14} + c_{12} x_2 + c_{27} x_{13}, \\
    \dot x_{14} &= c_{24} x_{14}^3 + c_{23} x_{13} x_{14}^2 + c_{22} x_{13}^2 x_{14} + c_{26} x_{14}^2 + c_{25} x_{13} x_{14} + c_{12} x_3 - c_{12} x_{11} + c_{28} x_{14},
    \end{align}
\end{subequations}
where $I_\mathrm{ac}$ and $S_\mathrm{H^+}$ are defined as  
\small
\begin{align}
I_\mathrm{ac} &= \frac{c_3}{c_3 + S_\mathrm{H^+}^{c_2}} \frac{x_4}{x_4 + c_8} \frac{\theta_7}{\theta_7 + x_{12}}, \label{eq:R3-Core:Iac} \\
% 
% S_\mathrm{H^+} &= -\frac{\Phi}{2} + \frac{1}{2} \sqrt{\Phi^2 + c_4}  \text{ , where } \label{R3-Core-SH+-x}\\
% 
S_\mathrm{H^+}(x,\theta) &= -\frac{\theta_8}{2} + \frac{x_{12} - x_4}{34} + \frac{x_{11}}{88} + \frac{x_{10}}{120} + \frac{1}{2} \sqrt{\left(\theta_8 + \frac{x_4 - x_{12}}{17} - \frac{x_{11}}{44} - \frac{x_{10}}{60}\right)^2 + c_4}.  %\text{ , where }
\label{eq:R3-Core:SHPlus}
% 
% \Phi &= \theta_8 + \frac{x_4 - x_{12}}{17} - \frac{x_{11}}{44} - \frac{x_{10}}{60}.\label{eq:R3-Core:Phi}
\end{align}
\normalsize
%
% \subsection{Time-invariant Parameters}
%
$x$ and $\xi$ denote states and corresponding influent concentrations, $u=\dot V_\mathrm{f}$ denotes the input (feed volume flow), $a_{ij}$ denote stoichiometric coefficients, and $c_i$ time-invariant parameters. Computation of $c_i$ from the typical ADM1-specific variables is summarized in Table~\ref{tab:R3-param-nomenclature}, with parameters therein as per Table~\ref{tab:model_parameters_soeren}. Table~\ref{tab:petersen-ADM1-R3-Core} shows the Petersen matrix which contains the stoichiometric coefficients $a_{ij}$. In Eq.~\eqref{eq:R3-Core:x}, only those values $a_{ij}$ with an absolute value \textit{other} than 0 or 1 need to be taken from Table~\ref{tab:petersen-ADM1-R3-Core}. Note that in $a_{ij}$ $i$ denotes the column and $j$ the row.

Corresponding output equations read
\begin{subequations}
    \label{eq:R3-Core:y}
	\begin{align}
	y_1 &= \dot V_\mathrm{gas} = c_{13} x_{13}^2 + c_{14} x_{13} x_{14} + c_{15} x_{14}^2 + c_{16} x_{13} + c_{17} x_{14} + c_{18}, \label{eq:R3-Core:y_vgas} \\
	y_2 &= p_\mathrm{ch4} = c_{19} x_{13}, \label{eq:R3-Core:y_pch4}  \\
	y_3 &= p_\mathrm{co2} = c_{20} x_{14}, \label{eq:R3-Core:y_pco2} \\
	y_4 &= pH = -\log_{10} S_\mathrm{H^+} (x,\theta), \label{eq:R3-Core:y_ph} \\
	y_5 &= S_\mathrm{IN} = x_4, \label{eq:R3-Core:y_IN} \\
	y_6 &= S_\mathrm{ac} = x_1, \label{eq:R3-Core:y_AC} 
	\end{align}
\end{subequations}
with $S_\mathrm{H^+}$ from Eq.~\eqref{eq:R3-Core:SHPlus}.

Table~\ref{tab:inlet_concentrations} summarizes the influent concentrations of individual substrates and the constant mixture according to Fig.~\ref{fig:feeding_pattern}. 

\begin{table}[htb]
\centering
\begin{threeparttable}
    \renewcommand{\arraystretch}{1.3}
    \caption{Influent concentrations$^{\mathrm{a}}$ $\xi$ for all used substrates in [\si{\kilogram\per\cubic\meter}]. Values of components not listed were assumed to be zero.}
    \label{tab:inlet_concentrations}
    \begin{tabularx}{0.6\linewidth}{*{2}{Y}*{4}{d{2}}}
        \toprule
        Value & Comp.$^{\mathrm{b}}$ & \multicolumn{1}{c}{MS$^{\mathrm{b}}$} & \multicolumn{1}{c}{GS$^{\mathrm{b}}$} & \multicolumn{1}{c}{CM$^{\mathrm{b}}$} & \multicolumn{1}{c}{mix} \\
        \midrule
        $\xi_1$ & $S_\mathrm{ac}$ & 10.27   & 12.33   & 4.99 & 7.64 \\ 
        $\xi_4$ & $S_\mathrm{IN}$ & 0.76    & 0.86    & 1.71 & 1.27 \\
        $\xi_5$ & $X_\mathrm{ch}$ & 304.50   &205.31   & 17.00 & 144.19 \\
        $\xi_6$ & $X_\mathrm{pr}$ & 24.20   & 42.27   & 10.80 & 18.54 \\
        $\xi_7$ & $X_\mathrm{li}$ & 18.10   & 15.42   & 1.40 & 9.03 \\
        % $\xi_8$ & $X_\mathrm{bac}$ & 0      & XY        & XY \\
        % $\xi_9$ & $X_\mathrm{ac}$ & 0       & XY        & XY \\
        % $\xi_{10}$&$S_\mathrm{ac^-}$ & 1.53 & 5.33    & 4.53 \\
        % $\xi_{12}$ & $S_\mathrm{nh3}$ & 0   & XY    & XY \\
        %
        \bottomrule
    \end{tabularx}
    \begin{tablenotes}
    \footnotesize
        \item[${\mathrm{a}}$] Computation based on \cite{Hellmann.2025}
        \item[${\mathrm{b}}$] Comp.: Computation, MS: maize silage, GS: grass silage, CM: cattle manure.
    \end{tablenotes}
\end{threeparttable}
\end{table}
\begin{table}[htb]
\centering
\small
    \renewcommand{\arraystretch}{1.3}
    \caption{Initial conditions $x_{0,\mathrm{ss}}$ for transition into steady-state (ss). Values are given as mass concentrations in [\si{\kilogram\per\cubic\meter}].}
    \label{tab:init_condition_ss}
    \begin{tabularx}{0.6\linewidth}{*{6}{Y}}
        \toprule
        Index $i$ & State & $x_{0,\mathrm{ss},i}$ & Index $i$ & State & $x_{0,\mathrm{ss},i}$ \\
        \midrule
        1 & $S_\mathrm{ac}$ & 0.049     & 8     & $X_\mathrm{bac}$ & 1.926 \\ 
        2 & $S_\mathrm{ch4}$ & 0.012    & 9     & $X_\mathrm{ac}$ & 0.552 \\ 
        3 & $S_\mathrm{IC}$ & 4.975     & 10    & $S_\mathrm{ac^-}$ & 0.049 \\
        4 & $S_\mathrm{IN}$ & 0.964     & 11    & $S_\mathrm{hco3^-}$ & 4.546 \\
        5 & $X_\mathrm{ch}$ & 2.962     & 12    & $S_\mathrm{nh3}$ & 0.022 \\
        6 & $X_\mathrm{pr}$ & 0.949     & 13    & $S_\mathrm{ch4,gas}$ & 0.358 \\
        7 & $X_\mathrm{li}$ & 0.412     & 14    & $S_\mathrm{co2,gas}$ & 0.660 \\
        \bottomrule
    \end{tabularx} 
\end{table}

Values of the normalization matrices $T$ in Table~\ref{tab:normalization} were obtained with default model parameter values $\theta$ (cf. Table~\ref{tab:meth:time_variant_params}) and average feed volume flow of the dynamic feeding schedule during the synthetic simulation scenario.
\begin{table}
\centering
\small
\renewcommand{\arraystretch}{1.3}
\caption{Diagonal entries of normalization matrices.}
\label{tab:normalization}
\begin{tabularx}{0.6\linewidth}{llYlXl}
    \toprule
    Entry & Value & Unit & Entry & Value & Unit \\
    \midrule
    $T_{\mathrm{x},1}$ & 0.182 & [\si{\kilogram\per\cubic\meter}] 
        & $T_{\mathrm{x},12}$ & 0.167 & [\si{\kilogram\per\cubic\meter}] \\ 
    $T_{\mathrm{x},2}$ & 0.014 & [\si{\kilogram\per\cubic\meter}] 
        & $T_{\mathrm{x},13}$ & 0.387 & [\si{\kilogram\per\cubic\meter}] \\ 
    $T_{\mathrm{x},3}$ & 11.011 & [\si{\kilogram\per\cubic\meter}] 
        & $T_{\mathrm{x},14}$ & 0.914 & [\si{\kilogram\per\cubic\meter}] \\ 
    \cdashline{4-6}
    $T_{\mathrm{x},4}$ & 3.371 & [\si{\kilogram\per\cubic\meter}] 
        & $T_{\mathrm{y},1}$ & \num[exponent-product = \times]{4.209E3} & [\si{\cubic\meter\per\day}] \\ 
    $T_{\mathrm{x},5}$ & 1.819 & [\si{\kilogram\per\cubic\meter}] 
        & $T_{\mathrm{y},2}$ & 0.550 & [\si{bar}] \\ 
    $T_{\mathrm{x},6}$ & 2.576 & [\si{\kilogram\per\cubic\meter}] 
        & $T_{\mathrm{y},3}$ & 0.472 & [\si{bar}] \\ 
    $T_{\mathrm{x},7}$ & 0.869 & [\si{\kilogram\per\cubic\meter}] 
        & $T_{\mathrm{y},4}$ & 7.588 & [-] \\ 
    $T_{\mathrm{x},8}$ & 9.712 & [\si{\kilogram\per\cubic\meter}] 
        & $T_{\mathrm{y},5}$ & 3.371 & [\si{\kilogram\per\cubic\meter}] \\ 
    $T_{\mathrm{x},9}$ & 2.453 & [\si{\kilogram\per\cubic\meter}] 
        & $T_{\mathrm{y},6}$ & 0.182 & [\si{\kilogram\per\cubic\meter}] \\ 
    \cdashline{4-6}
    $T_{\mathrm{x},10}$ & 0.181 & [\si{\kilogram\per\cubic\meter}] 
        & $T_\mathrm{u}$   & 29.821 & [\si{\cubic\meter\per\day}] \\
    $T_{\mathrm{x},11}$ & 10.483 & [\si{\kilogram\per\cubic\meter}] 
        & & & \\
    \bottomrule
\end{tabularx} 
\end{table}
\begin{table*}
\centering
\centering
\begin{threeparttable}
\footnotesize
\caption{Aggregation of time-invariant model parameters of ADM1-R3 and notation of Weinrich and Nelles \cite{Weinrich2021b}.} %
\begin{tabular}{clrl}
    \toprule
    \label{tab:R3-param-nomenclature}
    Parameter & Notation of \cite{Weinrich2021b} & Value$^{\mathrm{a}}$ & Unit \\
    \midrule
    $c_1$ & $V_\mathrm{liq}^{-1}$ & 5E-4 & \si{\per\cubic\meter} \\ 
    $c_2$ & $n_\mathrm{ac} = 3 \, \left(pH_{\mathrm{UL,ac}} - pH_{\mathrm{LL,ac}}\right)^{-1}$ & 3 & $-$ \\
    $c_3$ & $10^{- \frac{3}{2} \frac{pH_{\mathrm{UL,ac}} + pH_{\mathrm{LL,ac}}}{pH_{\mathrm{UL,ac}} - pH_{\mathrm{LL,ac}}}}$ & 3.162E-20 & $-$ \\
    $c_4$ & $4 K_\mathrm{W}$ & 8.315E-14 & \si{\kilo\mole\per\cubic\meter} \\
    $c_5$ & $k_\mathrm{L} a$ & 2E2 & \si{\per\day} \\
    $c_6$ & $k_\mathrm{L} a \, K_{\mathrm{H,ch4}} \bar{R} T$ & 4.997 & \si{\per\day} \\ 
    $c_7$ & $k_\mathrm{L} a \, K_{\mathrm{H,co2}} \bar{R} T$ & 1.136E2 & \si{\per\day} \\	
    $c_8$ & $K_{\mathrm{S,IN}}$ & 1.7E-3 & \si{\kilogram\per\cubic\meter} \\  
    $c_9$ & $k_{\mathrm{AB,ac}}$ & 1E10 & \si{\cubic\meter\per\kilo\mole\day} \\ 		
    $c_{10}$ & $k_{\mathrm{AB,co2}}$ & 1E10 & \si{\cubic\meter\per\kilo\mole\day} \\
    $c_{11}$ & $k_{\mathrm{AB,IN}}$ & 1E10 & \si{\cubic\meter\per\kilo\mole\day} \\		
    $c_{12}$ & $k_\mathrm{L} a \, V_\mathrm{liq} V_\mathrm{gas}^{-1}$ & 1.333E3 & \si{\per\day} \\
    $c_{13}$ & $k_\mathrm{p} p_0^{-1} \left(\bar R T \bar{M}_\mathrm{ch4}^{-1}\right)^2$ & 9.944E5 & $\textrm{m}^9\, \textrm{kg}^{-2} \textrm{d}^{-1}$ \\
    $c_{14}$ & $2 k_\mathrm{p} p_0^{-1} \left(\bar R T\right)^2 \bar{M}_\mathrm{ch4}^{-1} \bar{M}_\mathrm{co2}^{-1}$ & 7.232E5 & $\textrm{m}^9\, \textrm{kg}^{-2} \textrm{d}^{-1}$ \\	
    $c_{15}$ & $k_\mathrm{p} p_0^{-1} \left(\bar R T \bar{M}_\mathrm{co2}^{-1}\right)^2$ & 1.315E5 & $\textrm{m}^9\, \textrm{kg}^{-2} \textrm{d}^{-1}$ \\
    $c_{16}$ & $k_\mathrm{p} p_0^{-1} \bar R T \bar{M}_\mathrm{ch4}^{-1} \left(2 p_\mathrm{h2o} - p_0\right)$ & -7.098E5 & $\textrm{m}^6\, \textrm{kg}^{-1} \textrm{d}^{-1}$ \\
    $c_{17}$ & $k_\mathrm{p} p_0^{-1} \bar R T \bar{M}_\mathrm{co2}^{-1} \left(2 p_\mathrm{h2o} - p_0\right)$ & -2.581E5 & $\textrm{m}^6\, \textrm{kg}^{-1} \textrm{d}^{-1}$ \\
    $c_{18}$ & $k_\mathrm{p} p_0^{-1} \left(p_\mathrm{h2o} - p_0\right) p_\mathrm{h2o}$ & 0 & \si{\cubic\meter\per\day}{} \\
    $c_{19}$ & $\bar R T \bar{M}_\mathrm{ch4}^{-1}$ & 1.420 & \si{\bar\cubic\meter\per\kilogram} \\
    $c_{20}$ & $\bar R T \bar{M}_\mathrm{co2}^{-1}$ & 0.516 & \si{\bar\cubic\meter\per\kilogram} \\
    $c_{21}$ & $-k_\mathrm{p} V_\mathrm{gas}^{-1} p_0^{-1} \left(\bar R T \bar{M}_\mathrm{ch4}^{-1}\right)^2 $ & -3.315E3 & $\textrm{m}^6\, \textrm{kg}^{-2} \textrm{d}^{-1}$ \\
    $c_{22}$ & $-2 k_\mathrm{p} V_\mathrm{gas}^{-1} p_0^{-1} \left(\bar R T\right)^2 \bar{M}_\mathrm{ch4}^{-1} \bar{M}_\mathrm{co2}^{-1}$ & -2.411E3 & $\textrm{m}^6\, \textrm{kg}^{-2} \textrm{d}^{-1}$ \\  
    $c_{23}$ & $-k_\mathrm{p} V_\mathrm{gas}^{-1} p_0^{-1} \left(\bar R T \bar{M}_\mathrm{co2}^{-1}\right)^2$ & -4.383E2 & $\textrm{m}^6\, \textrm{kg}^{-2} \textrm{d}^{-1}$ \\ 
    $c_{24}$ & $-k_\mathrm{p} V_\mathrm{gas}^{-1} p_0^{-1} \bar R T \bar{M}_\mathrm{ch4}^{-1} \left(2 p_\mathrm{h2o} - p_0\right)$ & 2.366E3 & \si{\cubic\meter\per\kilogram\day} \\ 		
    $c_{25}$ & $-k_\mathrm{p} V_\mathrm{gas}^{-1} p_0^{-1} \bar R T \bar{M}_\mathrm{co2}^{-1} \left(2 p_\mathrm{h2o} - p_0\right)$ & 8.603E2 & \si{\cubic\meter\per\kilogram\day} \\ 
    $c_{26}$ & $- k_\mathrm{L} a \, V_\mathrm{liq} V_\mathrm{gas}^{-1} K_{\mathrm{H,ch4}} \bar{R} T - k_\mathrm{p} V_\mathrm{gas}^{-1} p_0^{-1} \left(p_\mathrm{h2o} - p_0\right) p_\mathrm{h2o}$ & -3.331E1 & \si{\per\day} \\ 
    $c_{27}$ & $- k_\mathrm{L} a \, V_\mathrm{liq} V_\mathrm{gas}^{-1} K_{\mathrm{H,co2}} \bar{R} T - k_\mathrm{p} V_\mathrm{gas}^{-1} p_0^{-1} \left(p_\mathrm{h2o} - p_0\right) p_\mathrm{h2o}$ & -7.571E2 & \si{\per\day} \\ 
    $c_{28}$ & $k_{\mathrm{AB,ac}} K_{\mathrm{a,ac}}$ & 1.738E5 & \si{\per\day} \\ 
    $c_{29}$ & $k_{\mathrm{AB,co2}} K_{\mathrm{a,co2}}$ & 5.129E3 & \si{\per\day} \\  
    $c_{30}$ & $k_{\mathrm{AB,IN}} K_{\mathrm{a,IN}}$ & 1.349E1 & \si{\per\day} \\ 
    $c_{31}$ & $V_\mathrm{liq} V_\mathrm{gas}^{-1}$ & 6.667 & $-$ \\ 
    \bottomrule
\end{tabular}
    \begin{tablenotes}
    \footnotesize
        \item[${\mathrm{a}}$] Scientific E notation applies, i.e., E$\pm$n means $\times 10^{\pm n}$.
    \end{tablenotes}
\end{threeparttable}
\end{table*}
\begin{table*}
\centering
\footnotesize
    \renewcommand{\arraystretch}{1.3}
    \caption{Time-invariant model parameters in typical ADM1 notation of \cite{Weinrich2021b}.}
    \label{tab:model_parameters_soeren}
    \begin{tabular}{*{6}{l}}
        \toprule
        Variable & Value & Unit & Variable & Value & Unit \\
        \midrule
        \multicolumn{6}{l}{Inhibition parameters} \\
        \cdashline{1-6}     
        $pH_\mathrm{LL,ac}$ & 6 & [-] & $K_{\mathrm{S,IN}}$ & 0.0017 & [\si{\kilogram\per\cubic\meter}] \\
        $pH_\mathrm{UL,ac}$ & 7 & [-] \\ 
        \midrule
        \multicolumn{6}{l}{Dissociation parameters} \\
        \cdashline{1-6}       
        $K_{\mathrm{a,ac}}$ & $1\cdot 10^{-4.76}$ & [\si{\kilo\mole\per\cubic\meter}] & $k_{\mathrm{AB,ac}}$ & $1\cdot 10^{10}$ & [\si{\cubic\meter\per\kilo\mole\day}] \\    
        $K_{\mathrm{a,co2}}$ & $1\cdot 10^{-6.29}$ & [\si{\kilo\mole\per\cubic\meter}] & $k_{\mathrm{AB,co2}}$ & $1\cdot 10^{10}$ & [\si{\cubic\meter\per\kilo\mole\day}] \\ 
        $K_{\mathrm{a,IN}}$ & $1\cdot 10^{-8.87}$ & [\si{\kilo\mole\per\cubic\meter}] & $k_{\mathrm{AB,IN}}$ & $1\cdot 10^{10}$ & [\si{\cubic\meter\per\kilo\mole\day}] \\ 
        $K_\mathrm{W}$ & $1\cdot 10^{-13.7}$ & [\si{\kilo\mole\per\cubic\meter}] &&& \\
        \midrule
        \multicolumn{6}{l}{Phase transition parameters} \\
        \cdashline{1-6} 
        $K_{\mathrm{H,ch4}}$ & $1.1\cdot 10^{-5}$ & [\si{\kilo\mole\per\kilo\joule}] & $k_\mathrm{L} a$ & 200 & [\si{\per\day}] \\
        $K_{\mathrm{H,co2}}$ & $2.5\cdot 10^{-4}$ & [\si{\kilo\mole\per\kilo\joule}] & $k_\mathrm{p}$ & $5\cdot 10^5$ & [\si{\cubic\meter\per\bar\day}] \\
        $\bar{M}_\mathrm{ch4}$ & 16 & [\si{\kilogram\per\kilo\mole}] & $\bar R$ & 8.314 & [\si{\kilo\joule\per\kilo\mole\kelvin}] \\ 
        $\bar{M}_\mathrm{co2}$ & 44 & [\si{\kilogram\per\kilo\mole}] &&& \\
        \bottomrule
    \end{tabular} 
\end{table*}
\begin{landscape}
\begin{center}
    \captionof{table}{Petersen matrix of ADM1-R3-Core, derived from \cite{SorenWeinrich2017}.}
    \label{tab:petersen-ADM1-R3-Core}
    \renewcommand{\arraystretch}{1.2}
    \setlength{\tabcolsep}{1pt}
    \vspace{1em}
    \begin{tabular}{ll*{9}{C{1.3cm}}l} 
    \toprule 
    \multicolumn{2}{l}{\textbf{Component i $\rightarrow$}}  & 1  & 2 & 3 & 4 & 5 & 6 & 7 & 8 & 9 \\
    
    \textbf{j} & \textbf{Reaction $\downarrow$} & $S_\mathrm{ac}$ & $S_\mathrm{ch4}$ & $S_\mathrm{IC}$ & $S_\mathrm{IN}$ & $X_\mathrm{ch}$ & $X_\mathrm{pr}$ & $X_\mathrm{li}$ & $X_\mathrm{bac}$ & $X_\mathrm{ac}$ &  \textbf{Reaction rate} $r_j$\\ 
    \midrule 
    1 & Fermentation $X_\mathrm{ch}$ & 0.6555 & 0.0818 & 0.2245 & -0.0169 & -1 & & & 0.1125 & & $\theta_1 \,  x_5$\\ 
    2 & Fermentation $X_\mathrm{pr}$ & 0.9947 & 0.0696 & 0.1029 & 0.1746 & & -1 & & 0.1349 & & $\theta_2 \,  x_6$\\
    3 & Fermentation $X_\mathrm{li}$ & 1.7651 & 0.1913 & -0.6472 & -0.0244 & & & -1 & 0.1621 & & $\theta_3 \,  x_7$\\ 
    \hdashline[0.6pt/1.8pt] 
    4 & Methanogenesis $S_\mathrm{ac}$ & -26.5447 & 6.7367 & 18.4808 & -0.1506 & & & & & 1 & $\theta_5 \, \frac{x_1}{\theta_6 + x_1} \, x_9 \, I_{ac}$\\ 
    \hdashline[0.6pt/1.8pt] 
    5 & Decay $X_\mathrm{bac}$ & & & & & 0.18 & 0.77 & 0.05 & -1 & & $\theta_4 \,  x_8$\\
    6 & Decay $X_\mathrm{ac}$ & & & & & 0.18 & 0.77 & 0.05 & & -1 & $\theta_4 \,  x_9$\\
    \midrule  
    \midrule 
    & & 2 & 3 & $\ldots$ & $\ldots$ & 10 & 11 & 12 & 13 & 14 & \\
    & & $S_\mathrm{ch4}$ & $S_\mathrm{IC}$ & &  & $S_\mathrm{ac^-}$ & $S_\mathrm{hco3^-}$ & $S_\mathrm{nh3}$ & $S_\mathrm{ch4,gas}$ & $S_\mathrm{co2,gas}$ & \\
    \midrule 
    7 & Dissociation $S_\mathrm{ac}$ & & & & & -1 & & & & & $c_{29} \, (x_{10} - x_1) + c_9 \, x_{10} \, S_\mathrm{H^+}$\\
    8 & Dissociation $S_\mathrm{IC}$ & & & & & & -1 & & & & $c_{30} \, (x_{11} - x_3) + c_{10} \, x_{11} \, S_\mathrm{H^+}$\\ 
    9 & Dissociation $S_\mathrm{IN}$ & & & & & & & -1 & & & $c_{31} \, (x_{12} - x_4) + c_{11} \, x_{12} \, S_\mathrm{H^+}$\\
    \hdashline[0.6pt/1.8pt]
    10 & Phase transition $S_\mathrm{ch4}$& -1 & & & & & & & $c_{31}$ & & $c_5 \, x_2 - c_6 \, x_{13}$\\
    11 & Phase transition $S_\mathrm{co2}$& & -1 & & & & & & & $c_{31}$ & $c_5 \, (x_3 - x_{11}) - c_7 \, x_{14}$\\ 
    \bottomrule
    \end{tabular}
\end{center}
\end{landscape}
\section{Alternative approaches to address multirate measurements}
\label{sec:multirate_alternatives}
\begin{figure}[htb]
    \centering
    \includegraphics[width=0.95\linewidth]{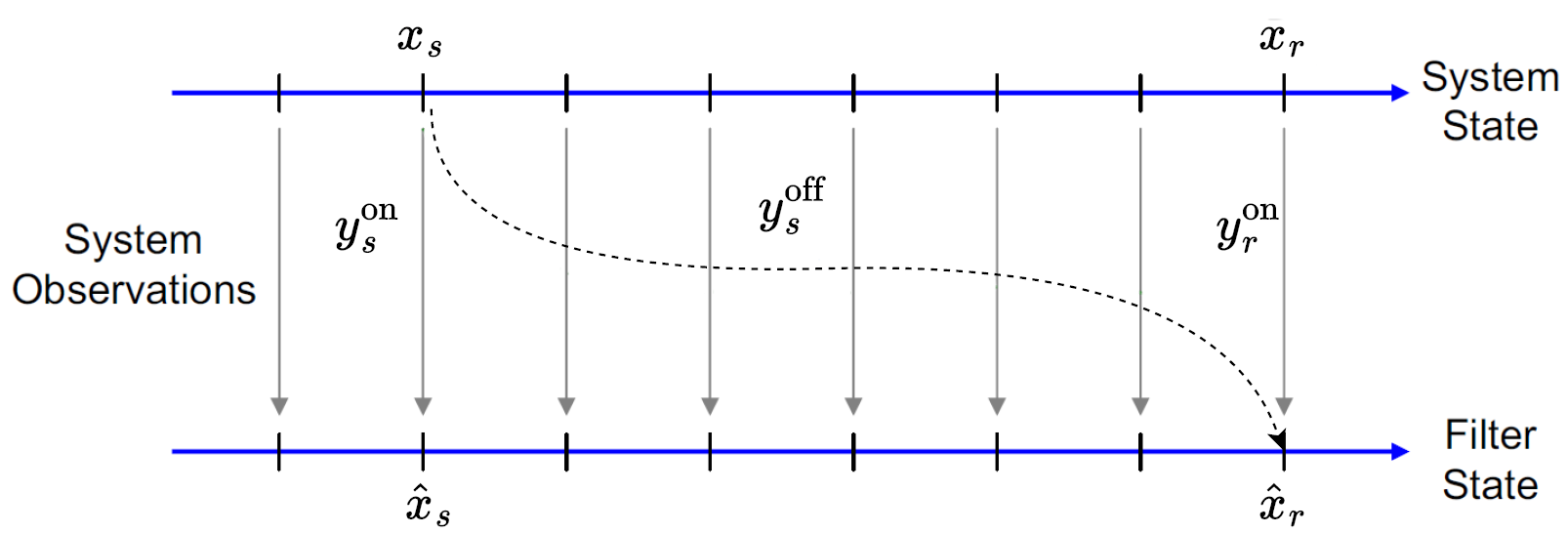}
    \caption{Synchronous online and delayed offline measurements considered in multirate Kalman filtering.%Note that in the figure, the number of delay time steps is denoted as $n$ instead of $N_\mathrm{d}$, and the delayed offline measurement sampled at time $s$ is denoted as $y_{k-n}^\mathrm{off}$.
    }
\label{fig:asynchronous_measurements_setup}
\end{figure}

Delayed offline measurements can be dealt with in a number of ways, only one of which is sample-state augmentation as described in the present paper. In this context, offline measurements are commonly assumed to be sampled at time $s$ but will only return $N_\mathrm{d}$ discrete time steps later at time $r=s+N_\mathrm{d}$, while online measurements are assumed to arrive instantaneously and at high measurement frequency. Therefore, the time period $s \le k \le r$ is referred to as the delay period. The setup is illustrated in Fig.~\ref{fig:asynchronous_measurements_setup}.
\subsection{Filter recalculation}
Perhaps the most intuitive approach to deal with delayed offline measurements is \textit{filter recalculation} as explained by Gopalakrishnan et al (2011) \cite{Gopalakrishnan.2011}. It can generally be applied to linear and nonlinear systems using arbitrary Kalman filters (linear KF, EKF, UKF etc.). Here the entire delay period is passed twice, i.e. once before the delayed offline measurement is available and then a second time after it became available. During the first pass, the a priori estimate $\hat x_s^-$ at time $s$ is saved. Only the online measurements up to time $k=r$ are used during the measurement updates, and all online measurements during the delay period are stored as well. Once the offline measurement of time $s$ becomes available at time $r$, the filter jumps back to the previously stored a-priori estimate $\hat x_s^-$ and performs the measurement update using the full measurement array of time $s$ (which is now available). This delivers the a-posteriori estimate $\hat x_s$. Then the delay period is passed again starting with $\hat x_s$ and using the stored online measurements up to (and including) time $k=r$. The resulting a-posteriori estimate $\hat x_r$ optimally used all possibly available measurements. 

The downside of this approach is twofold: first, it is memory-intense since all online measurements of the delay period must be stored; and second, the filter recalculation might pose a challenge since it must be finished before the next online measurement arrives. This can be especially challenging with a high online sampling rate. 
\subsection{Linear multirate systems}
For linear time-invariant systems  Alexander (1991) proposed a method to account for delayed measurements in linear Kalman filters \cite{Alexander.1991}. The considered systems are denoted in discrete time and of the form: 
\begin{subequations}
\begin{align}
    x_{k+1} &= A_k x_k + B_k u_k + w_k \\
    y_k &= \begin{cases} \quad h^\mathrm{on} (x_k) + v_k^\mathrm{on} & , \,\, k \ne r \\
        \left[ \begin{array}{c} h^\mathrm{on} (x_k) + v_k^\mathrm{on} \\
        h^\mathrm{off} (x_s) + v_s^\mathrm{off}      
        \end{array} \right] & , \,\, k=r, 
    \end{cases} %
    \notag \\
    &= \begin{cases} \quad C_k^\mathrm{on} x_k + v_k^\mathrm{on} & , \quad k \ne r \\
        \left[ \begin{array}{c} C_k^\mathrm{on} x_k + v_k^\mathrm{on} \\
        C_s^\mathrm{off} x_s + v_s^\mathrm{off}      
        \end{array} \right] & , \quad k=r, 
    \end{cases} \label{eq:linear_output_model}
\end{align} 
\end{subequations}
with $r = s+N_\mathrm{d}$. The method of Alexander was simplified and extended by Larsen and Poulsen (1998) \cite{Larsen.1998}, where the measurement noise covariance matrix $R^\mathrm{off}$ of the delayed offline measurements is assumed to be known already at the sample time $s$. For many scenarios with time-invariant (and thus a priori known) measurement covariances, this is a reasonable assumption. The simplified method is still referred to as Alexander's method in the following. The measurement scenario considered for these derivations is illustrated in Figure~\ref{fig:asynchronous_measurements_setup}.
\subsubsection{Alexander's Method}
The central notion in Alexander's method is that at each measurement update of a Kalman filter, the actual values of the measurements $y$ 
%$y^\mathrm{off}$ (sampled at time $s$, returned at time $r$) 
are only required for the measurement update of the state prior $\hat x^-$, but not for the P-matrix $P^-$. Moreover, the Kalman gain matrix $K$ can be computed based on the full array of output equations (i.e., online \textit{and} offline) $ \left[ {h^\mathrm{on}}^T , {h^\mathrm{off}}^T\right]^T$, as long as the values of the measurement noise covariance $R^\mathrm{on}$ and $R^\mathrm{off}$ are known at the time the Kalman gain is to be computed. 

This means that at sample time $s$, the Kalman gain can already be computed \textit{as if the full array of measurements %including the offline measurements 
were available at time $s$}, although the offline measurements won't be available for another $N_\mathrm{d}$ time steps. Alexander denotes this \textit{prematurely} computed Kalman gain as $K_s'$, as opposed to the Kalman gain $K_s$ that relies only on the actually available measurements at time $s$ (i.e., only the online measurements). 

Measurement updates relying on an \textit{incorrect} Kalman gain matrix result in suboptimal state estimates. Therefore, the estimates derived with Kalman gain matrices $K_k'$ during the delay period are corrected at the return time $k=r=s+N_\mathrm{d}$ \textit{after} the online measurements at time $r$ have been fused using the measurement update equations of an ordinary Kalman filter, Eq.~\eqref{eq:EKF_MU}. This is achieved by adding the correction $\delta \hat x_r$ with 
\begin{subequations}
\label{eq:alexander_correction_at_return}
\begin{align}
    \delta \hat x_r &= M_{\delta} K_s \left( y_s^\mathrm{off} - C_s^\mathrm{off} \hat x_s \right) \quad \text{with} \\
    M_{\delta} &= \prod_{i=0}^{N_\mathrm{d}-1} \left( I - K_{r-i}' \, C_{r-i}  \right) A_{r-i-1}.
\end{align}
\end{subequations}
Note that another correction of the P-matrix is not necessary as its value is independent of the (online and offline) measurements and only relies on the measurement model \eqref{eq:linear_output_model}, which is of course always available. However, it is implicitly assumed that the dynamics matrix $A_k$ is not re-evaluated during the delay period. The system is thereby assumed to be time-invariant.

After incorporating the correction \eqref{eq:alexander_correction_at_return} at the return time $r$, the state estimates are again optimal, i.e. they match with the values obtained through filter recalculation. 
\subsubsection{Parallel filters}
Larsen and Poulsen (1998) extended the method of Alexander through a second Kalman filter which during the delay period runs in parallel to the filter according to Alexander. The central notion is the following: 

Alexander's filter only delivers optimal estimates \textit{after} the offline measurement has returned at time~$r$. During all of the delay period, Alexander's filter performs measurement updates with an adapted (and thus suboptimal) Kalman gain.\footnote{Note that in this context a filter is considered as optimal if in an online application it used all measurements available up until the respective time points.} This can be mitigated by bridging the delay period with an ordinary, second Kalman filter which is initialized at the sample time $s$ and is only used to fuse the online measurements during the delay period. At the return time $r$, the estimates of the second Kalman filter are replaced with those of the first Kalman filter which is given by Alexander's filter.

At the expense of double the computational load during the delay period, this delivers optimal estimates at all times \cite{Gopalakrishnan.2011}.
\section{Synchronous and asynchronous extended Kalman filtering}
\subsection{Synchronous Kalman filter updates}
The governing equations of discrete time, synchronous extended Kalman filtering are thoroughly derived e.g. in \cite{Simon.2006}. They shall only be reviewed here for didactic reasons such that the differences with respect to asynchronous Kalman filtering in \ref{sec:asynchronour_KF_updates} become apparent. 
\paragraph{Initialization}
$\hat x_{k-1}$ and $P_{x_{k-1}}$ are initialized e.g. according to \cite{Schneider.2013} or used as tuning parameters.
\paragraph{Time Update}
Discrete-time system dynamics are used, where for ease of notation, dependency on inputs and noise is suppressed
\begin{align}
    \hat x_k^- &= f(\hat x_{k-1}) \\
    P_{x_k}^- &= F_{x_k} P_{x_{k-1}} F_{x_k}^T + Q_k \quad \text{with} \\
    F_{x_k} &= \nabla_x f(x_k) |_{\hat x_k^-}.
\end{align}
\paragraph{Measurement Update}
The Kalman gain $K_k$ and the output linearization $H_{x_k}$ can be computed solely given the information available after the time update. With this the algorithm can proceed with the measurement update 
\begin{align}
    \hat x_k &= \hat x_k^- + K_k \underbrace{(y_k - \hat y_k^-)}_{\tilde y_k} \\
    P_{x_k} &= P_{x_k}^- - K_k \underbrace{H_{x_k} P_{x_k}^-}_{P_{x_k \tilde y_k}^T} = \left[ I - K_k H_{x_k} \right] P_{x_k}^- \left[ I - K_k H_{x_k} \right]^T \label{eq:joseph-form}  \quad \text{with} \\
    K_k &= \underbrace{P_{x_k}^- H_{x_k}^T}_{P_{x_k \tilde y_k}} \underbrace{\left[ H_{x_k} P_{x_k}^- H_{x_k}^T + R_k \right]^{-1}}_{P_{\tilde y_k}^{-1}} \label{eq:synchronous_EKF_Kalman_gain} \\
    H_{x_k} &= \nabla_x h(x_k) |_{\hat x_k^-}.
\end{align} 
The second term of \eqref{eq:joseph-form} is called the Joseph form of the measurement update and ensures symmetry and positive definiteness of the P-matrix.  
\subsection{Asynchronous Kalman filter updates}
\label{sec:asynchronour_KF_updates}
Asynchronous updates are more difficult to deal with as they involve to fuse measurements at time $k$ that were taken $N_\mathrm{d}$ time steps before. In this case, the correct update equations must be 
\begin{subequations}
\begin{align}
    \hat x_k &= \hat x_k^- + K_k^{\textbf{as}} \underbrace{(y_{k-N_\mathrm{d}} - \hat y_{k-N_\mathrm{d}}^-)}_{\tilde y_{k-N_\mathrm{d}}} \\
    P_{x_k} &= P_{x_k}^- - K_k^{\textbf{as}} P_{\tilde y_k \tilde y_{k-N_\mathrm{d}}} {K_k^{\textbf{as}}}^T, \quad \text{with} \\
    K_k^{\textbf{as}} &= P_{x_k \tilde y_{k-N_\mathrm{d}}} \, P_{\tilde y_{k-N_\mathrm{d}}}^{-1}. \label{eq:asynchronous_KF_kalman_gain} 
\end{align}
\end{subequations}
The term $P_{\tilde y_{k-N_\mathrm{d}}}$ in \eqref{eq:asynchronous_KF_kalman_gain} can be computed from standard EKF equations by consequently using values from time $k-N_\mathrm{d}$ in \eqref{eq:synchronous_EKF_Kalman_gain}. Nevertheless, the former term is difficult to compute since it contains cross covariances between the current state and the measurement taken at the sample time $k-N_\mathrm{d}$. For the EKF, it may be approximated as follows
\begin{align}
    P_{x_k \tilde y_k} &= E\{ (x_k - \hat x_k^-) \, (y_{k-N_\mathrm{d}} - \hat y_{k-N_\mathrm{d}}^-)^T \} \\
    &\approx E\{ (x_k - \hat x_k^-) \, (x_{k-N_\mathrm{d}} - \hat x_{k-N_\mathrm{d}}^-)^T \} H_{x_{k-N_\mathrm{d}}}^T  \\
    &= P_{x_k x_{k-N_\mathrm{d}}}^- H_{x_{k-N_\mathrm{d}}}^T.
\end{align}
The problem is still to get an expression for $P_{x_k x_{k-N_\mathrm{d}}}^-$ which involves the cross covariance between the states estimates at the current time $k$ and the sample time $k-N_\mathrm{d}$. It may be possible through application of a cumulative transformation matrix $\Phi$ as proposed by Larsen \cite{Larsen.1998} and as thoroughly described in \cite{vanderMerwe.2004b}. However, this is strictly valid for linear systems only, and thus only an approximation for nonlinear systems in the context of extended Kalman filtering. The same holds for the term $P_{\tilde y_k \tilde y_{k-N_\mathrm{d}}}$.
% 

%% References
\sloppy
\bibliography{AD_literature_2025}

\end{document}